\documentclass[a4paper,11pt]{article}

\pdfoutput=1

\usepackage{jheppub}

\usepackage[T1]{fontenc} 
\usepackage{longtable}
\usepackage{multirow}
\usepackage[small]{subfigure}

\title{\boldmath Bottom and Charm Mass Determinations with a Convergence Test}

\preprint{\begin{flushright} UWThPh-2015-09, LPN15-024\end{flushright}\vspace*{-1cm}}

\author[a]{Bahman Dehnadi,}
\author[a,b]{Andr\'e H. Hoang}
\author[a]{and Vicent Mateu}
\affiliation[a]{Fakult\"at f\"ur Physik, Universit\"at Wien,
Boltzmanngasse 5, 1090 Vienna, Austria}
\affiliation[b]{Erwin Schr\"odinger International Institute for Mathematical Physics,
University of Vienna, Boltzmanngasse 9, A-1090 Vienna, Austria}

\emailAdd{bahman.dehnadi@univie.ac.at}
\emailAdd{andre.hoang@univie.ac.at}
\emailAdd{vicent.mateu@univie.ac.at}

\abstract{
We present new determinations of the {$\overline{\rm MS}$} charm quark mass using relativistic
QCD sum rules at ${\cal O}(\alpha_s^3)$ from the moments of the vector and the pseudoscalar current 
correlators. We use available experimental measurements from $e^+\,e^-$ collisions and 
lattice simulation results, respectively. Our analysis of the theoretical uncertainties is based on 
different implementations of the perturbative series and on independent variations of the 
renormalization scales for the mass and the strong coupling. Taking into account the resulting
set of series to estimate perturbative uncertainties is crucial, since some ways to treat the 
perturbative expansion can exhibit extraordinarily small scale dependence when the two scales are 
set equal. As an additional refinement, we address the issue that double scale variation could 
overestimate the perturbative uncertainties. We supplement the analysis with a test that quantifies 
the convergence rate of each perturbative series by a single number. We find that this convergence 
test allows to determine an overall and average convergence rate that is characteristic for the 
series expansions of each moment, and to discard those series for which the convergence rate is 
significantly worse. We obtain \mbox{$\overline m_c(\overline m_c) = \,1.288\,\pm 0.020\,$GeV} from 
the vector correlator. The method is also applied to the extraction of the {$\overline{\rm MS}$} 
bottom quark mass from the vector correlator. We compute the experimental moments including a 
modeling uncertainty associated to the continuum region where no data is available. We obtain 
$\overline m_b(\overline m_b) = \,4.176 \, \pm 0.023\,$GeV.
}

\begin{document}
\maketitle
\flushbottom

\section{Introduction}
\label{sec:intro}

Precise and reliable determinations of the charm and bottom quark masses are an important input for 
a number of theoretical predictions, such as Higgs branching ratios to charm and bottom quarks or 
for the corresponding Yukawa couplings \cite{Heinemeyer:2013tqa, Petrov:2015jea}. They also affect 
the theoretical predictions of radiative and inclusive B decays, as well as rare kaon decays. For 
example, the inclusive semileptonic decay rate of B mesons depends on the fifth power of the bottom 
quark mass. These weak decays provide crucial methods to determine elements of the CKM matrix,
which in turn are important for testing the validity of the Standard Model, as well as for 
indirect searches of new physics. In this context, having a reliable estimate of uncertainties for 
the quark masses is as important as knowing their precise values~\cite{Antonelli:2009ws}. Due to 
confinement quark masses are not physical observables. Rather, they are scheme-dependent parameters 
of the QCD Lagrangian which have to be determined from quantities that strongly depend on them.

One of the most precise tools to determine the charm and bottom quark masses is the QCD sum rule 
method, where weighted averages of the normalized cross section $R_{e^+e^-\to\, q\bar{q}\,+X}$, 
with $q = c, b$,

\begin{align}
\label{eq:momentdefvector}
& M_n^V = \!
\int\!\dfrac{{\rm d}s}{s^{n+1}}R_{e^+e^-\to\, q\bar{q}\,+X}(s)\,,\qquad
R_{e^+e^-\to\, q\bar{q}\,+X}(s) = \dfrac{\sigma_{e^+e^-\to\,
q\bar{q}\,+X}(s)}{\sigma_{e^+e^-\to\,\mu^+\mu^-}(s)}\,, 
\end{align}

can be related to moments of the quark vector current correlator $\Pi_V$~\cite{Shifman:1978bx, 
Shifman:1978by}:

\begin{align}
&M_n^{V,\,\rm th} =\dfrac{12\pi^2 Q_q^2}{n!}\,\dfrac{{\rm d}^n}{{\rm 
d}s^n}\Pi_V(s)\Big|_{s=0}\,,\quad\,
j^{\mu}(x) = \bar{q}(x)\gamma^\mu q(x)
\,,\nonumber \\
&\big(g_{\mu\nu}\,s-q_{\mu}q_{\nu}\big)\Pi_V(s)
= -\, i\!\int\!\mathrm{d}x\, e^{iqx}\left\langle \,0\left|T\,
j_{\mu}(x)j_{\nu}(0)\right|0\,\right\rangle.
\end{align}

Here $Q_q$ is the quark electric charge and $\sqrt{s} = \sqrt{q^2}$ is the $e^+e^-$ center-of-mass 
energy. Given that the integration over the experimental $R$-ratio extends from the quark pair 
threshold up to infinity {\it but} experimental measurements only exist for energies up to around 
$11$\,GeV, one relies on using theory input for energies above that scale (which we call the 
``continuum'' region). For the charm moments, the combination of all available measurements 
is actually sufficient to render the experimental moments essentially independent of uncertainties 
one may assign to the theory input for the continuum region~\cite{Dehnadi:2011gc}. For the bottom 
moments, the dependence on the continuum theory input is very large, and the dependence of the 
low-$n$ experimental moments on unavoidable assumptions about the continuum uncertainty can be the 
most important component of the error budget, see e.g.~\cite{Corcella:2002uu}. In fact, the use of 
the first moment $M_1^V$ to determine the bottom mass appears to be excluded until more 
experimental data becomes available for higher energies.

Alternatively one can also consider moments of the pseudoscalar current correlator to extract the 
heavy quark masses. Experimental information on the pseudoscalar correlator $\Pi_P$ is not 
available in a form useful for quark mass determinations, but for the charm quark very precise 
lattice calculations have become available recently~\cite{Allison:2008xk}. For the pseudoscalar 
correlator it turns out that the first two Taylor coefficients in the small-$q^2$ expansion need to 
be regularized and defined in a given scheme, and that the third term (which we will denote by 
$M_0^P$) is hardly sensitive to $m_q$. We adopt the definitions

\begin{align}
\label{eq:momentdefpseudo}
&\Pi_P(s) = i\!\int\!\mathrm{d}x\, e^{iqx}\left\langle \,0\left|T\,
j_P(x)j_P(0)\right|0\,\right\rangle ,\quad j_P(x) = 2\,m_q\,i\,\bar{q}(x)\gamma_5 q(x)\,,\\
&M_n^{P,\,\rm th} =\dfrac{12\pi^2 Q_q^2}{n!}\,\dfrac{{\rm d}^n}{{\rm d}s^n} P(s)\Big|_{s=0}\,, 
\qquad P(s) = \dfrac{\Pi_P(s) - \Pi_P(0) - \Pi^\prime_P(0)\,s }{s^2}\,,\nonumber
\end{align}

where the explicit mass factor in the definition of the pseudoscalar current ensures it is 
renormalization-scheme independent.

For small values of $n$ such that $m_q/n\gtrsim\Lambda_{{\rm QCD}}$, the theoretical moments for 
the vector and pseudoscalar correlators can be computed  in the framework of the OPE, i.e.\ as an 
expansion in vacuum matrix elements involving operators of increasing 
dimension~\cite{Shifman:1978bx, Shifman:1978by}. The leading matrix element corresponds to the 
perturbative QCD computations, which greatly dominates the series. Nonperturbative corrections are 
parametrized by vacuum condensates, and we find that even the leading correction, given by the gluon 
condensate, has a very small effect for low $n$, particularly so for the bottom correlator. For 
moments at low values of $n$, it is mandatory to employ a short-distance mass scheme such as 
$\overline{\rm MS}$~\cite{Kuhn:2001dm}, which renders the quark mass $\overline m_q(\mu_m)$ 
dependent on its renormalization scale $\mu_m$, similar to the strong coupling 
$\alpha_s(\mu_\alpha)$, which depends on $\mu_\alpha$. This method of determining the heavy quark 
masses with high precision is frequently called relativistic charmonium/bottomonium sum rules.

For the perturbative term, the exact analytic expressions for the $\Pi$ functions are known at 
${\mathcal O}(\alpha_s^0)$ and ${\mathcal O}(\alpha_s)$, \cite{Kallen:1955fb}. Therefore any 
moment can be obtained simply by Taylor expanding around $q^2 = 0$. At ${\mathcal O}(\alpha_s^2)$ 
moments are known to up to $n=30$~\cite{Chetyrkin:1995ii, Chetyrkin:1996cf, Boughezal:2006uu, 
Czakon:2007qi,Maier:2007yn}. At ${\mathcal O}(\alpha_s^3)$ they are known analytically for $n=1$ 
\cite{Chetyrkin:2006xg, Boughezal:2006px, Sturm:2008eb}, $n=2$ and $n=3$ (and even $n=4$ for the 
pseudoscalar correlator) \cite{Maier:2008he, Maier:2009fz}. Higher moments at ${\cal O}(\alpha_s^3)$ 
have been determined by a semi-analytical procedure \cite{Hoang:2008qy, Kiyo:2009gb, Greynat:2010kx, 
Greynat:2011zp}. The Wilson coefficient of the gluon condensate contribution is known to ${\mathcal 
O}(\alpha_{s})$ \cite{Broadhurst:1994qj}.

The most recent determinations of the $\overline{\rm MS}$ charm mass from charmonium sum rules for 
the vector correlator~\cite{Dehnadi:2011gc, Bodenstein:2011ma, Chetyrkin:2009fv} obtain very 
accurate results, but differ in the way they estimate theoretical uncertainties, and also in the 
computation of the moments from experimental data. Concerning the estimate of the perturbative 
uncertainties, Ref.~\cite{Dehnadi:2011gc} obtained $19$\,MeV compared to $1$ and $2$\,MeV obtained 
in Refs.~\cite{Bodenstein:2011ma} and \cite{Chetyrkin:2009fv}, respectively. The discrepancy arises 
from two differences. First, in Refs.~\cite{Bodenstein:2011ma, Chetyrkin:2009fv} the renormalization 
scales $\mu_m$, and $\mu_\alpha$ were set equal, while in Ref.~\cite{Dehnadi:2011gc} it was argued 
that they should be varied independently. Second, in Refs.~\cite{Bodenstein:2011ma, 
Chetyrkin:2009fv} the lowest renormalization scale was chosen to be $2$\,GeV, while in 
Ref.~\cite{Dehnadi:2011gc} variations down to the charm mass value were considered. For the case of 
the pseudoscalar moments, the HPQCD collaboration has made a number of very accurate predictions 
for charm and bottom masses \cite{Allison:2008xk, McNeile:2010ji, Chakraborty:2014aca, 
Colquhoun:2014ica}, the last of which has the smallest uncertainty claimed so far for the charm 
mass. In all these analyses the renormalization scales $\mu_m$, and $\mu_\alpha$ are set equal when 
estimating the truncation uncertainty. A detailed discussion on the estimates of theoretical and 
experimental uncertainties can be found in Secs.~\ref{sec:previous-results} and 
\ref{sec:comparison-masses} of this article (see also Ref.~\cite{Dehnadi:2011gc}).

Similarly, bottomonium sum rules have been used to determine the bottom mass from low-$n$ moments. 
To the best of our knowledge, the most recent and precise determinations are from 
Refs.~\cite{Bodenstein:2012, Chetyrkin:2009fv}. These two analyses estimate their perturbative 
uncertainties in the same way as the corresponding charm mass extractions from the same 
collaborations~\cite{Bodenstein:2011ma,Chetyrkin:2009fv}. Furthermore, when it comes to compute 
the experimental moments, they use theoretical input at $\mathcal{O}(\alpha_s^3)$ with perturbative 
uncertainties to model the high-energy region (continuum region) of the spectrum. As we discuss in 
this work, similar caveats as for their charm analyses can be argued to also affect their bottom 
quark results.

In this work we revisit the charmonium sum rules for the vector correlator, refining our 
perturbative error estimate from Ref.~\cite{Dehnadi:2011gc} by incorporating a convergence test.
The convergence test addresses the issue that the independent variation of $\mu_m$ and $\mu_\alpha$ 
together with the relatively large value of the $\alpha_s$ close to the charm mass scale, might 
lead to an overestimate of the perturbative uncertainty. The convergence test allows to quantify 
the convergence property of each perturbative series with a single parameter and to discard series 
for which the convergence is substantially worse then for the rest of the series. We show that this 
procedure is meaningful, since the complete set of series for the moments shows a strongly peaked 
distribution in these convergence values, which allows to define an overall convergence for the set 
of perturbative series. This leads to a reduction of the perturbative uncertainty from $19$ to 
$14$\,MeV, and the corresponding result for the $\overline{\rm MS}$ charm mass supersedes the 
main result given in Ref.~\cite{Dehnadi:2011gc}. We also apply this improved method of estimating 
theory uncertainties to obtain a new charm mass determination from the pseudoscalar correlator, and 
to extract the bottom mass from the vector correlator. For the latter, we compute the bottom 
experimental moments by combining contributions from narrow resonances, experimental data taken in 
the continuum, and a theoretical model for the continuum region. We carefully study the 
assignment of adequate uncertainties to this last contribution, to make sure that the the model 
dependence is reduced to an acceptable level.

This paper is organized as follows: In Sec.~\ref{sec:theory} we summarize the theoretical framework 
introduced in \cite{Dehnadi:2011gc}, and adapt it to cover the case of the pseudoscalar moments. 
We also introduce the ratios of moments, also used before in Ref.~\cite{Kuhn:2001dm}, and the 
perturbative expansions associated to them. 
Sec.~\ref{sec:previous-results} contains a brief summary of the results obtained in 
\cite{Dehnadi:2011gc}, and the discussion is extended to the case of the pseudoscalar correlator 
and the bottom mass. In Sec.~\ref{sec:convergence} we introduce the convergence test, and discuss 
how it allows to identify and discard series with a bad convergence. Sec.~\ref{sec:lattice-data} 
contains a discussion on the lattice simulation results we use for our analysis. In 
Sec.~\ref{sec:exp} we present our computation of the experimental moments for the bottom 
correlator. The results are compared to previous determinations in Sec.~\ref{sec:comp-exp}. The 
computation of the ratio of experimental moments is presented in Sec.~\ref{sec:exp-ratio}. Our final 
results for the quark masses are given in Sec.~\ref{sec:results}. The results are compared to 
previous charm and bottom mass analyses in Sec.~\ref{sec:comparison-masses}. We present our 
conclusions in Sec.~\ref{sec:conclusions}. In Appendix~\ref{app:coefs} the numerical values of the 
perturbative coefficients that enter into our analysis and are not yet provided by 
Ref.~\cite{Dehnadi:2011gc} are collected for the convenience of the reader.

\section{Theoretical Input}
\label{sec:theory}

\subsection{Perturbative Contribution}
\label{sec:perturbative}

The moments of the vector and pseudoscalar current correlators are defined in
Eqs.~(\ref{eq:momentdefvector}) and (\ref{eq:momentdefpseudo}), respectively. In the OPE framework 
they are dominated by the perturbative contribution (that is, a partonic computation), which 
exhibits a nonlinear dependence on the mass. Within perturbation theory one can decide to manipulate 
the series expansion to get a more linear dependence on the mass. Conceptually there is no 
preference. As advocated in our previous analyses~\cite{Dehnadi:2011gc}, one might consider various 
versions of the expansion to reliably estimate the perturbative uncertainties. Four types of 
expansion were suggested in Ref.~\cite{Dehnadi:2011gc}, which we briefly review below.

\vskip 5mm
\noindent
{\bf (a) Standard fixed-order expansion}\\

We write the perturbative vacuum polarization function as

\begin{eqnarray}
\label{eq:Mnpertfixedorder1}
\widehat\Pi_X(s, n_f, \alpha_s^{(n_f)}(\mu_\alpha), \overline m_q(\mu_m), 
\mu_\alpha, \mu_m) 
\, = \, \dfrac{1}{12\pi^2 Q_q^2}\sum_{n=0}^\infty s^{n} \hat M^X_n
\,,
\end{eqnarray}

where $X = V, P$ for vector and pseudoscalar currents, respectively. Note that for notation 
reasons, in Eqs.~(\ref{eq:Mnpertfixedorder1}, \ref{eq:Mnpertcontour1}, \ref{eq:Mnpertcontour2}, 
\ref{eq:Pi0msbar}) we use $\Pi_P(q^2) = P(q^2)$, where $P$ is the twice-subtracted pseudoscalar 
correlator defined in Eq.~(\ref{eq:momentdefpseudo}). Here $Q_q$ is the quark electric charge with 
$q = c, b$, and $n_f = 4,5$ for charm and bottom, respectively. In full generality, the 
perturbative moments $\hat M_n$ can be expressed as the following sum:

\begin{eqnarray}
\label{eq:Mn-theo-FO}
\hat M^X_n & = &
\frac{1}{[4\,\overline m^{\,2}_q(\mu_m)]^n}
\sum_{i,a,b} \bigg(\frac{\alpha^{(n_f)}_s(\mu_\alpha)}{\pi}\bigg)^{\!i}
[C_X(n_f)]^{a,b}_{n,i}\,\ln^a\!\bigg(\frac{\overline{m}^{\,2}_q(\mu_m)}{\mu^2_m}\bigg)\!
\ln^b\!\bigg(\frac{\overline{m}^{\,2}_q(\mu_m)}{\mu^2_\alpha}\bigg).
\end{eqnarray}

This is the standard fixed-order expression for the perturbative moments. The numerical values for 
the $[C_V(n_f = 5)]^{a,b}_{n,i}$ coefficients are given in Table~\ref{tab:cfixedorder} (for the 
vector current with $n_f = 4$, these coefficients can be found in Table~1 of 
Ref.~\cite{Dehnadi:2011gc}). Likewise, $[C_P(n_f = 4)]^{a,b}_{n,i}$ are collected in a numerical 
form in Table~\ref{tab:cPfixedorder}.

The expression in Eq.~(\ref{eq:Mn-theo-FO}) is the common way to write the perturbative series of 
the moments. However, as noted in Ref.~\cite{Dehnadi:2011gc}, the nonlinear dependence on 
$\overline m_q$ of the standard fixed-order expansion has the disadvantage that for charm (bottom) 
quarks there are frequently no solutions for the mass in the sum rule mass determination, for 
moments higher than the first (second), for some set of values of the renormalization scales.

\vskip 5mm
\noindent
{\bf (b) Linearized expansion}\\

One can linearize the the fixed-order form expansion of Eq.~(\ref{eq:Mn-theo-FO}) with respect to 
the exponent of the quark mass pre-factor by taking the \mbox{$2n$-th} root. This choice is e.g.\ 
made in Ref.~\cite{McNeile:2010ji}, and in general one can write:

\begin{align}
\label{eq:Mn-theo-exp}
\Big(\hat M^X_n\Big)^{\!1/2n} =
\frac{1}{2\,\overline{m}_q(\mu_m)} 
\,\sum_{i,a,b}\bigg(\frac{\alpha^{(n_f)}_s(\mu_\alpha)}{\pi}\bigg)^{\!i}
[\bar C_X(n_f)]_{n,i}^{a,b}\,\ln^a\!\bigg(\frac{\overline{m}^{\,2}_q(\mu_m)}{\mu^2_m}\bigg)\!
\ln^b\!\bigg(\frac{\overline{m}^{\,2}_q(\mu_m)}{\mu^2_\alpha}\bigg).
\end{align}

The coefficients $[\bar C_V(n_f = 5)]_{n,i}^{a,b}$ and $[\bar C_P(n_f = 4)]_{n,i}^{a,b}$ are given 
in Tables~\ref{tab:ctildefixedorder} and \ref{tab:cPtildefixedorder}, respectively (for $n_f = 4$ 
the coefficients for the vector current can be found in Table~2 of Ref.~\cite{Dehnadi:2011gc}). 
Even though relation (\ref{eq:Mn-theo-exp}) still exhibits some nonlinear dependence on ${\overline 
m}_q$ through perturbative logarithms, we find that it always has a solution for the quark mass.

\vskip 5mm
\noindent
{\bf (c) Iterative linearized expansion}\\

For the expansion methods (a) and (b) shown in Eqs.~(\ref{eq:Mn-theo-FO}) and 
(\ref{eq:Mn-theo-exp}), one solves for the quark masses $\overline m_{c,b}(\mu_m)$ numerically 
keeping the exact mass dependence on the theory side of the equation. Alternatively, one can solve 
for $\overline m_{c,b}(\mu_m)$ iteratively order by order, which is perturbatively equivalent to 
the exact numerical solution, but gives different numerical results. The method consists of 
inserting the lower order values for $\overline m_{c,b}(\mu_m)$ in the higher order perturbative 
coefficients, and re-expanding consistently. This method has been explained in detail in Sec. 2.1(c) 
of 
Ref.~\cite{Dehnadi:2011gc}, and we only quote the final results here:

\begin{align}\label{eq:iterative-general}
{\overline m}_q(\mu_m) & =\,{\overline m}_q^{(0)}
\sum_{i,a,b}\bigg(\frac{\alpha^{(n_f)}_s(\mu_\alpha)}{\pi}\bigg)^{\!i} [\tilde 
C_X(n_f)]_{n,i}^{a,b}\,
\ln^a\!\bigg(\frac{{\overline m}_q^{(0)\,2}}{\mu^2_m}\bigg)  \! 
\ln^b\!\bigg(\frac{{\overline m}_q^{(0)\,2}}{\mu^2_\alpha}\bigg),\\
\overline m_q^{(0)}(\mu_m) &=
\frac{1}{2\big(M^X_n\big)^{1/2n}} \, [\tilde C_X(n_f)]_{n,0}^{0,0}\,,\nonumber
\end{align}

where the numerical value of the coefficients $[\tilde C_V(n_f=5)]_{n,i}^{a,b}$ and $[\tilde 
C_P(n_f=4)]_{n,i}^{a,b}$ are collected in Tables~\ref{tab:vector-it} and \ref{tab:cPhat}, and the 
values for the vector current with $n_f = 4$ can be found in Table~3 of Ref.~\cite{Dehnadi:2011gc}.

By construction, the iterative expansion always has a solution for the quark mass. Accordingly, 
potential biases on the numerical analysis related to any possible nonlinear dependence are 
eliminated.

\vskip 5mm
\noindent
{\bf (d) Contour-improved expansion}\\

\begin{figure}[t]
\center
 \includegraphics[width=0.3\textwidth]{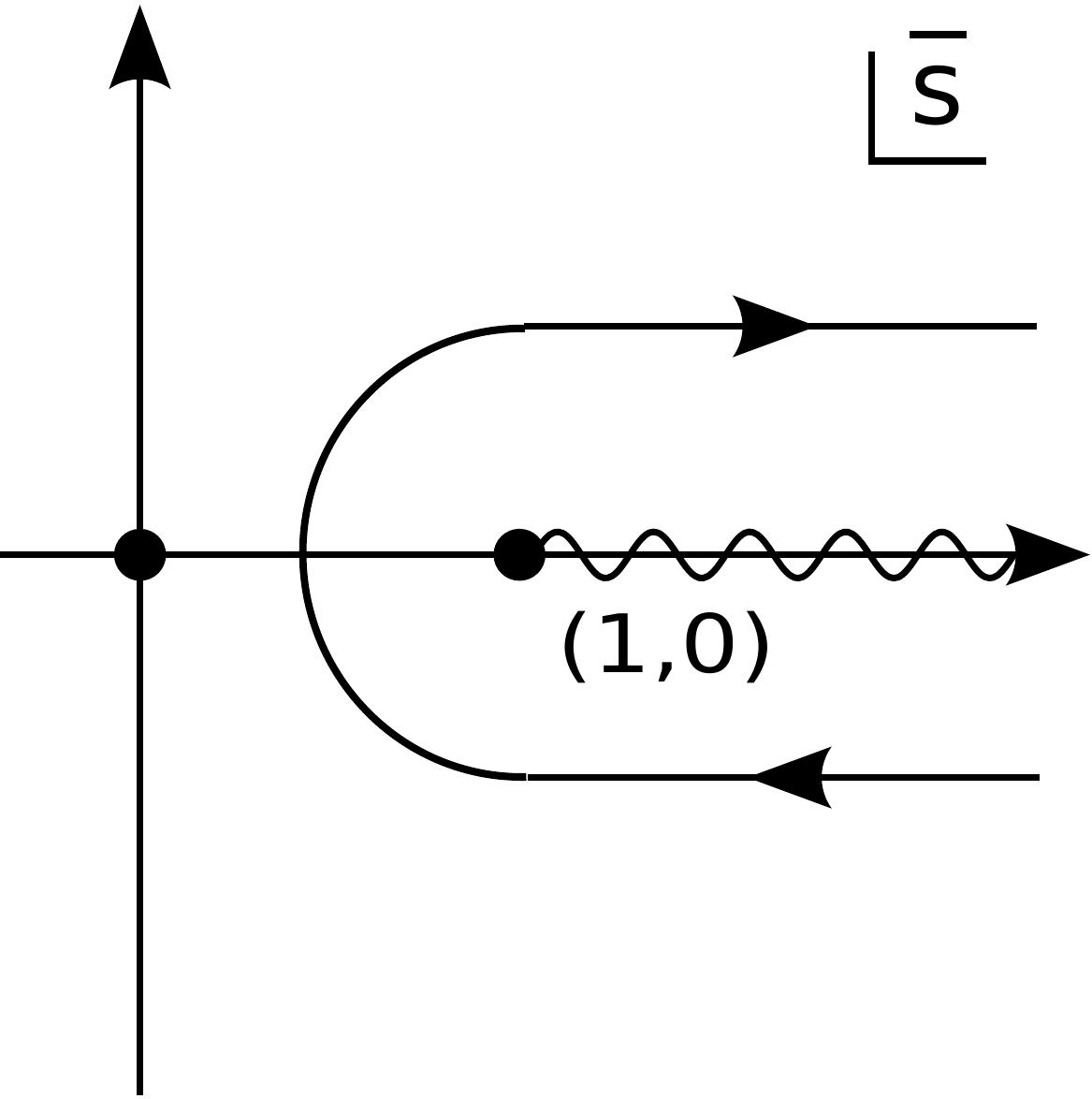}
 \caption{
One possible integration path in the complex \mbox{$\bar s$-plane} for the computation of the 
contour-improved moments.\label{fig:contour} }
\end{figure}

For the expansions (a), (b) and (c) the moments and the quark masses are computed for a fixed 
value of the renormalization scale $\mu_\alpha$. Using the analytic properties of the vacuum 
polarization function, one can rewrite the fixed-order moments as integrals in the complex plane. 
This opens the possibility of making $\mu_\alpha$ dependent on the integration variable, in analogy 
to the contour-improved methods used for \mbox{$\tau$-decays} (see e.g.\ 
Refs.~\cite{LeDiberder:1992te, Pivovarov:1991rh, Braaten:1991qm, Narison:1988ni, Braaten:1988ea, 
Braaten:1988hc}). Therefore we define the contour-improved moments~\cite{Hoang:2004xm} as (see 
Fig.~\ref{fig:contour}),

\begin{eqnarray}
\label{eq:Mnpertcontour1}
\hat M_n^{X,\mathcal{C}} & = &
\frac{6\pi Q_q^2}{i}\,\int_\mathcal{C}\,\frac{{\rm d}s}{s^{n+1}}
\widehat\Pi_X(s, n_f, \alpha_s^{(n_f)}(\mu_\alpha^c(s,\overline m_q^{\,2})), \overline{m}_q(\mu_m),
\mu_\alpha^q(s,\overline m_q^{\,2}), \mu_m)
\,,
\end{eqnarray}

and we employ the following path-dependent $\mu_\alpha^c$, first used in Ref.~\cite{Hoang:2004xm}

\begin{eqnarray}
\label{mualphacontour}
(\mu_\alpha^q)^2(s,\overline m_q^{\,2}) & = &
\mu_\alpha^2\,\bigg(\,1-\frac{s}{4\,\overline m_q^{\,2}(\mu_m)}\,\bigg) 
\,.
\end{eqnarray}

It weights in a different way the threshold versus the high energy parts of the spectrum. It was 
shown in Ref.~\cite{Dehnadi:2011gc} that the resulting moments $\hat M_n^{X,\mathcal{C}}$ can be 
obtained analytically from the Taylor expansion around $s=0$ of the vacuum polarization function 
using an $s$-dependent $\mu_\alpha^c(s,\overline m_q^{\,2})$:

\begin{eqnarray}
\label{eq:Mnpertcontour2}
\widehat\Pi_X^{\overline {\rm MS}}\Big(s,\alpha^{(n_f)}_s(\mu_\alpha^c(s,\overline m_q^{\,2})),
\overline m_q(\mu_m), \mu_\alpha^c(s,\overline m_q^{\,2}),
\mu_m\Big) & = & \sum\limits_{n=0}^\infty \, s^{n}\,\hat M_n^{X,\mathcal{C}}
\,.
\end{eqnarray}

This trick works because $\alpha_s(\mu_\alpha^c(s,\overline m_q^{\,2}))$ has the same cut as the 
fixed-order expression for $\widehat\Pi_X$. Other choices could spoil this property. Expanding the 
analytic expression for $\hat M_n^{X,\mathcal{C}}$ on $\alpha_s$ at a given finite order, one 
recovers the fixed-order moments $\hat M_n^X$. This shows that the dependence on the contour is 
only residual and represents an effect from higher order terms from beyond the order one employs 
for the calculation.

The contour-improved moments have a residual sensitivity to the value of the vacuum polarization 
function at zero momentum transfer.\footnote{This means that the dependence vanishes in the 
large-order limit.} For the case of the vector correlator this value depends on the UV-subtraction 
scheme and corresponds to $\widehat\Pi(0)=\hat M_0^V$. For the case of the pseudoscalar correlator, 
$\hat M_0^P$ is scheme-independent, since $P(q^2)$ already includes two UV subtractions. However, 
one could as well define a three-times-subtracted pseudoscalar correlator of the form $\overline 
P(q^2) = P(q^2) -  P(0)$. Slightly abusing notation, we denote $\overline P$ as the ``on-shell'' 
scheme for $P(q^2)$, and the twice subtracted (original) definition as the $\overline{\rm MS}$ 
scheme for $P(q^2)$. Using the OS scheme with $\widehat\Pi^X(0)=0$ for either vector or 
pseudoscalar correlator, we find that the first moment for the contour-improved expansion gives 
exactly the first fixed-order moment, $\hat M_1^{X,\mathcal{C}} = \hat M_1^X$. Thus, in order to 
implement a non-trivial modification, and following Ref.~\cite{Dehnadi:2011gc}, we employ the 
$\overline{\rm MS}$ scheme for $\widehat\Pi_V(0)$ defined for $\mu={\overline m}_q({\overline 
m}_q)$, and the twice-subtracted expression for $P(q^2)$. Generically it can be written as

\begin{eqnarray}
\label{eq:Pi0msbar}
\widehat\Pi_X^{\overline{\rm MS}}(0,n_f) & = &
\sum_{i,a,b} \bigg(\frac{\alpha^{(n_f)}_s(\mu_\alpha)}{\pi}\bigg)^{\!i}
[C_X(n_f)]^{a,b}_{0,i}\,\ln^a\!\bigg(\frac{\overline{m}^{\,2}_q(\mu_m)}{\mu^2_m}\bigg)\!
\ln^b\!\bigg(\frac{\overline{m}^{\,2}_q(\mu_m)}{\mu^2_\alpha}\bigg).
\end{eqnarray}

The numerical values for the coefficients $[C_X]^{a,b}_{0,i}$ are collected in 
Table~\ref{tab:cPi0vec} for the vector correlator with $5$ flavors and the pseudoscalar correlator 
with $4$ flavors. In Table~4 or Ref.~\cite{Dehnadi:2011gc} one finds the the numerical values of 
$[C_V(n_f = 4)]^{a,b}_{0,i}$.

\begin{figure*}[t]
\subfigure[]
{\includegraphics[width=0.5\textwidth]{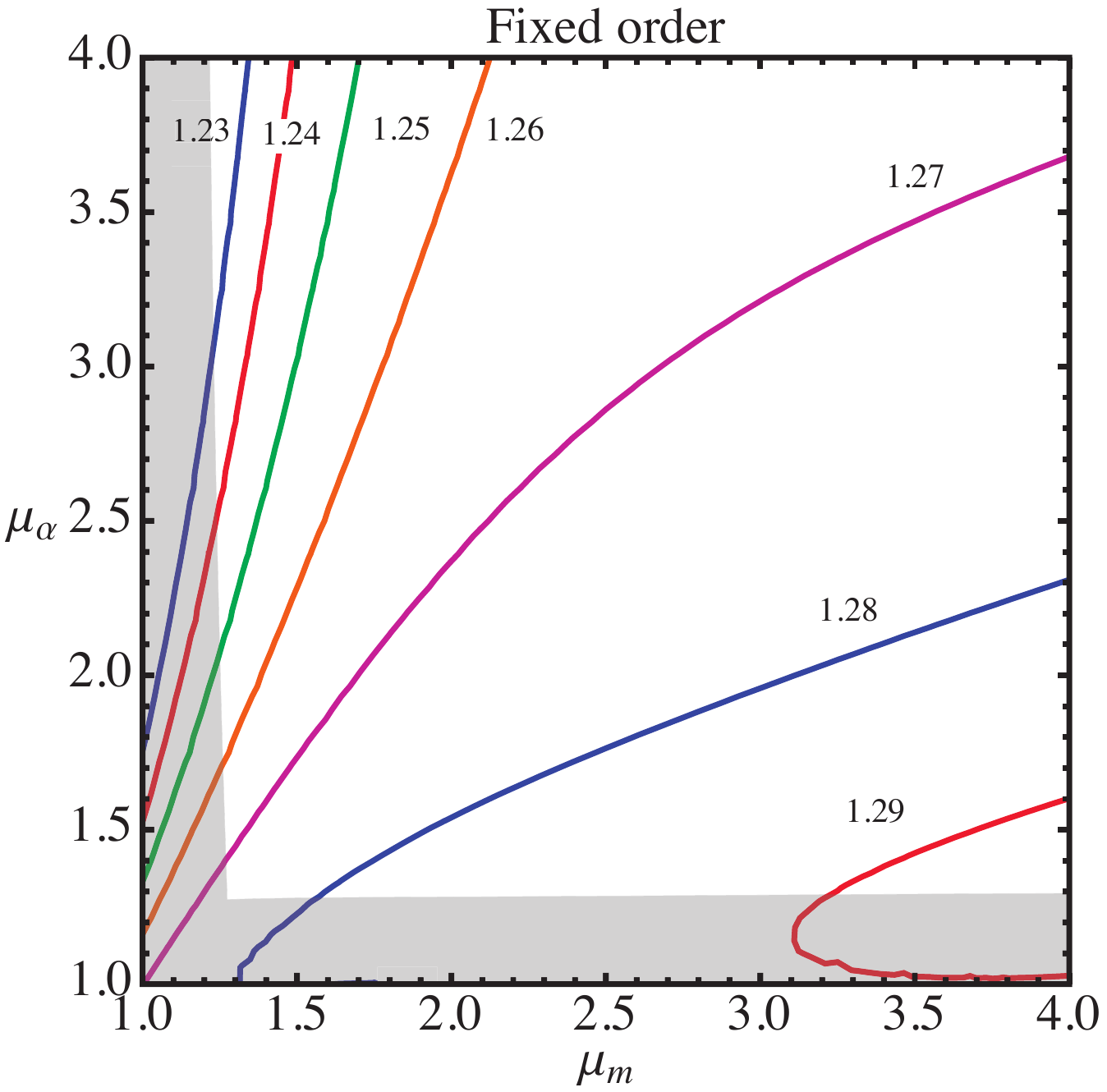}}
 \subfigure[]
{\includegraphics[width=0.5\textwidth]{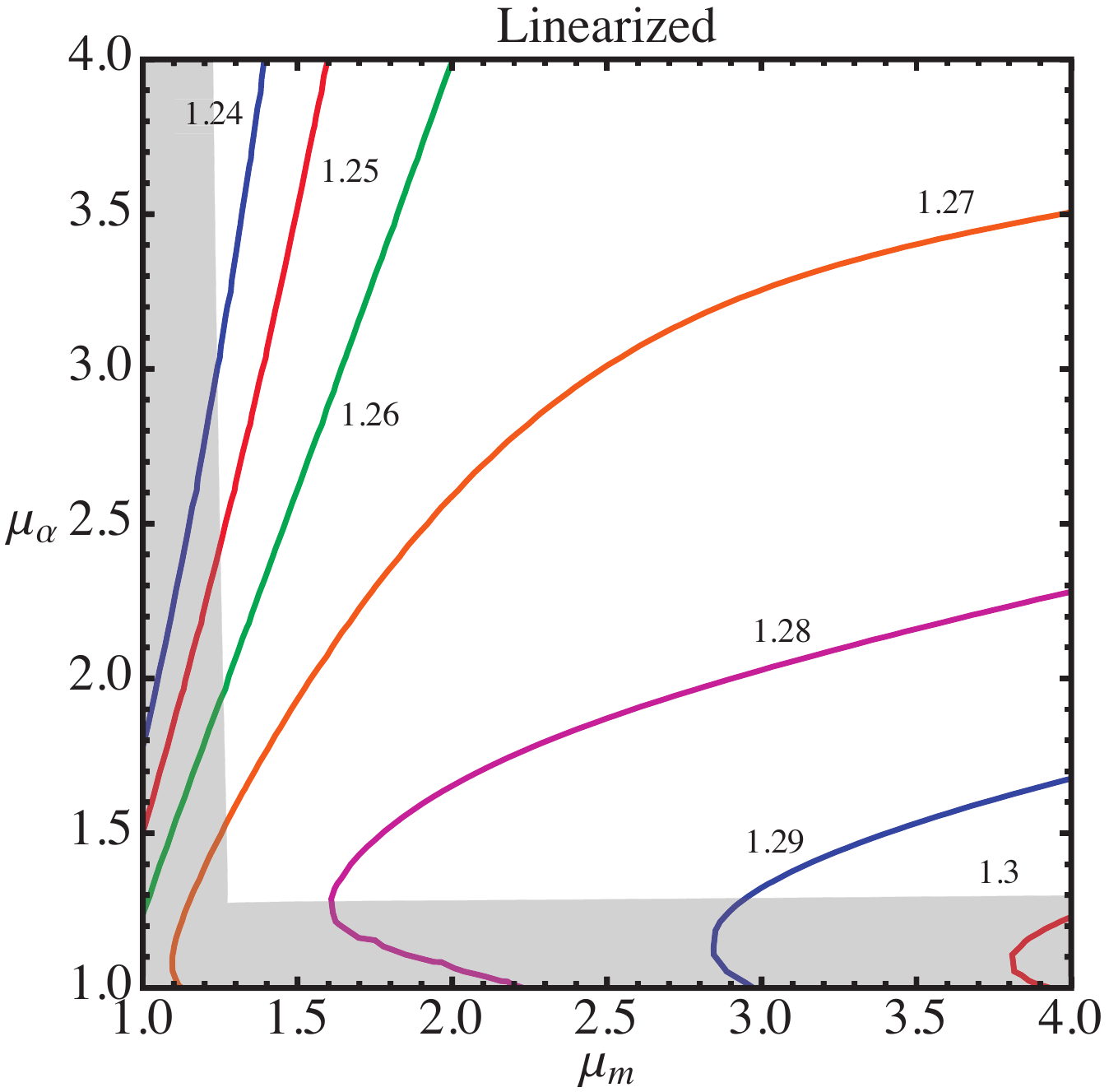}}
\subfigure[]
{\includegraphics[width=0.5\textwidth]{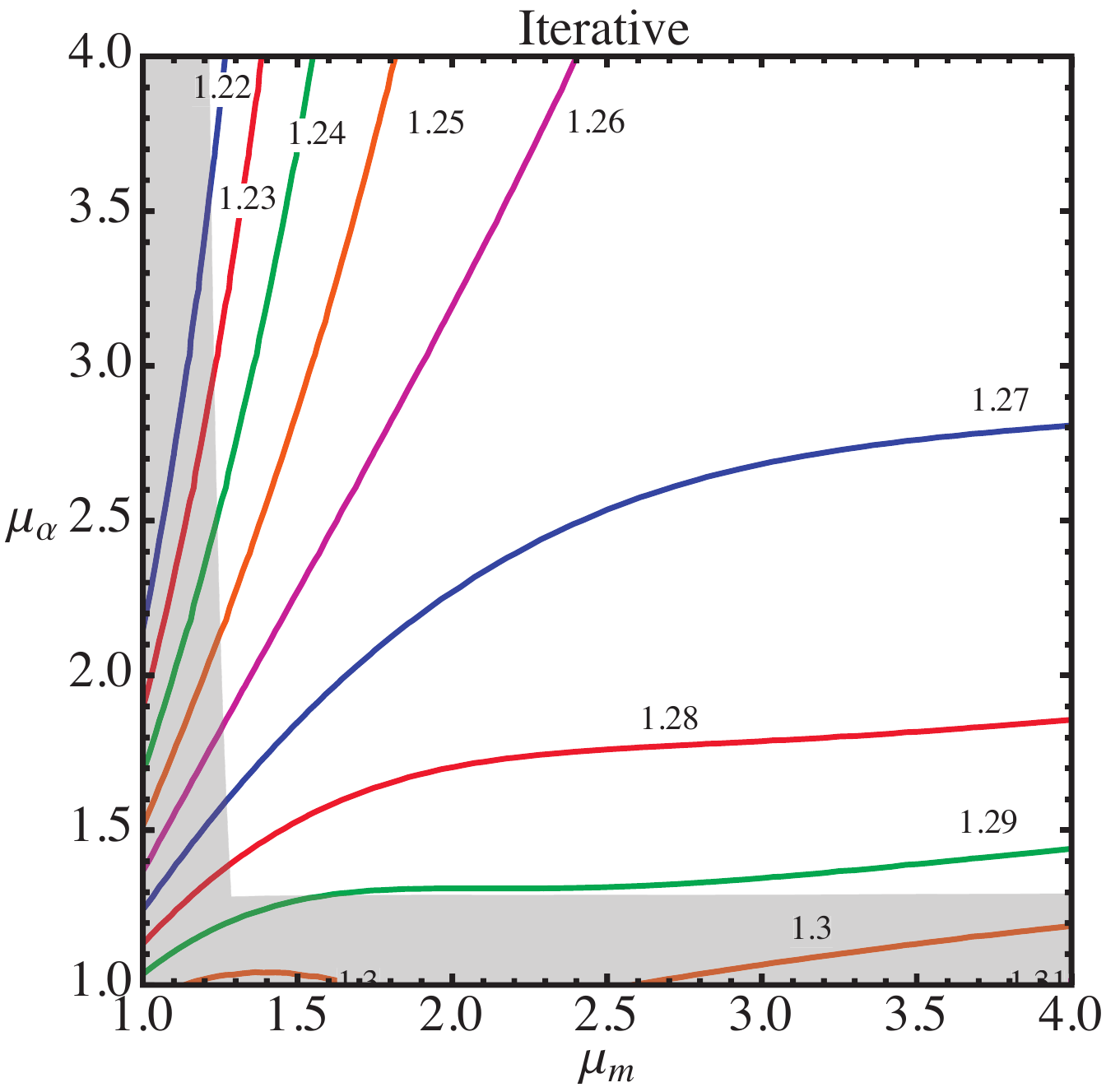}}
\subfigure[]
{\includegraphics[width=0.5\textwidth]{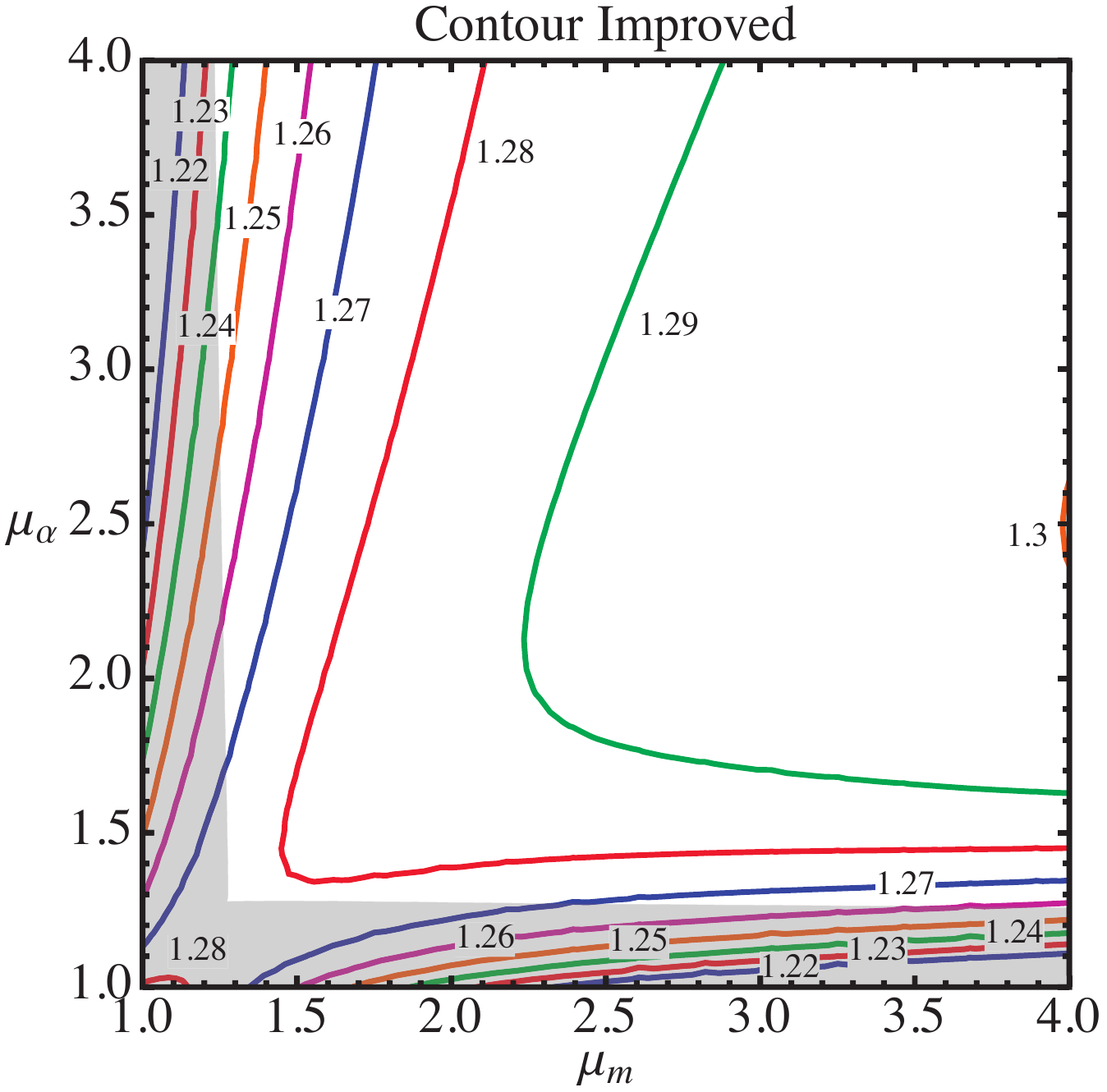}}
 \caption{Contour plots for $\overline{m}_c(\overline{m}_c)$ as obtained from the first moment of 
the pseudoscalar correlator $M_1^P$, as a function of $\mu_\alpha$ and $\mu_m$ at ${\mathcal 
O}(\alpha_s^3)$, for methods (a)\,--\,(d). The shaded areas represent regions with 
$\mu_m,\mu_\alpha < \overline{m}_c(\overline{m}_c)$, and are excluded of our analysis. For this 
plot we employ $\alpha_s(m_Z) = 0.118$.
\label{fig:mccontour1}}
\end{figure*}

\begin{figure*}[t]
\subfigure[]
{\includegraphics[width=0.5\textwidth]{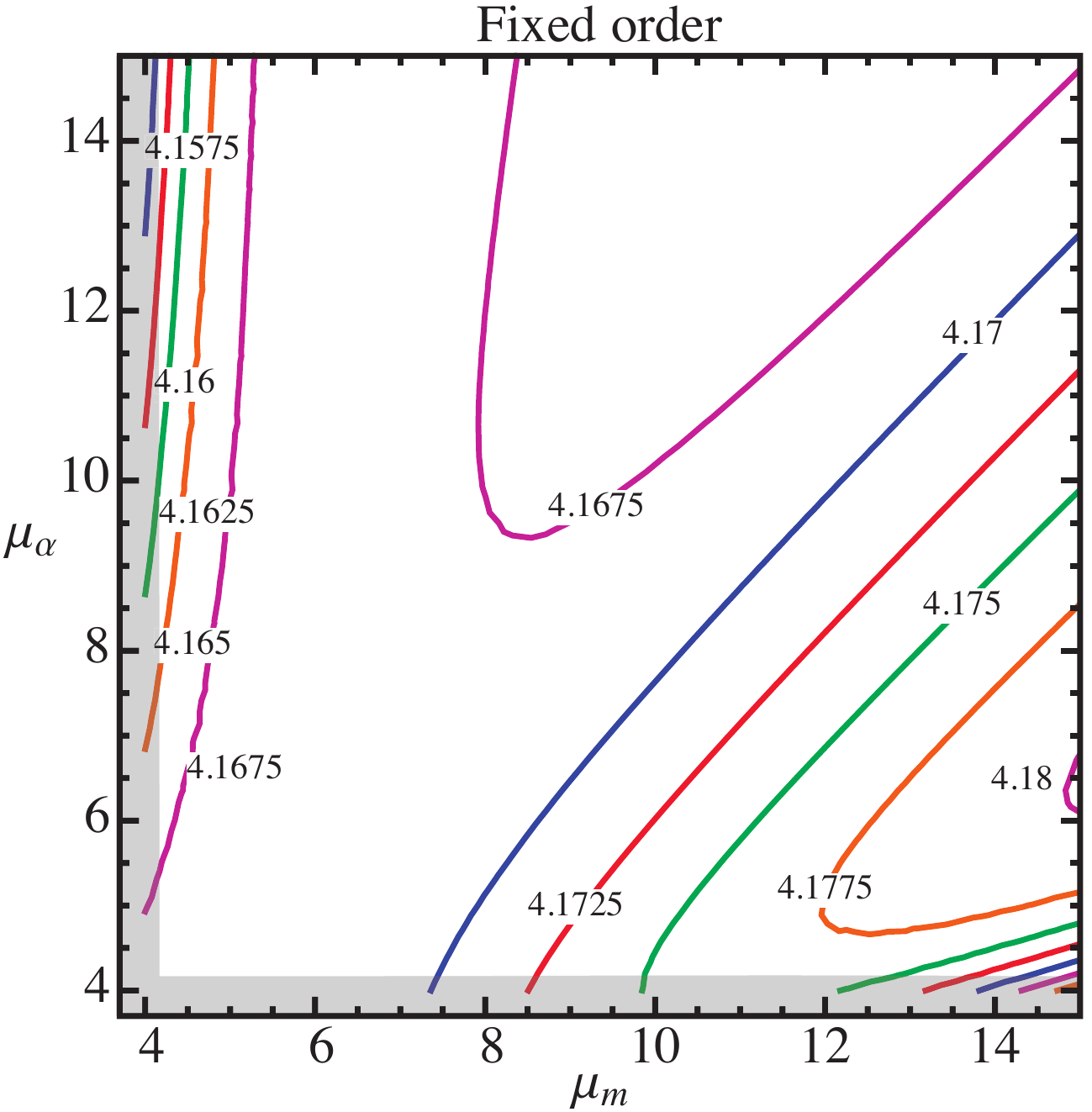}}
\subfigure[]
{\includegraphics[width=0.5\textwidth]{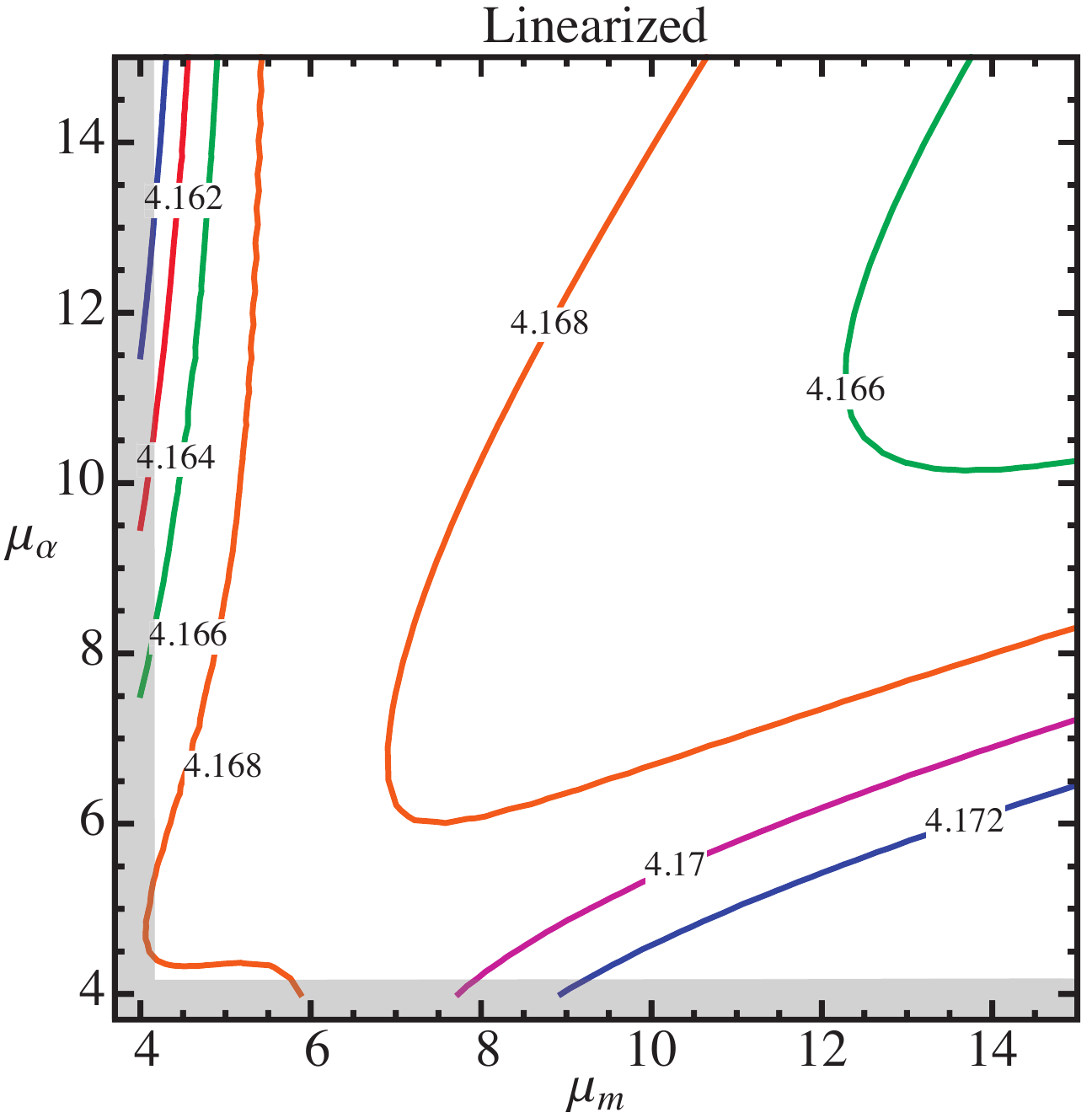}}
\subfigure[]
{\includegraphics[width=0.5\textwidth]{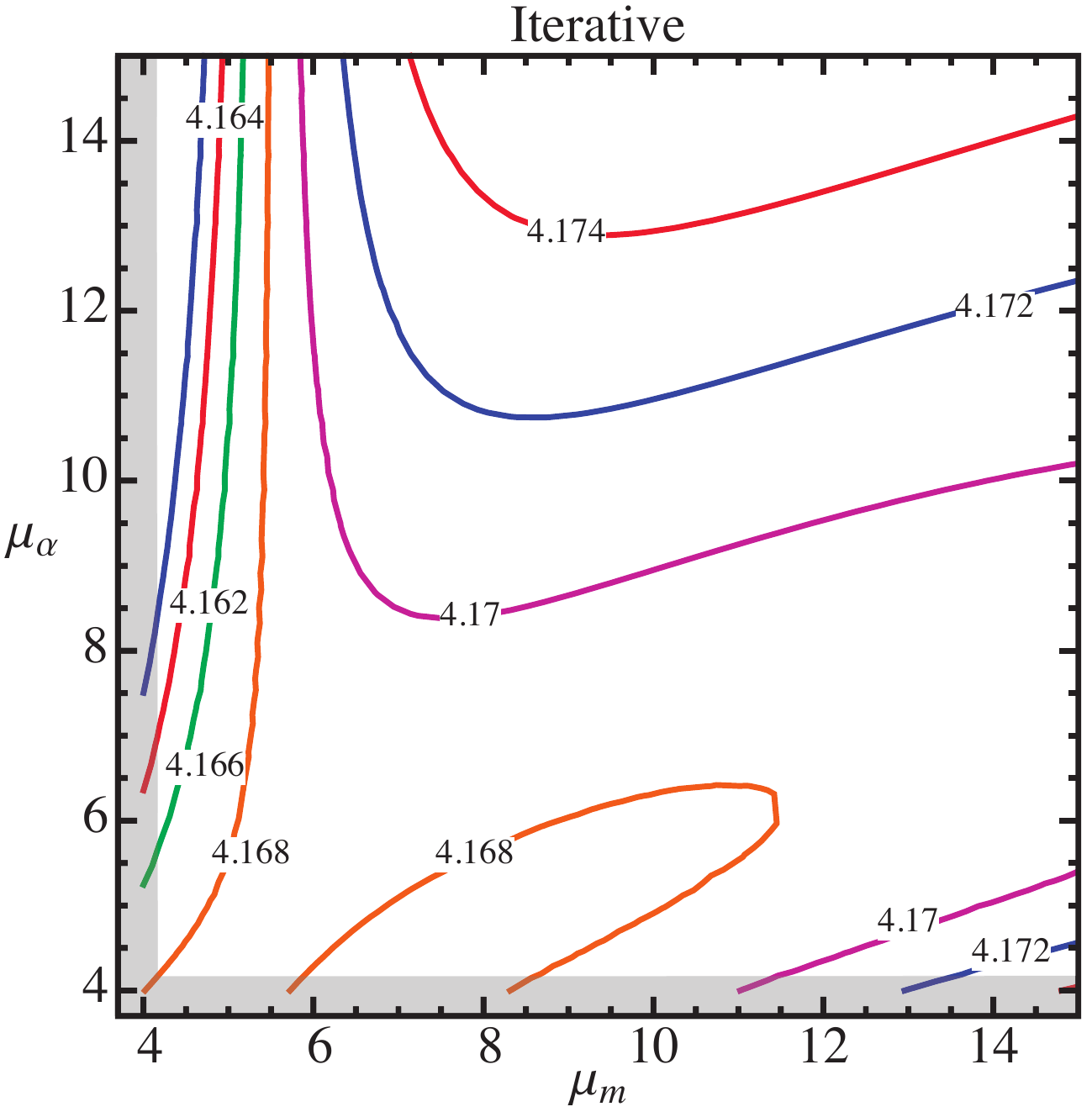}}
\subfigure[]
{\includegraphics[width=0.5\textwidth]{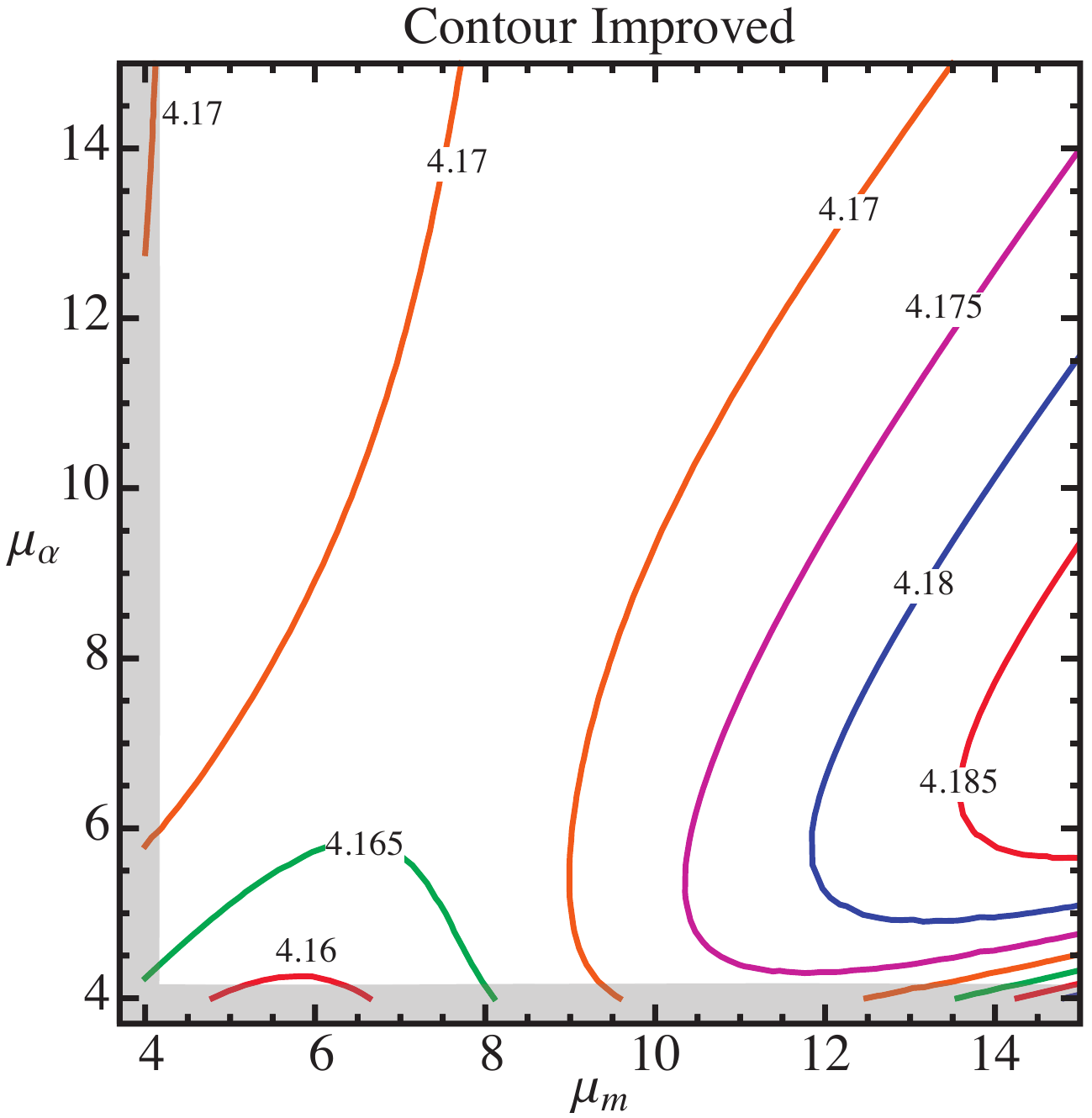}}
\caption{Contour plots for $\overline{m}_b(\overline{m}_b)$ as obtained from the second moment of 
the vector correlator $M_2^V$ with $n_f=5$, as a function of $\mu_\alpha$ and $\mu_m$ at ${\mathcal 
O}(\alpha_s^3)$, for methods (a)\,--\,(d). The shaded areas represent regions with 
$\mu_m,\mu_\alpha < \overline{m}_b(\overline{m}_b)$, and are excluded of our analysis. For this 
plot we employ $\alpha_s(m_Z) = 0.118$.
\label{fig:mbcontour1}}
\end{figure*}

\subsection{Gluon Condensate Contribution}
\label{subsectioncondensate}
We estimate nonperturbative power corrections by including the gluon condensate contribution. The 
gluon condensate is a dimension-4 matrix element and gives the leading power correction in the 
OPE for the moments~\cite{Novikov:1977dq,Baikov:1993kc}

\begin{eqnarray}
\label{MnOPE1} 
M^X_n & = & \hat M_n^X + \Delta M_n^{X,\,\langle G^2\rangle}\,+\,\ldots
\end{eqnarray}

Here the ellipses represent higher-order power corrections of the OPE involving condensates with 
dimensions bigger than $4$. The Wilson coefficients of the gluon condensate corrections are known 
to $\mathcal{O}(\alpha_s)$ accuracy~\cite{Broadhurst:1994qj}. Following 
Ref.~\cite{Chetyrkin:2010ic}, we express the Wilson coefficient of the gluon condensate in terms of 
the pole mass, since in this way the correction is numerically more stable for higher moments. 
However, as we did in Ref.~\cite{Dehnadi:2011gc}, we still write the pole mass in terms of the 
$\overline{\mbox{MS}}$ quark mass at one loop. The resulting expression reads

\begin{eqnarray}
\label{eq:GG}\Delta M_n^{X,\,\langle G^2\rangle}
& = & \dfrac{1}{(4M_q^2)^{n+2}}\Big\langle\frac{\alpha_s}{\pi} G^2\Big\rangle_{\rm RGI}  \left[
   [a_X(n_f)]^0_n+\dfrac{\alpha^{(n_f)}_s(\mu_\alpha)}{\pi}\,[a_X(n_f)]^1_n\right],\\
M_q & = & \overline{m}_q(\mu_m)\left\{1 + \dfrac{\alpha^{(n_f)}_s(\mu_\alpha)}{\pi}
\left[\dfrac{4}{3} - 
\ln\left(\frac{\overline{m}^{\,2}_q(\mu_m)}{\mu^2_m}\right)\right]\right\}.\nonumber
\end{eqnarray}

We use the renormalization group invariant (RGI) scheme for the gluon condensate 
\cite{Narison:1983kn}. The numerical value of the $[a_V(n_f = 5)]^a_n$ and $[a_P(n_f = 4)]^a_n$ 
coefficients are collected in Table~\ref{tab:gluoncondensate}. The values for $[a_V(n_f = 4)]^a_n$ 
can be found in Table~5 of Ref.~\cite{Dehnadi:2011gc}. For methods (b) and (c) one can obtain the 
gluon condensate contribution by performing simple algebra operations and re-expansions in 
$\alpha_s^{(n_f)}$ and $\langle G^2\rangle$. For method (d) we employ Eqs.~(\ref{MnOPE1}) and 
(\ref{eq:GG}) as shown. For the RGI gluon condensate we adopt~\cite{Ioffe:2005ym}

\begin{eqnarray}
\label{condensatevalue1}
\Big\langle\frac{\alpha_s}{\pi} G^2\Big\rangle_{\rm RGI} & = & 0.006\pm0.012\;\mathrm{GeV}^4\,.
\end{eqnarray}

\subsection{Ratios of Moments}
\label{sec:ratios}

An alternative set of observables which are also highly sensitive to the quark masses are the 
ratios of consecutive moments. To that end we define $R_n^X(n_f) \equiv M^X_{n+1}(n_f)/M^X_n(n_f)$. 
Such ratios are proportional to the inverse square of the quark mass for any value of $n$. 
Their perturbative series can be expressed as an expansion in powers of $\alpha_s^{(n_f)}$ 
analogous to Eq.~(\ref{eq:Mn-theo-FO}), with the replacements 
$[4\,\overline{m}_q^{\,2}(\mu_m)]^n\to4\,\overline{m}_q^{\,2}(\mu_m)$ and 
$[C_X(n_f)]_{a,b}^{i,j} \to [R_X(n_f)]_{a,b}^{i,j}$. Their computation is trivial, as one only 
needs to take the ratio of the two consecutive theoretical moments and re-expand as a series in 
$\alpha_s^{(n_f)}$. We call this the standard fixed-order expansion, analogous to the expansion (a) 
of Sec.~\ref{sec:perturbative}. The numerical expressions for the $[R_X(n_f)]_{a,b}^{i,j}$ 
coefficients for the vector correlator with $n_f = 4, 5$ are given in Table~\ref{tab:Rfixedorder}, 
and for the pseudoscalar correlator with $n_f = 4$ in Table~\ref{tab:RPFOorder}. As for the regular 
moments, we find that the nonlinear dependence of $R_n^X$ on the quark mass sometimes causes that 
there is no numerical solution for $\overline m_q$.

By taking the square root of the ratio of two consecutive moments one gets a linear dependence on 
the inverse of the quark mass. The corresponding theoretical expression is obtained by re-expanding 
the perturbative expansion of $\sqrt{R_n^X(n_f)}$ as a series in powers of  $\alpha_s^{(n_f)}$.
Thus we obtain an expression of the form of Eq.~(\ref{eq:Mn-theo-exp}) with the replacement $[\bar 
C_X(n_f)]_{a,b}^{i,j} \to [\bar R_X(n_f)]_{a,b}^{i,j}$. This is referred to as the linearized 
expansion, in analogy to the expansion (b) of Sec.~\ref{sec:perturbative}. The numerical values for 
the $[\bar R_X(n_f)]_{a,b}^{i,j}$ coefficients are collected for the vector correlator with $n_f = 
4, 5$ in Table~\ref{tab:Rexporder}, and for the pseudoscalar correlator with $n_f = 4$ in 
Table~\ref{tab:RPexporder}.

Finally, one can use $\sqrt{R_n^X(n_f)}$ to solve for $\overline m_q(\mu_m)$ in an iterative way, 
exactly as explained in Sec.~\ref{sec:perturbative} for expansion (c). The theoretical expression 
is analogous to Eq.~(\ref{eq:iterative-general})  with the replacement $[\tilde 
C_X(n_f)]_{a,b}^{i,j} \to [\tilde R_X(n_f)]_{a,b}^{i,j}$. We collect the numerical values for the 
$[\tilde R_X(n_f)]_{a,b}^{i,j}$ coefficients, in Tabs.~\ref{tab:RITorder} and \ref{tab:RPITorder} 
for the vector ($n_f = 4, 5$) and pseudoscalar ($n_f = 4$) correlators, respectively. We call this 
the iterative linearized expansion.

One cannot implement a contour-improved expression for the ratios of moments, as they cannot be 
computed as the contour integral of a correlator. For the ratios of moments, in any of the three 
expansions, one can include non-perturbative corrections in the form of a gluon condensate OPE 
term, just using Eq.~(\ref{eq:GG}) and performing simple algebra operations and re-expansions in 
$\alpha_s^{(n_f)}$ and $\langle G^2\rangle$.

\section{Previous Results and Scale Variations}
\label{sec:previous-results}

In a number of recent low-$n$ sum rule analyses~\cite{Bodenstein:2012, Bodenstein:2011ma, 
McNeile:2010ji, Chetyrkin:2009fv, Allison:2008xk, Kuhn:2007vp, Boughezal:2006px, 
Chakraborty:2014aca}, which determined the charm and bottom quark masses with very small 
uncertainties using ${\cal O}(\alpha_s^3)$ theoretical computations for the 
moments~\cite{Hoang:2008qy, Kiyo:2009gb, Greynat:2010kx, Greynat:2011zp}, the theory uncertainties 
from the truncation of the perturbative series have been estimated with the scale setting 
$\mu_m=\mu_\alpha$ based on just one type of expansion, which was either the fixed-order [expansion 
(a)] for the vector correlator, or the linearized [expansion (b)] for the pseudoscalar correlator. 
In Ref.~\cite{Dehnadi:2011gc} we analyzed the perturbative series for the moments $M^V_{1,2,3,4}$ 
of the charm vector correlator at ${\cal O}(\alpha_s^3)$ using an alternative way to estimate the 
perturbative uncertainties, based on the four different expansion methods (a)\,--\,(d), as 
explained in Sec.~\ref{sec:perturbative}. We also focused on the question whether renormalization 
scale variation restricted to $\mu_m=\mu_\alpha$ leads to a compatible estimate of the perturbative 
uncertainties. From our analysis we found:

\begin{itemize}

\item The extractions for the $\overline{\rm MS}$ charm mass using the expansions (a)\,--\,(d) with 
correlated variations of $\mu_m$ and $\mu_\alpha$ (e.g.\ $\mu_m=\mu_\alpha$) for the vector 
correlator can lead to very small scale variations, which can be very different depending on the 
method.\footnote{We judge the compatibility of the perturbative error estimates based on the size of 
scale variations alone, i.e.\ without accounting at the same time for other sources of uncertainties 
such as experimental errors or the uncertainty in the value of the strong coupling.} Moreover, for 
some expansions also the results from the different orders can be incompatible to each other. It was 
therefore concluded that using correlated scale variation and one type of expansion can lead to an 
underestimate of the perturbative uncertainty.

\item Uncorrelated (i.e.\ independent) variation of $\mu_m$ and $\mu_\alpha$ leads to charm mass 
extractions with perturbative uncertainty estimates that are in general larger, but fully 
compatible among the expansions (a)\,--\,(d) and for the different orders. It was 
therefore concluded that $\mu_m$ and $\mu_\alpha$ should be varied independently to obtain a 
reliable estimate of the perturbative uncertainty.

\item The size of the charm mass perturbative uncertainty has a significant dependence on the value 
of the lower bound of the range of the scale variation. The choice of the upper bound has a 
marginal impact.

\item The pattern of size of the correlated scale variations for the different expansions can be 
traced back to the form of the contours of constant charm mass in the $\mu_m$\,--\,$\mu_\alpha$ 
plane, which happens to be located along the diagonal $\mu_m\sim\mu_\alpha$ for expansions (a) and 
(b), but roughly orthogonal for expansions (c) and (d), see Fig.~6 of Ref.~\cite{Dehnadi:2011gc}.
\end{itemize}

\begin{figure*}[t!]
\subfigure[]
{
\includegraphics[width=0.31\textwidth]{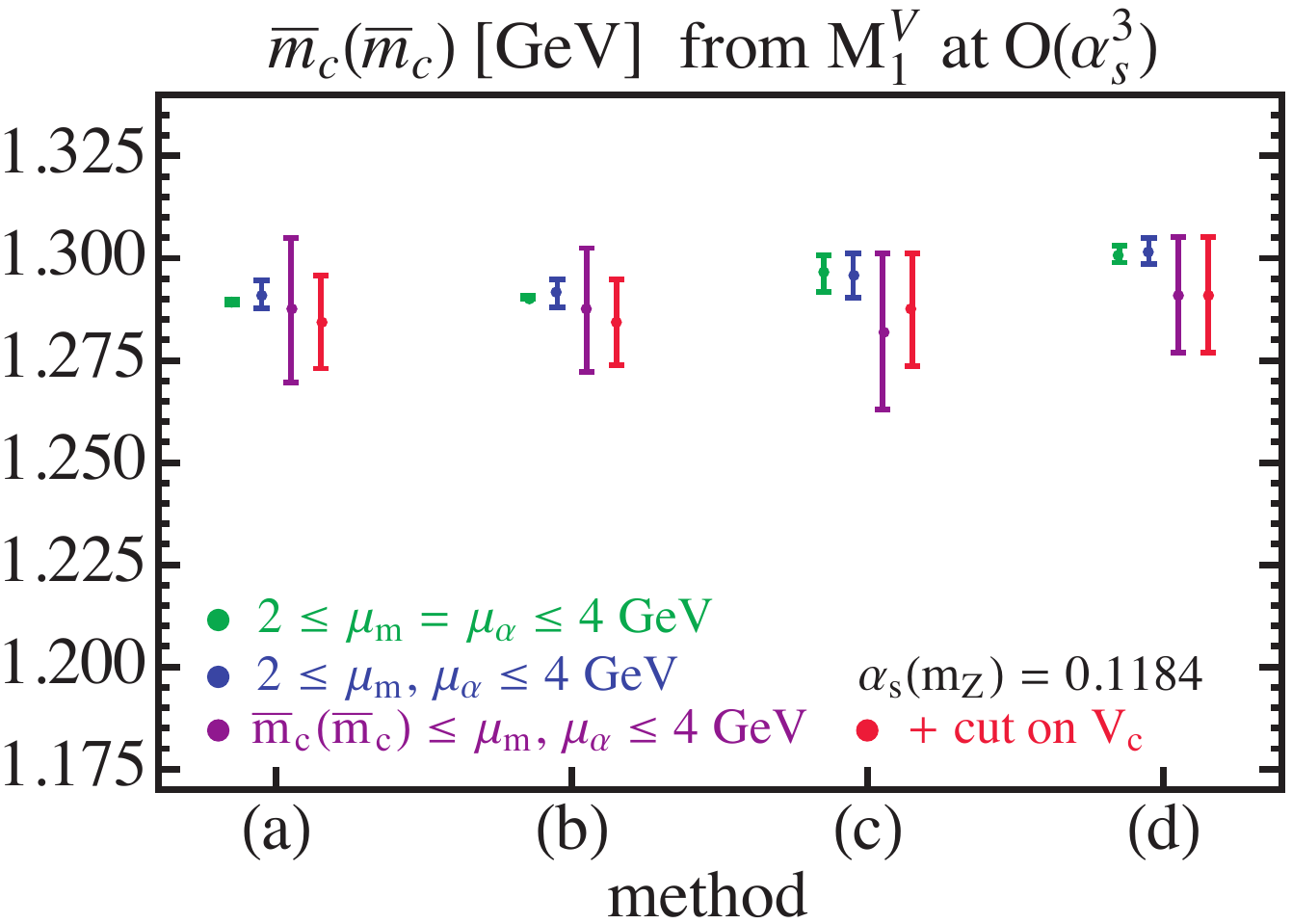}
\label{fig:charm-variations}}
\subfigure[]
{
\includegraphics[width=0.31\textwidth]{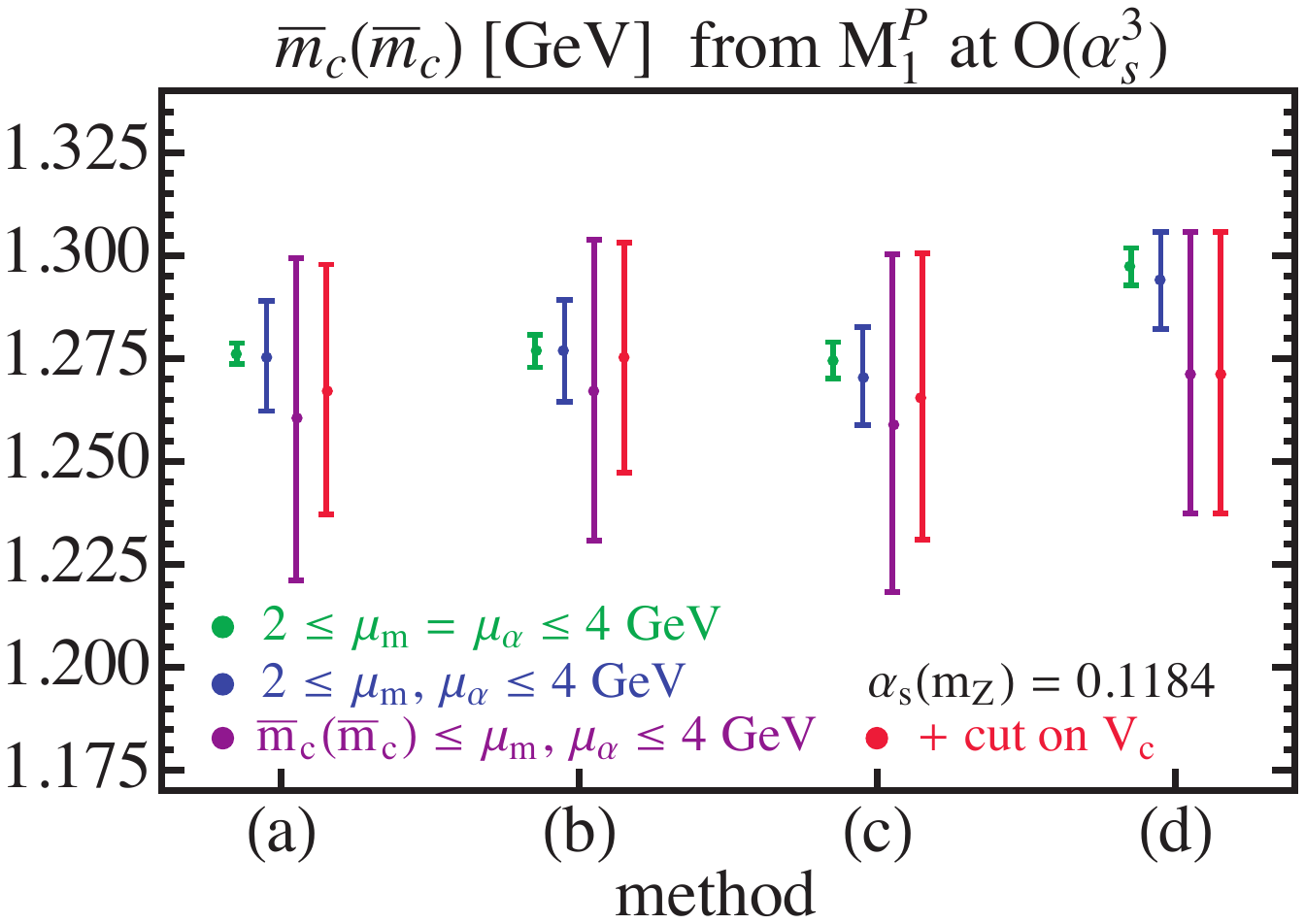}
\label{fig:pseudo-variations}}
\subfigure[]
{
\includegraphics[width=0.31\textwidth]{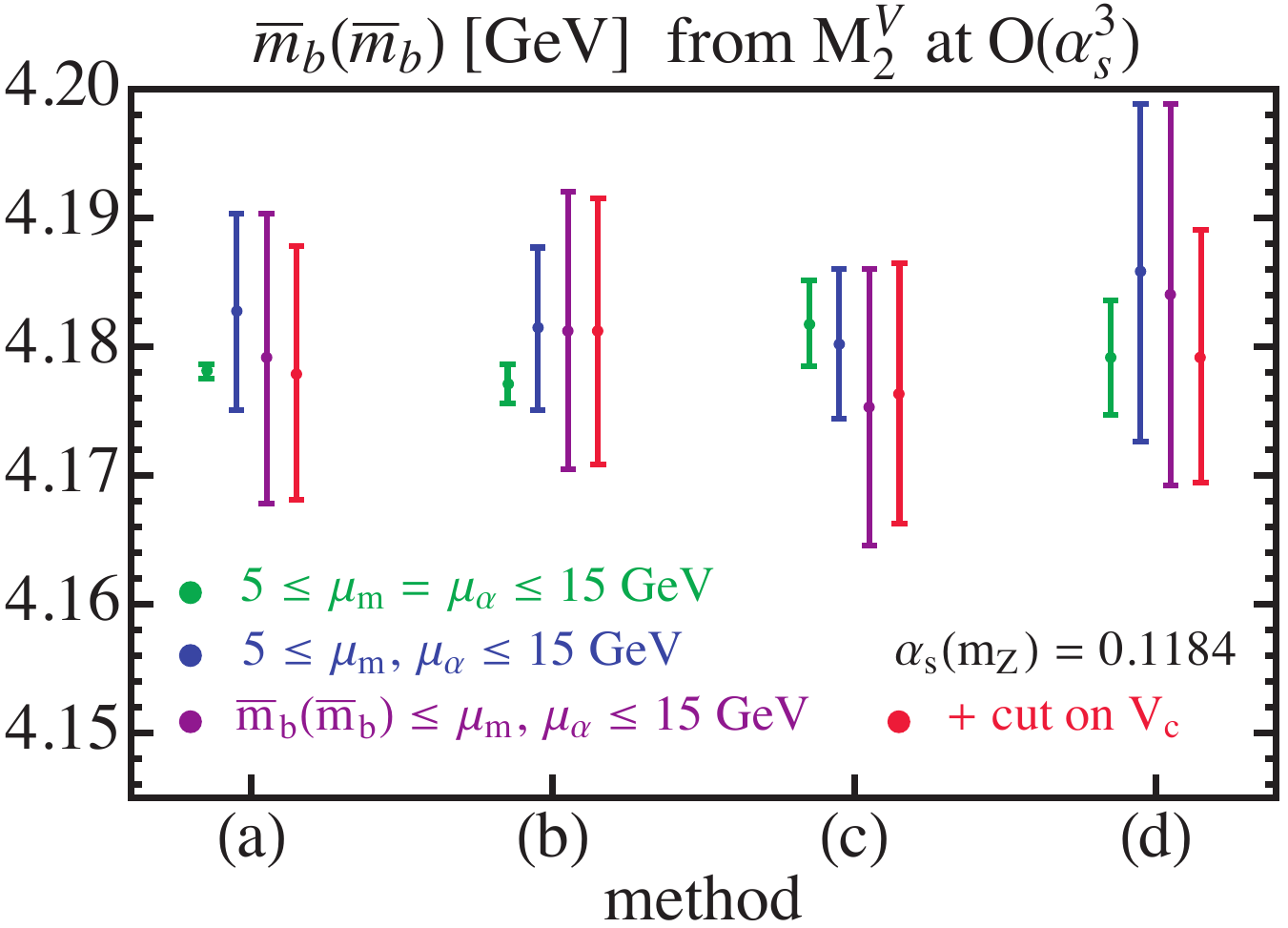}
\label{fig:bottom-variations}}
\subfigure[]
{
\includegraphics[width=0.31\textwidth]{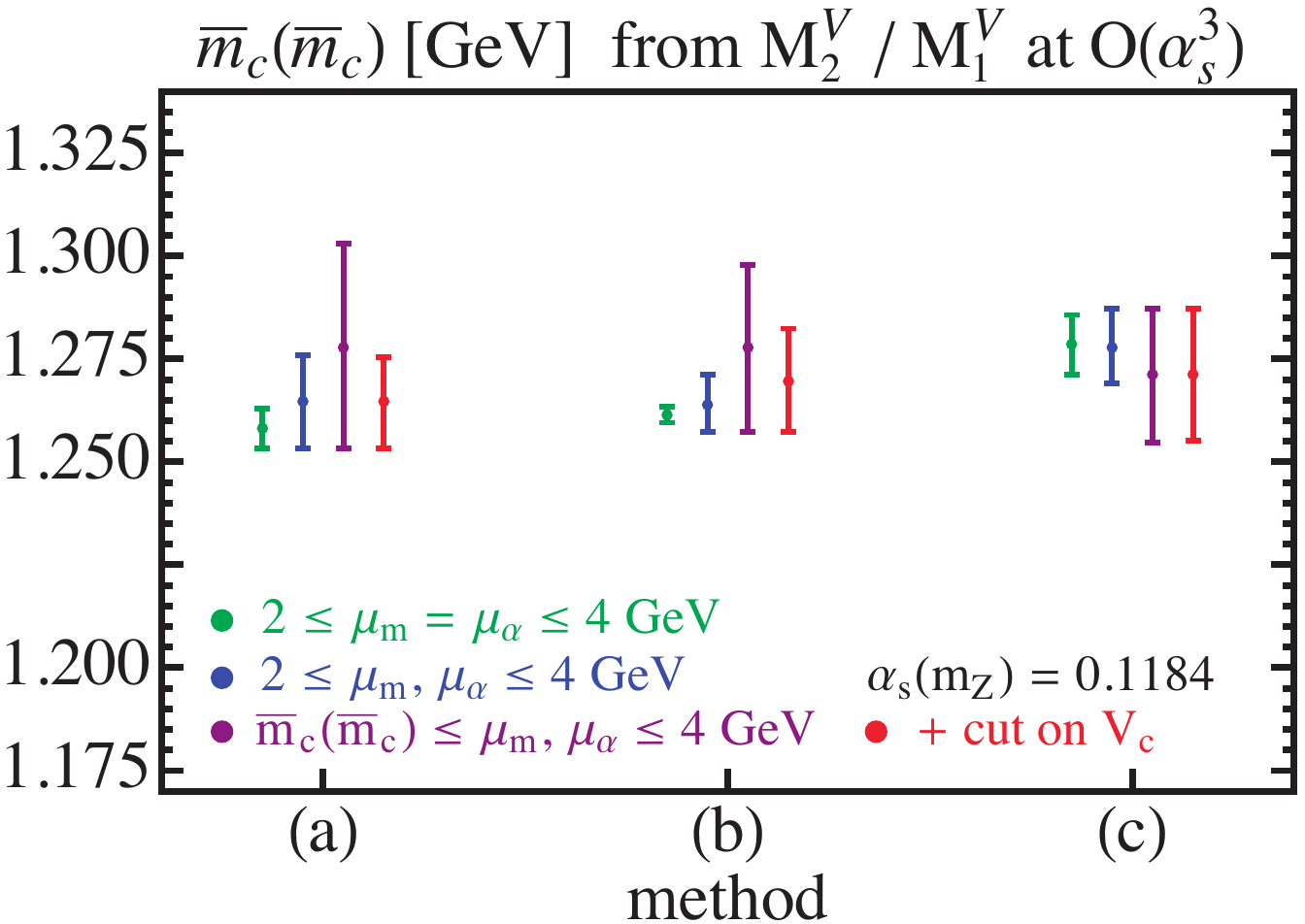}
\label{fig:charm-ratio-variations}}
\subfigure[]
{
\includegraphics[width=0.31\textwidth]{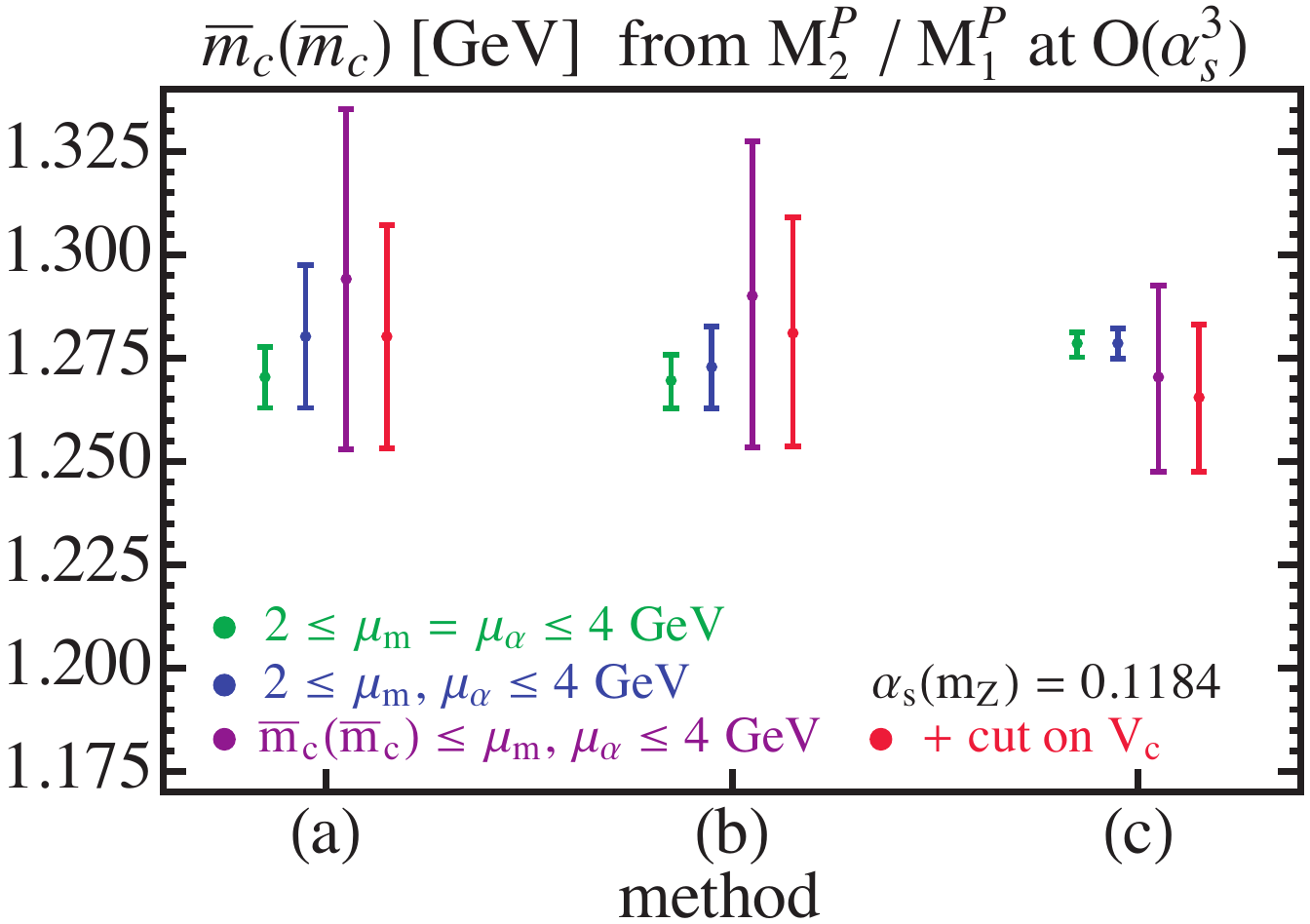}
\label{fig:pseudo-ratio-variations}}
\subfigure[]
{
\includegraphics[width=0.31\textwidth]{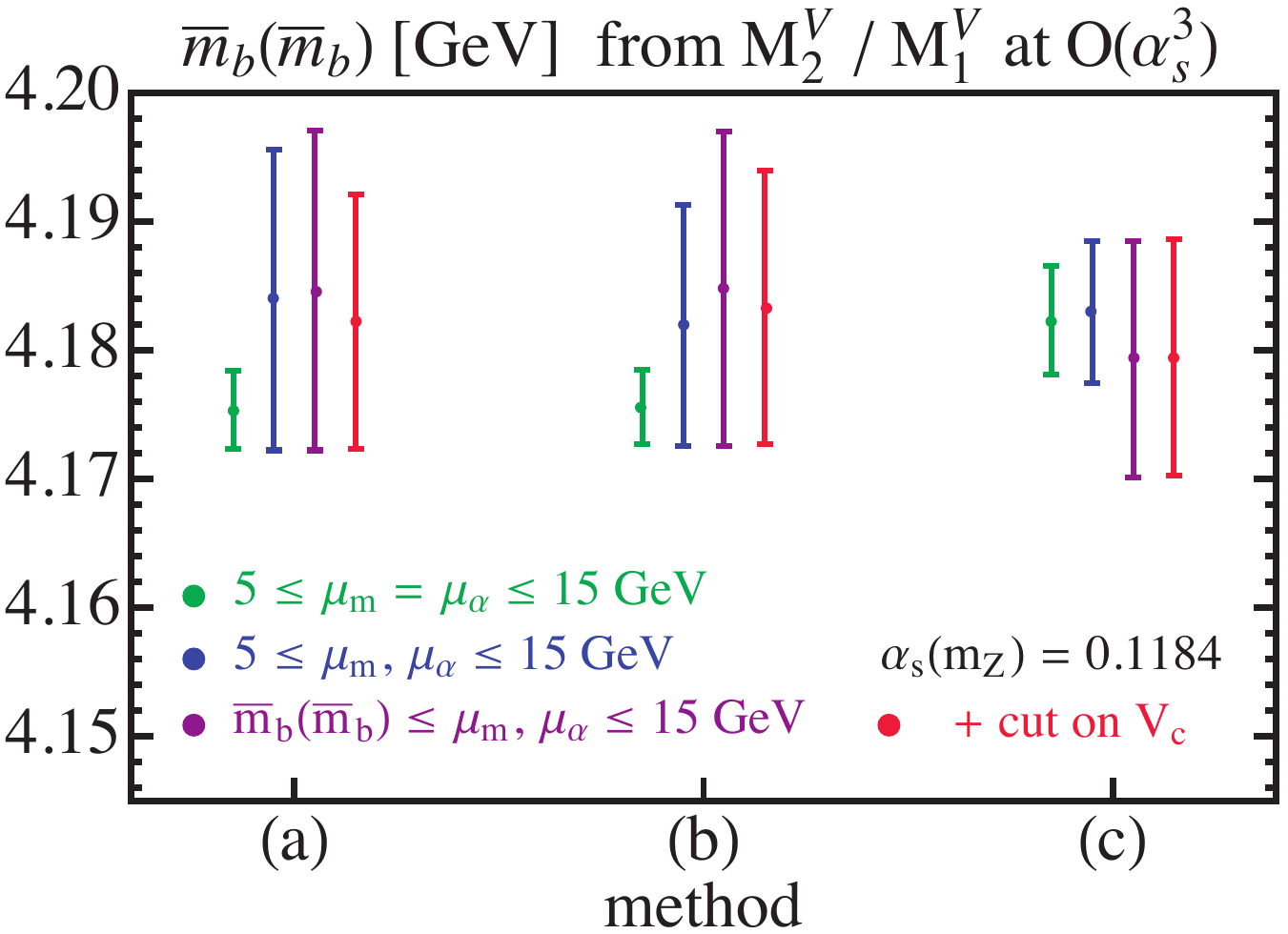}
\label{fig:bottom-ratio-variations}}
\caption{Charm and bottom mass values from the first [second] moment of the vector (a) for charm 
[(c) for bottom] and pseudoscalar [(b), charm] currents at $\mathcal{O}(\alpha_s^3)$; and for the 
ratio of the second over the first moment for the vector [(d) for charm, (g) for bottom] and 
pseudoscalar [(e), charm] correlators. We show the outcome of various scale variations for the
perturbative expansions (a)\,--\,(d) [(a)\,--\,(c) for ratios], where green (rightmost) 
corresponds to $2\,{\rm GeV}\le\mu_m=\mu_\alpha\le4\,{\rm GeV}$ [$5\,{\rm 
GeV}\le\mu_m=\mu_\alpha\le15\,{\rm GeV}$ for bottom], blue (second from the left) $2\,{\rm 
GeV}\le\mu_m,\mu_\alpha\le4\,{\rm GeV}$ [$5\,{\rm GeV}\le\mu_m,\mu_\alpha\le15\,{\rm GeV}$ for 
bottom], purple (second from the right) $\overline m_c(\overline m_c)\le\mu_m,\mu_\alpha\le\,4{\rm 
GeV}$ [$\overline m_b(\overline m_b)\le\mu_m,\mu_\alpha\le15\,{\rm GeV}$ for bottom] and in red 
(rightmost) we supplement the latter variation with a cut on the series with larger values of 
$V_c$.}\label{fig:variations}
\end{figure*}

For example,\footnote{The size of the scale variations quoted in this paragraph applies to 
$\overline m_c(\overline m_c)$ as well as to $\overline m_c(3\,{\rm GeV})$, and all numerical 
results are obtained at ${\cal O}(\alpha_s^3)$. We also stress that there are no perturbative 
instabilities concerning the use of the RGE down to the scale $\overline m_c(\overline m_c)$.} in 
Ref.~\cite{Chetyrkin:2009fv} method (a) has been used for $M_1^V$ with $\mu_m=\mu_\alpha$ varied 
between $2$ and $4$\,GeV, quoting a perturbative error estimate of $2$\,MeV. For the expansion 
methods\footnote{To compute the charm and bottom masses in this section we use $M_1^{V,\,\rm 
exp}=0.2121\,{\rm GeV}^{-2}$ for the charm vector correlator, result obtained in 
Ref.~\cite{Dehnadi:2011gc}, $M_1^{P,\,\rm latt}=0.1402\,{\rm GeV}^{-2}$ for the charm pseudoscalar 
correlator, from Ref.~\cite{Allison:2008xk}, and our own computation $M_2^{V,\,{\rm exp}} = 
2.834\times10^{-5}\,{\rm GeV}^{-4}$ for the bottom vector correlator, see Sec.~\ref{sec:exp}. We 
also use $\alpha_s(m_Z) = 0.1184.$} (a)\,--\,(d) we obtain for the same scale variation $1.2893 \pm 
0.0007$, $1.2904\pm0.0004$, $1.2963\pm0.0045$ and $1.3009\pm0.0020$\,GeV, respectively, for 
$\overline m_c(\overline m_c)$, which are inconsistent. This can be compared to the corresponding 
results using independent variations as suggested in Ref.~\cite{Dehnadi:2011gc}. Using $2\,{\rm 
GeV}\le \mu_\alpha,\mu_m\le4\,{\rm GeV}$ we obtain $1.291\pm0.003$, $1.291\pm0.003$, $1.296\pm0.005$ 
and $1.302\pm0.003$\,GeV, respectively, for expansions (a)\,--\,(d). These results are not 
consistent either. It was furthermore argued in Ref.~\cite{Dehnadi:2011gc} that an adequate 
variation range should include the charm mass itself (after all, that is the scale that governs the 
series), motivated by the range $2\,m_c\,\pm\, m_c$ around the pair production threshold. Thus, 
adopting independent scale variation in the range $\overline{m}_c(\overline{m}_c) \le \mu_m, 
\mu_\alpha \le 4$\,GeV one obtains $1.287\pm0.018$, $1.287\pm0.015$, $1.282\pm0.019$ and 
$1.291\pm0.014$\,GeV respectively. The results show consistency and demonstrate the strong 
dependence on the lower bound of the renormalization scale variation. The outcome is illustrated 
graphically in Fig.~\ref{fig:charm-variations}, and the order-by-order dependence in Fig.~1 of 
Ref.~\cite{Dehnadi:2014kya}. In Ref.~\cite{Dehnadi:2011gc} we also explored scale setting in which 
$\mu_m$ was fixed to $\overline m_c(\overline m_c)$ and only $\mu_\alpha$ was varied. The outcome is 
shown in Figs.~4 and 5 of that reference. The contour lines in the  $\mu_m$\,--\,$\mu_\alpha$ plane 
for the mass extraction from the first moment of the vector correlator for all methods are shown in 
Fig.~6 of Ref.~\cite{Dehnadi:2011gc}. The final result quoted in Ref.~\cite{Dehnadi:2011gc}, using 
$\alpha_s(m_Z)=0.1184\, \pm \,0.0021$, was 
$\overline m_c(\overline m_c) = 1.282 \, \pm \, (0.006)_{\rm stat}\, \pm \, (0.009)_{\rm syst} \, 
\pm \, (0.019)_{\rm pert}\, \pm \, (0.010)_{\alpha_s} \, \pm \, (0.002)_{\langle GG\rangle}\,$GeV 
based on the iterative expansion method (c).

We have repeated this analysis for the first moment of the pseudoscalar correlator $M_1^P$. 
Ref.~\cite{Allison:2008xk} uses method (b) with the same scale variation as 
Ref.~\cite{Chetyrkin:2009fv}, quoting $4$\,MeV for the truncation error. For methods (a)\,--\,(d) 
and using $2\,{\rm GeV}\le\mu_m=\mu_\alpha\le4\,{\rm GeV}$ we obtain $1.276 \pm 0.003$, 
$1.277\pm0.004$, $1.275\pm0.005$ and $1.297\pm0.004$\,GeV, respectively. For independent double 
scale variation between $2$ and $4$\,GeV we obtain, $1.276 \pm 0.013$, $1.277\pm0.012$, 
$1.271\pm0.012$ and $1.294\pm0.012$\,GeV, and if we use $\overline m_c(\overline m_c)$ as the lower 
bound to we obtain $1.260 \pm 0.039$, $1.267\pm0.037$\,GeV, $1.259\pm0.041$ and $1.272\pm0.034$. 
These results are displayed graphically in Fig.~\ref{fig:pseudo-variations}. The contour lines for 
the mass extraction from the first moment of the pseudoscalar correlator for all methods are shown 
in Fig.~\ref{fig:mccontour1}. We see that the results show a qualitative agreement with the 
situation for the vector current, but at a level of perturbative scale variations that are in 
general roughly larger by a factor of two.

A similar study can be performed for the extraction of the bottom mass $\overline m_b(\overline 
m_b)$ from the second moment of the vector correlator $M_2^V$. Ref.~\cite{Chetyrkin:2009fv} uses 
the fixed-order expansion [method~(a)] and correlated scale variation between 
$5\,{\rm GeV}\le\mu_m=\mu_\alpha\le 15\,{\rm GeV}$, quoting a perturbative error of $3$\,MeV. 
For the same variation we obtain $4.1781\pm0.0005$, $4.1771\pm0.0015$, $4.1818\pm0.0034$ and 
$4.1792\pm0.0044$\,GeV for methods (a) and (d), respectively. As in the charm case the results are 
not consistent, but the variations of the results have a much smaller size, as is expected from the 
fact that for the bottom the renormalization scales are much larger. For independent variation 
between the same values we get $4.183 \pm 0.008$, $4.181\pm0.006$, $4.180\pm0.006$ and 
$4.186\pm0.013$\,GeV. Finally, if the lower limit of the double variation starts at $\overline 
m_b(\overline m_b)$ we find $4.179 \pm 0.011$, $4.181\pm0.011$, $4.175\pm0.011$ and 
$4.184\pm0.0015$\,GeV. These results are collected in Fig.~\ref{fig:bottom-variations}. The 
corresponding $\overline m_b(\overline m_b)$ contours in the $\mu_m$\,--\,$\mu_\alpha$ plane are 
shown in Fig.~\ref{fig:mbcontour1}. As for the charm case, we find fully consistent results for the 
independent scale variation and using $\overline m_b(\overline m_b)$ as the lower bound.

We have also studied the ratio of the first over the second moments for the three cases, and 
observe a very similar pattern. We do not provide a detailed discussion in the text, but display 
the outcome graphically in Figs.~\ref{fig:charm-ratio-variations} to 
\ref{fig:bottom-ratio-variations}.

\section{Convergence Test}
\label{sec:convergence}

At this point it is useful to consider the perturbative series for all choices of $\mu_\alpha$ and 
$\mu_m$ as different perturbative expansions, which can have different convergence properties. 
To estimate the perturbative uncertainties one analyzes the outcome of this set of (truncated) 
series. While the uncorrelated scale variation certainly is a conservative method, one possible 
concern is that it might lead to an overestimate of the size of the perturbative error. For 
instance, this might arise for a non-vanishing value of $\ln(\mu_m/\mu_\alpha)$ in connection with 
sizeable values of $\alpha_s(\mu_\alpha)$ for $\mu_\alpha$ close to the charm mass scale, which 
might artificially spoil the convergence of the expansion. One possible resolution might be to 
simply reduce the range of scale variation (such as increasing the lower bound). However, this does 
not resolve the issue, since the resulting smaller variation merely represents a matter of choice. 
Furthermore, there is in general no guarantee that the series which are left have a better 
convergence despite the fact that the overall scale variation might become reduced. Preferably, the 
issue should be fixed from inherent properties of the perturbative series themselves. It is possible 
to address this issue by supplementing the uncorrelated scale variation method with a convergence 
test constraint, which we explain in the following.

We implement a finite-order version of the root convergence test. Let us recall that in 
mathematics, the root test (also known as Cauchy's radical test) states that for a series of terms 
$a_n$, $S[a]=\sum_{n} a_n$, if the quantity $V_\infty$ defined as

\begin{equation}\label{eq:cauchi}
V_\infty \equiv \limsup_{n\to\infty} (a_n)^{1/n}\,,
\end{equation}

is smaller (bigger) than $1$, the sum is absolutely convergent (divergent). If $V_\infty$ 
approaches $1$ from above then the series is still divergent, otherwise the test is not conclusive. 
In Eq.~(\ref{eq:cauchi}) $\limsup$ stands for the superior limit, which essentially means that in 
case of oscillating series, one takes the maximum value of the oscillation. In the context of our 
analysis with truncated series the relevant property is that a smaller $V_\infty$ implies a better 
convergent series. For the different expansion methods we use, it is simplest to apply the method 
directly to the sequence of quark masses that are extracted order by order, rewriting the results as 
a series expansion. Since we only know a finite number of coefficients of the perturbative series, 
we need to adapt the test. We now introduce $V_c$ and proceed as follows:\,\footnote{One could think 
of implementing the ratio test as well. However, since we only known a small number of terms, it is 
likely that one of them becomes close to zero, making one of the ratios blow up. This makes this 
test very unstable.}

\begin{itemize}
\item[(a)] For each pair of renormalization scales $(\mu_m,\mu_\alpha)$ we define the convergence 
parameter $V_c$ from the charm mass series $\overline{m}_c(\overline{m}_c)=m^{(0)}+\delta 
m^{(1)}+\delta m^{(2)}+\delta m^{(3)}$ resulting from the extractions at 
${\cal O}(\alpha_s^{0,1,2,3})$:

\begin{equation}
V_c = \max\!\bigg[\frac{\delta m^{(1)}}{m^{(0)}}\,,\Big(\frac{\delta 
m^{(2)}}{m^{(0)}}\Big)^{\!\!1/2},
\Big(\frac{\delta m^{(3)}}{m^{(0)}}\Big)^{\!\!1/3}\,\bigg].
\end{equation}

\item[(b)] The resulting distribution for $V_c$ values can be conveniently cast as a histogram, and 
the resulting distribution is a measure for the overall convergence of the perturbative expansion 
being employed. We apply the convergence analysis to the region 
\mbox{$\overline{m}_c(\overline{m}_c) 
\leq \mu_\alpha,\mu_m \le 4$\,GeV} for charm, and $\overline{m}_b(\overline{m}_b) \leq 
\mu_\alpha,\mu_m \le 15$\,GeV for bottom. If the distribution is peaked around the average $\langle 
V_c\rangle$ it has a well-defined convergence. Hence discarding series with $V_c\gg \langle 
V_c\rangle$ (particularly if they significantly enlarge the estimate of the perturbative error) 
is justified.

\end{itemize}

\begin{figure*}[t!]
\subfigure[]
{
\includegraphics[width=0.31\textwidth]{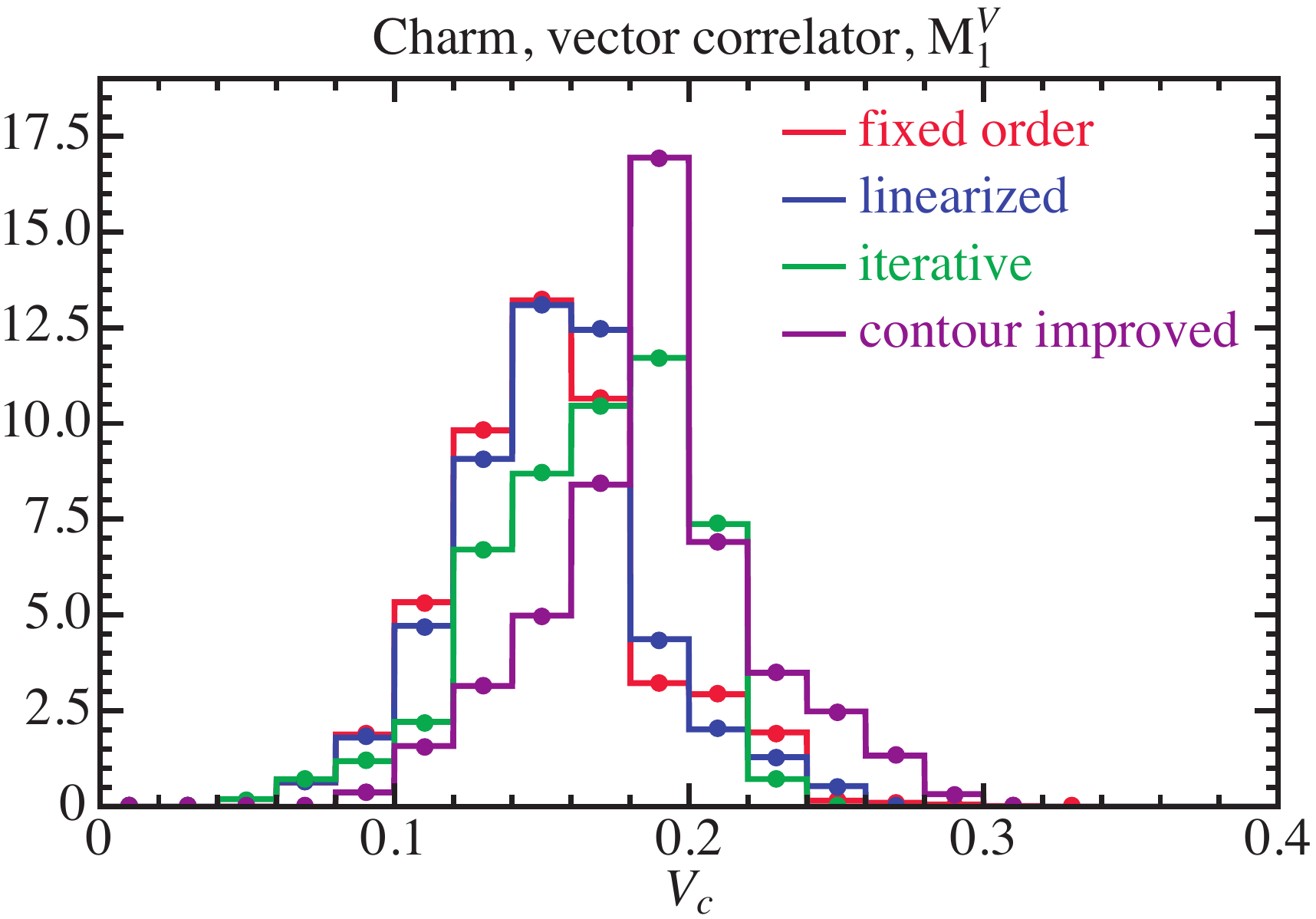}
\label{fig:histograms-vector}}
\subfigure[]
{
\includegraphics[width=0.31\textwidth]{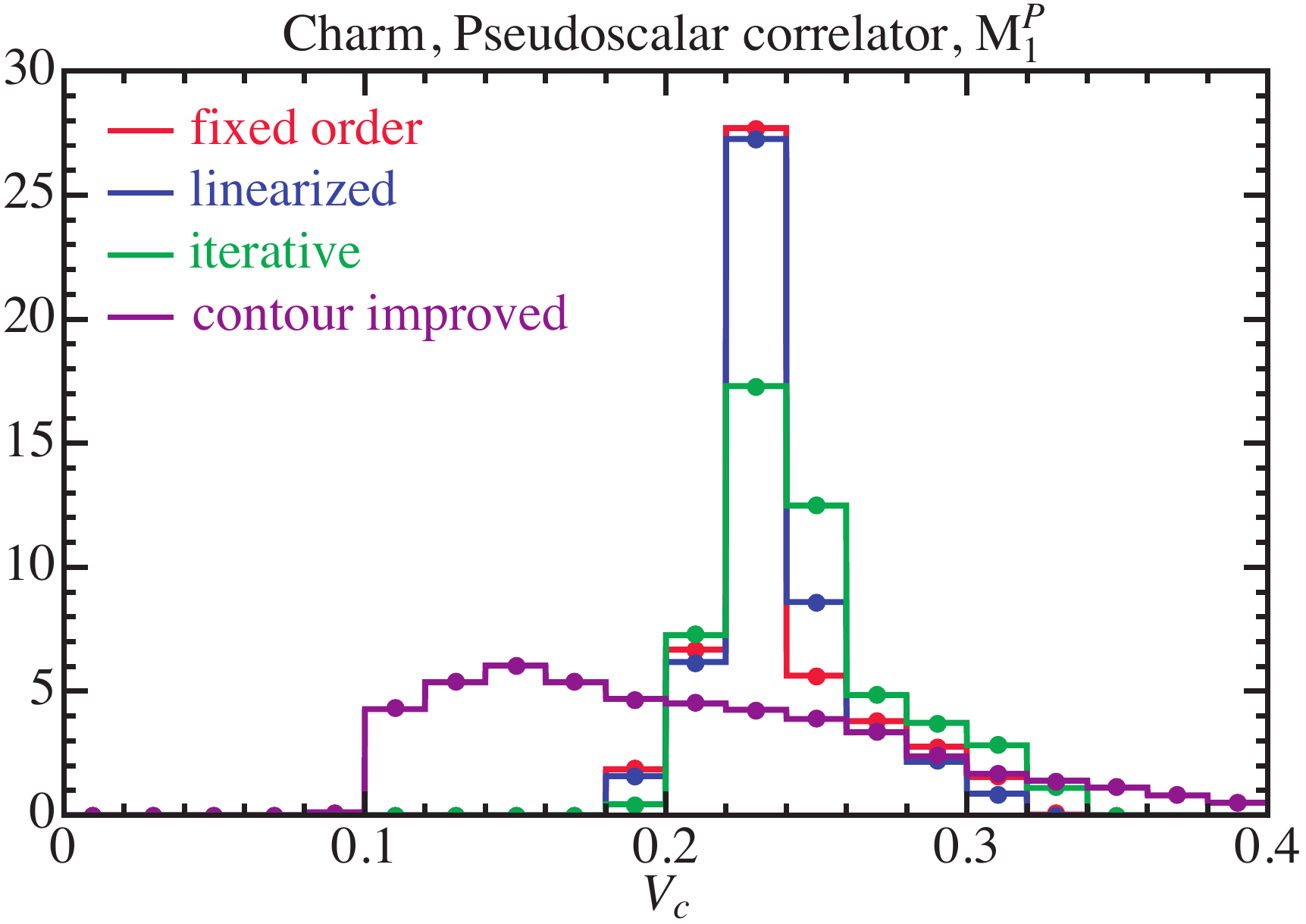}
\label{fig:histograms-pseudo}}
\subfigure[]
{
\includegraphics[width=0.31\textwidth]{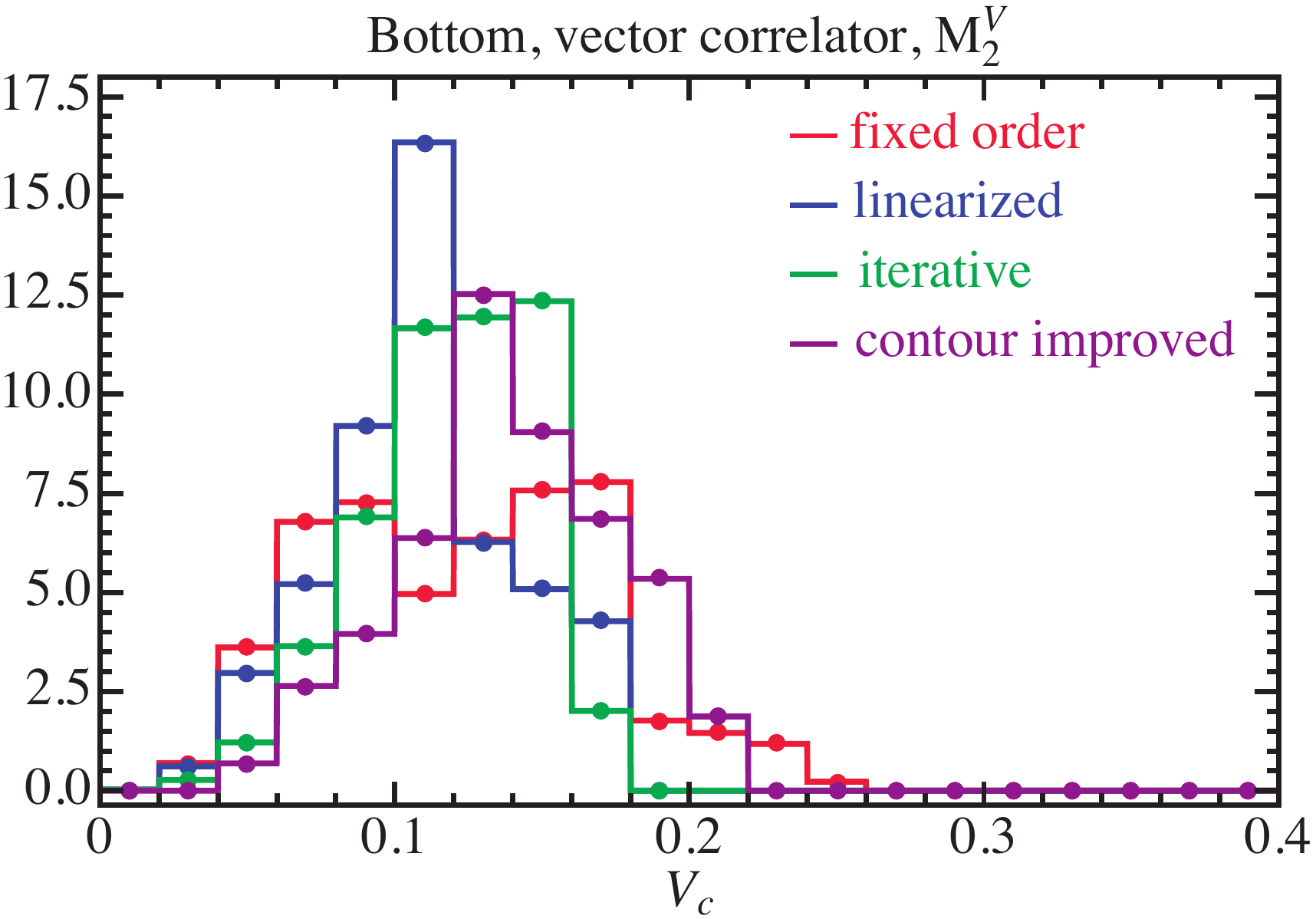}
\label{fig:histograms-bottom}}
\subfigure[]
{
\includegraphics[width=0.31\textwidth]{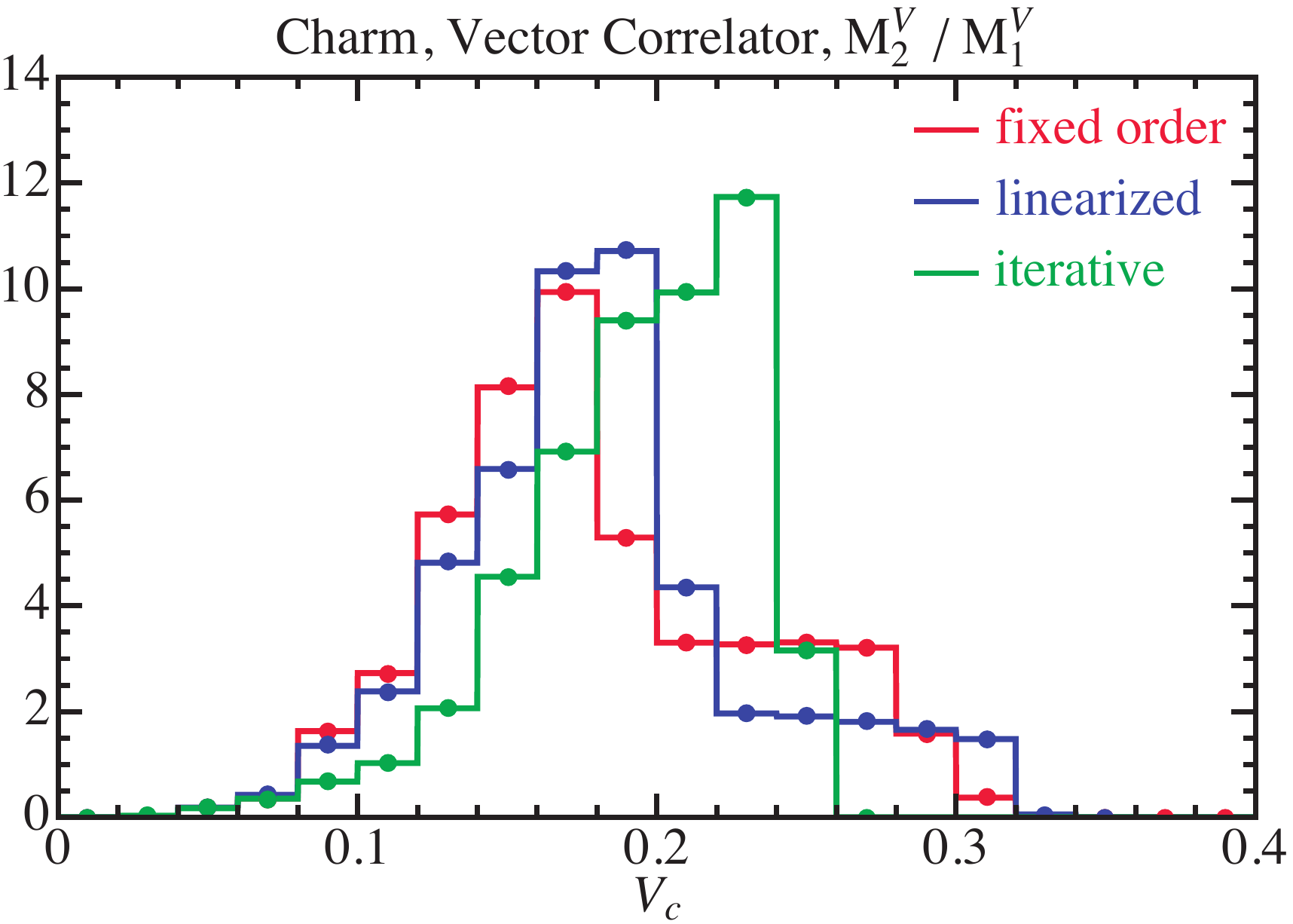}
\label{fig:histograms-vector-ratio}}
\subfigure[]
{
\includegraphics[width=0.31\textwidth]{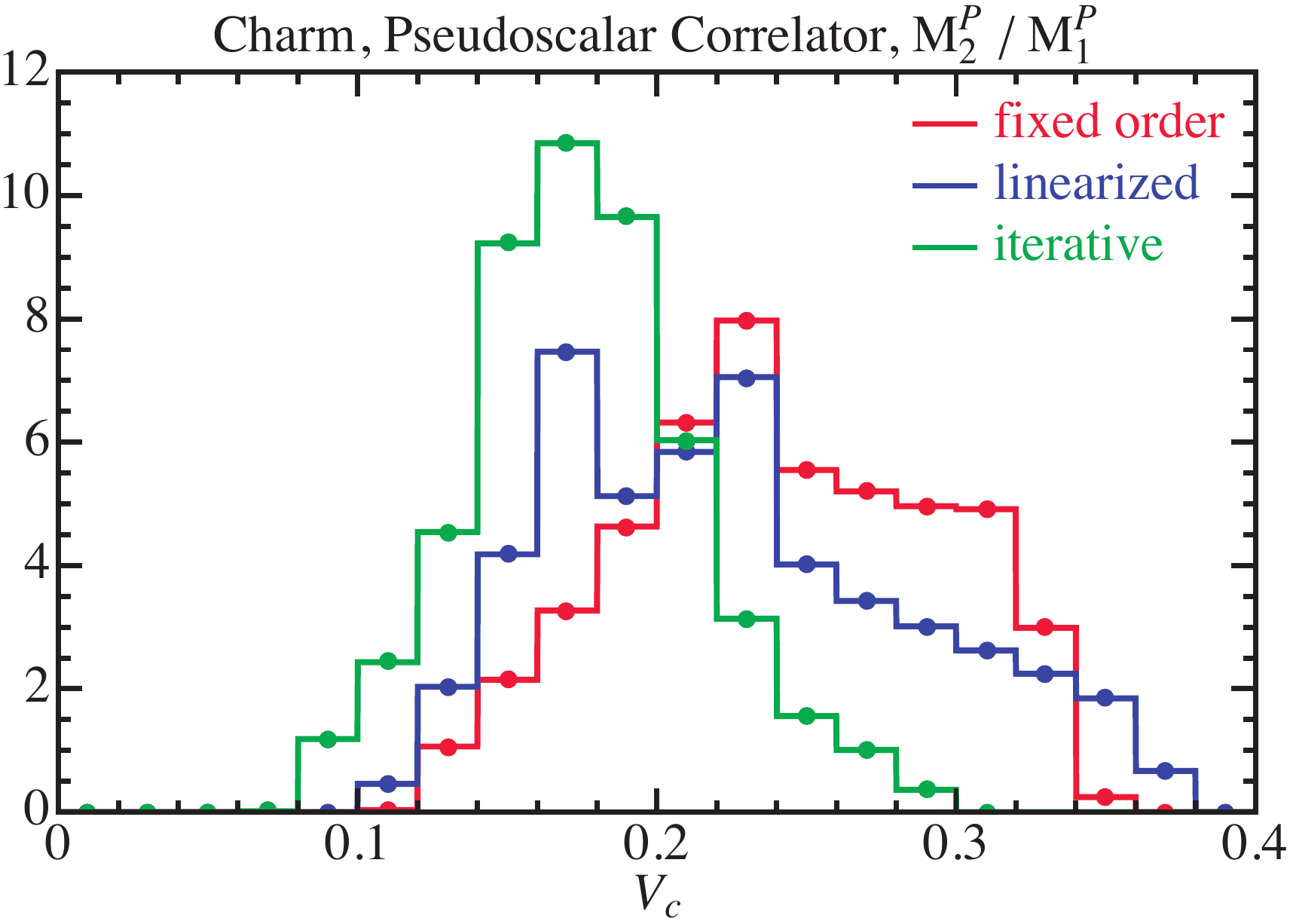}
\label{fig:histograms-pseudo-ratio}}
\subfigure[]
{
\includegraphics[width=0.31\textwidth]{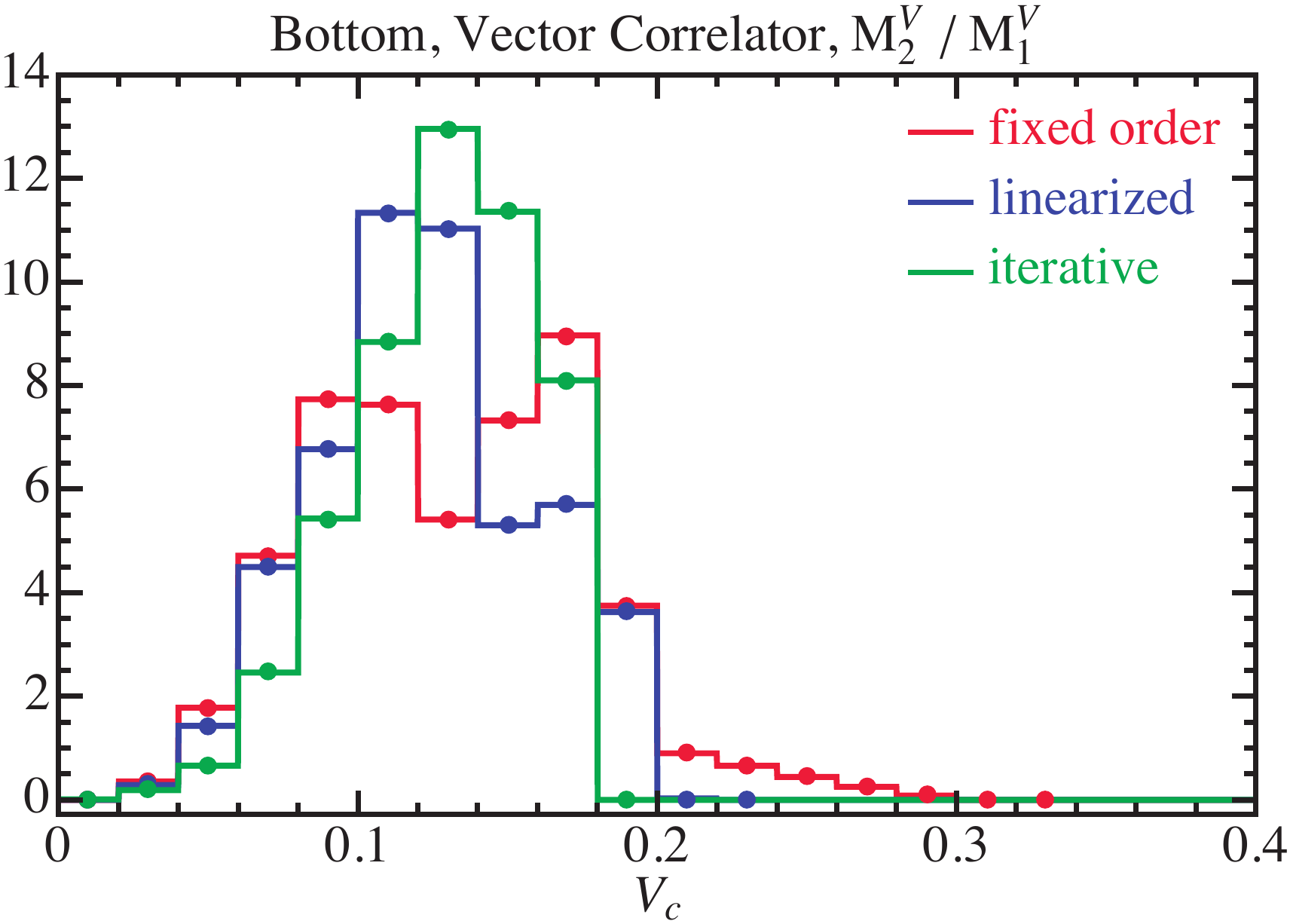}
\label{fig:histograms-bottom-ratio}}
\caption{$V_c$ distribution for $\overline{m}_c(\overline{m}_c)$ from the first moment of the 
vector (a) and pseudoscalar (b) correlator, and for $\overline{m}_b(\overline{m}_b)$ for the second 
moment of the vector correlator (c), for expansions (a)\,--\,(d). The three lower panels show the 
same for the ratio of the second over the first moment for expansions (a)\,--\,(c).}
\label{fig:histograms}
\end{figure*}

\begin{figure*}[t!]
\subfigure[]
{
\includegraphics[width=0.31\textwidth]{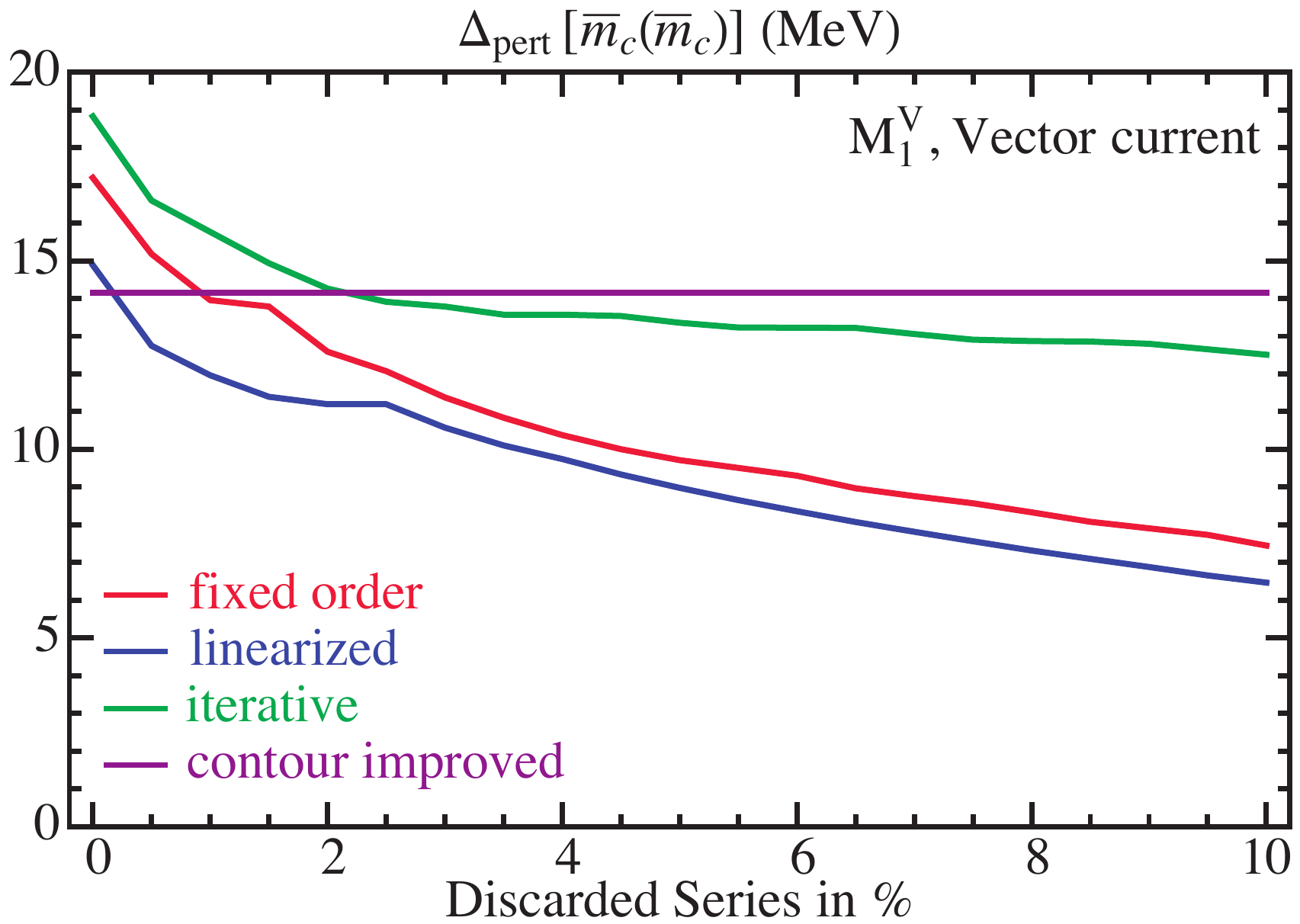}
\label{fig:trimming-vector}}
\subfigure[]
{
\includegraphics[width=0.31\textwidth]{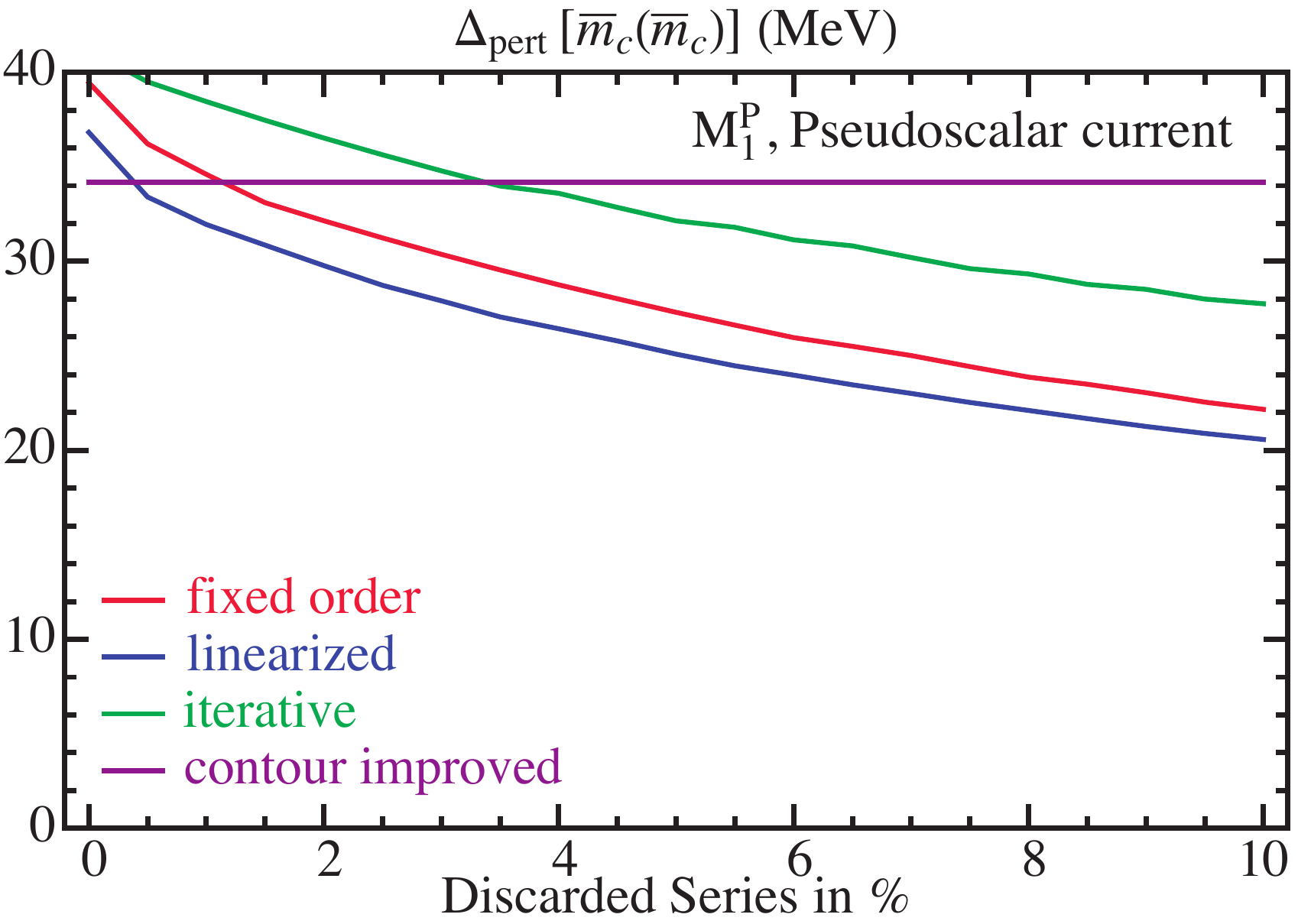}
\label{fig:trimming-pseudo}}
\subfigure[]
{
\includegraphics[width=0.31\textwidth]{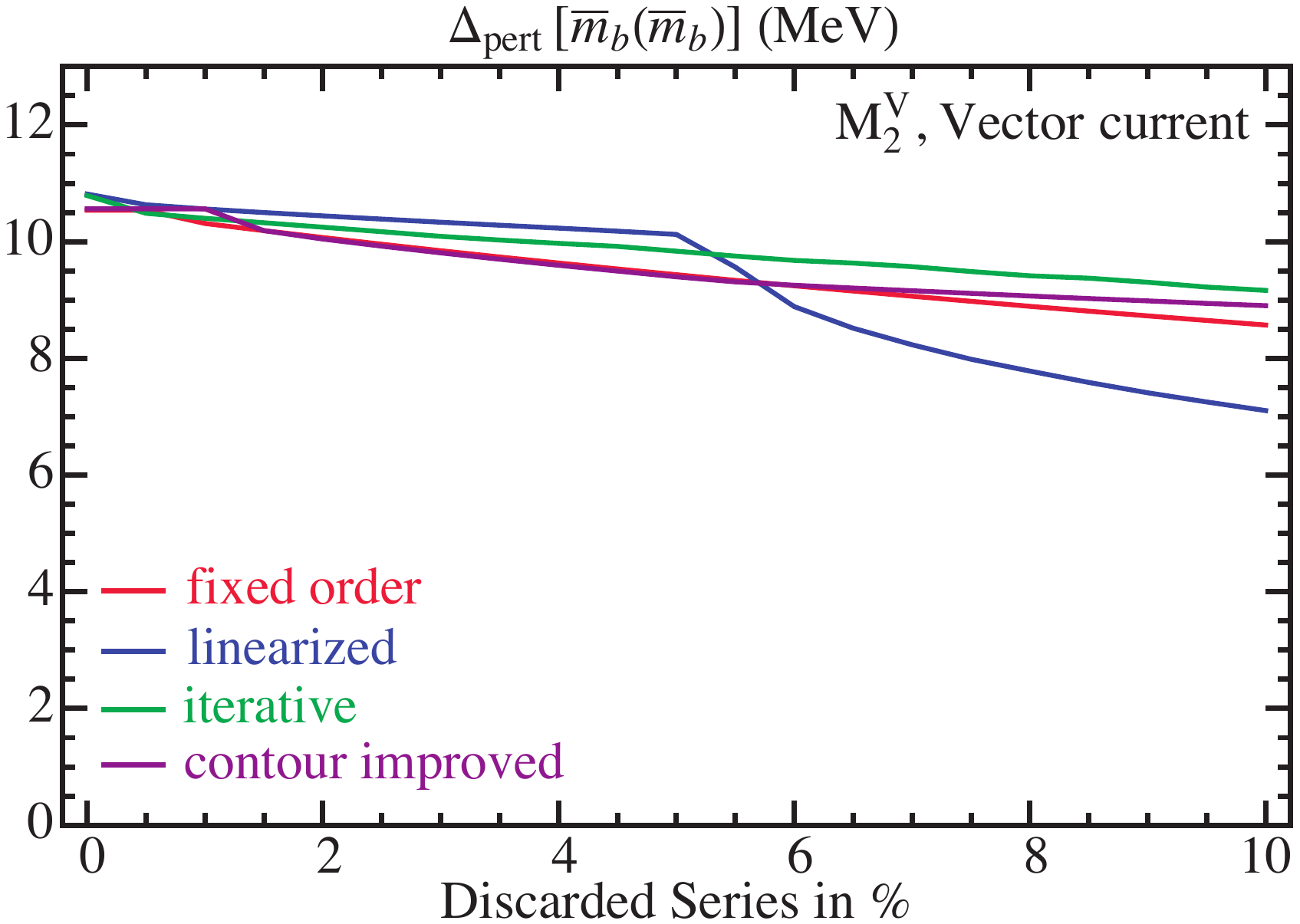}
\label{fig:trimming-bottom}}
\subfigure[]
{
\includegraphics[width=0.31\textwidth]{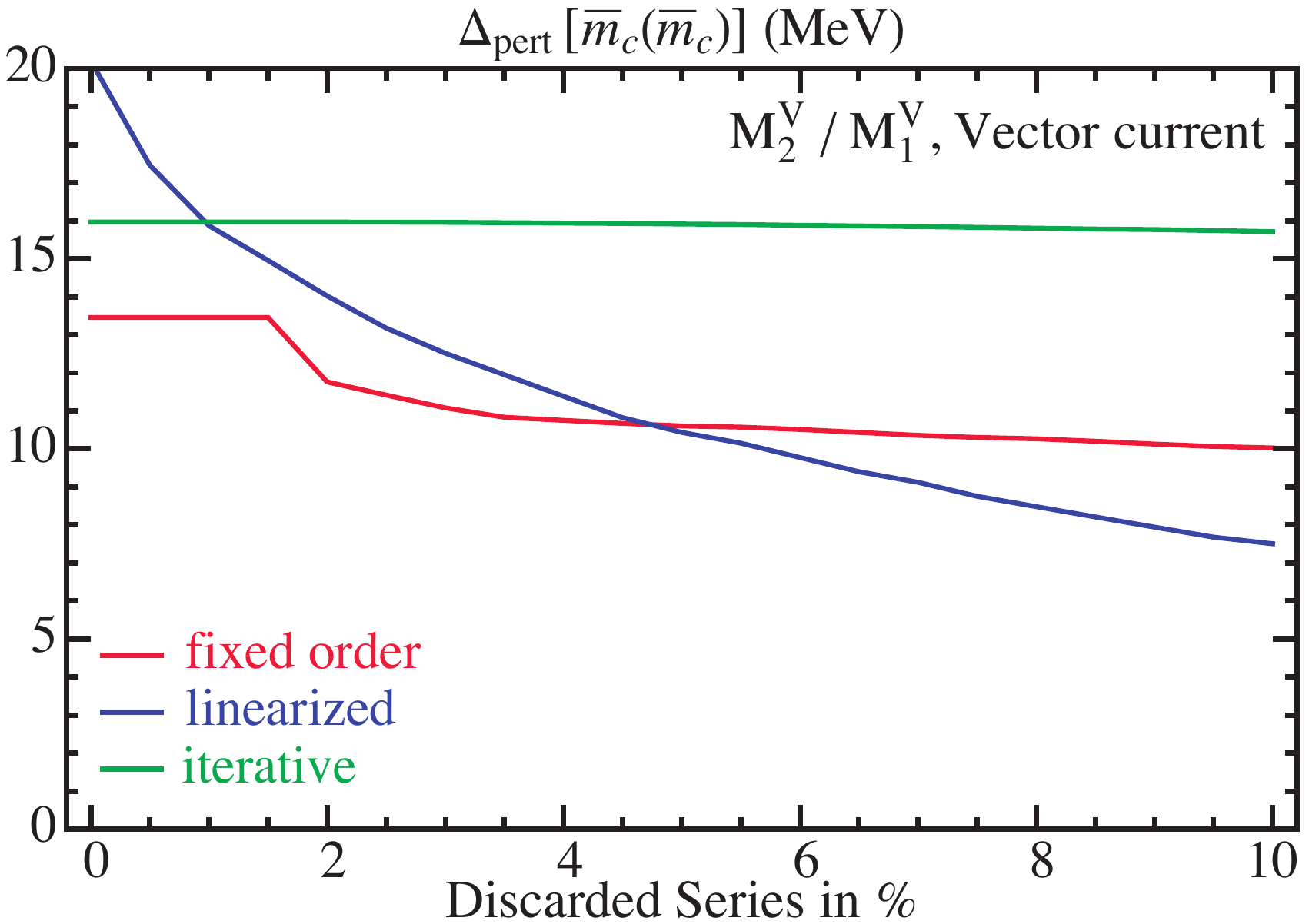}
\label{fig:trimming-vector-ratio}}
\subfigure[]
{
\includegraphics[width=0.31\textwidth]{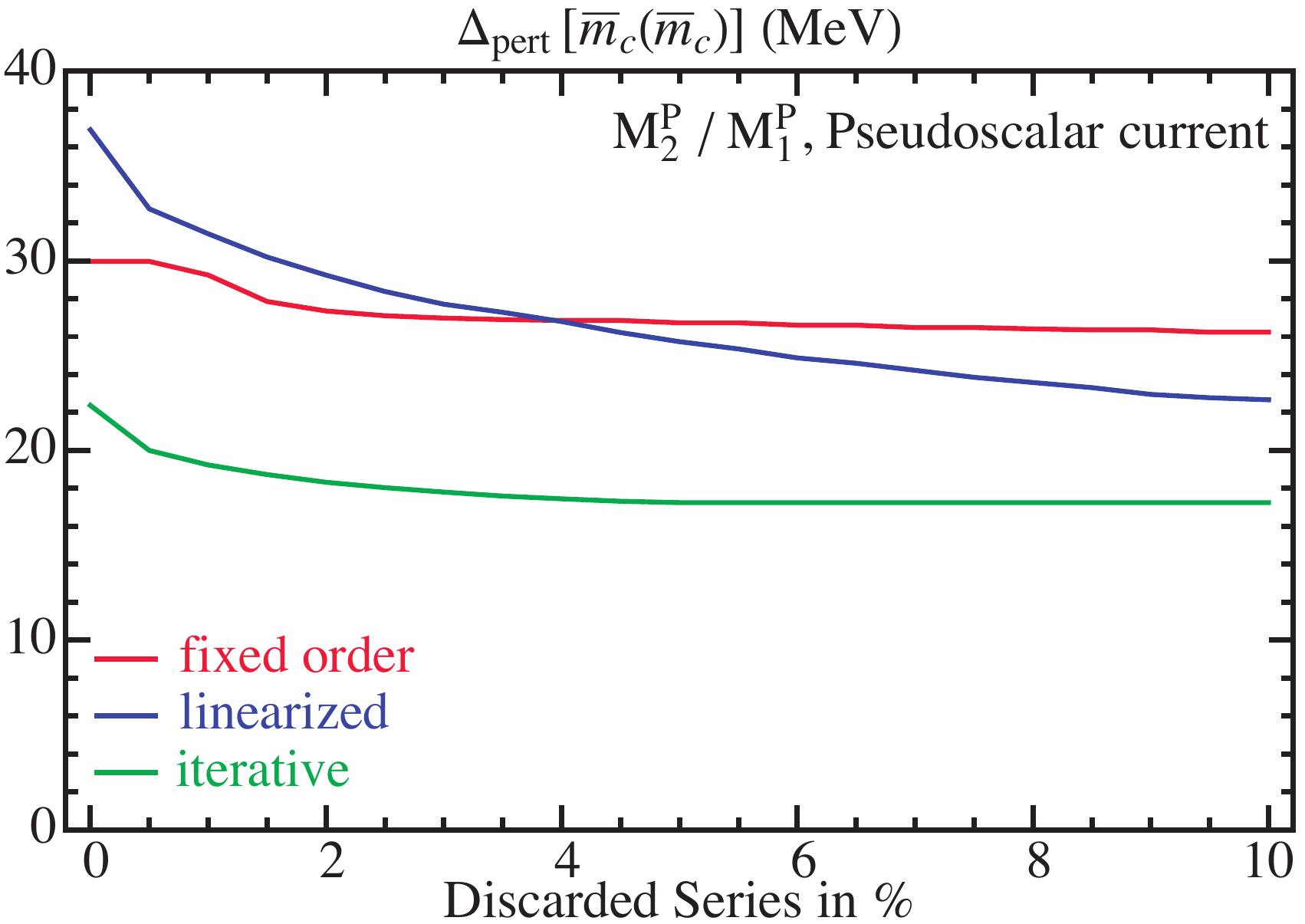}
\label{fig:trimming-pseudo-ratio}}
\subfigure[]
{
\includegraphics[width=0.31\textwidth]{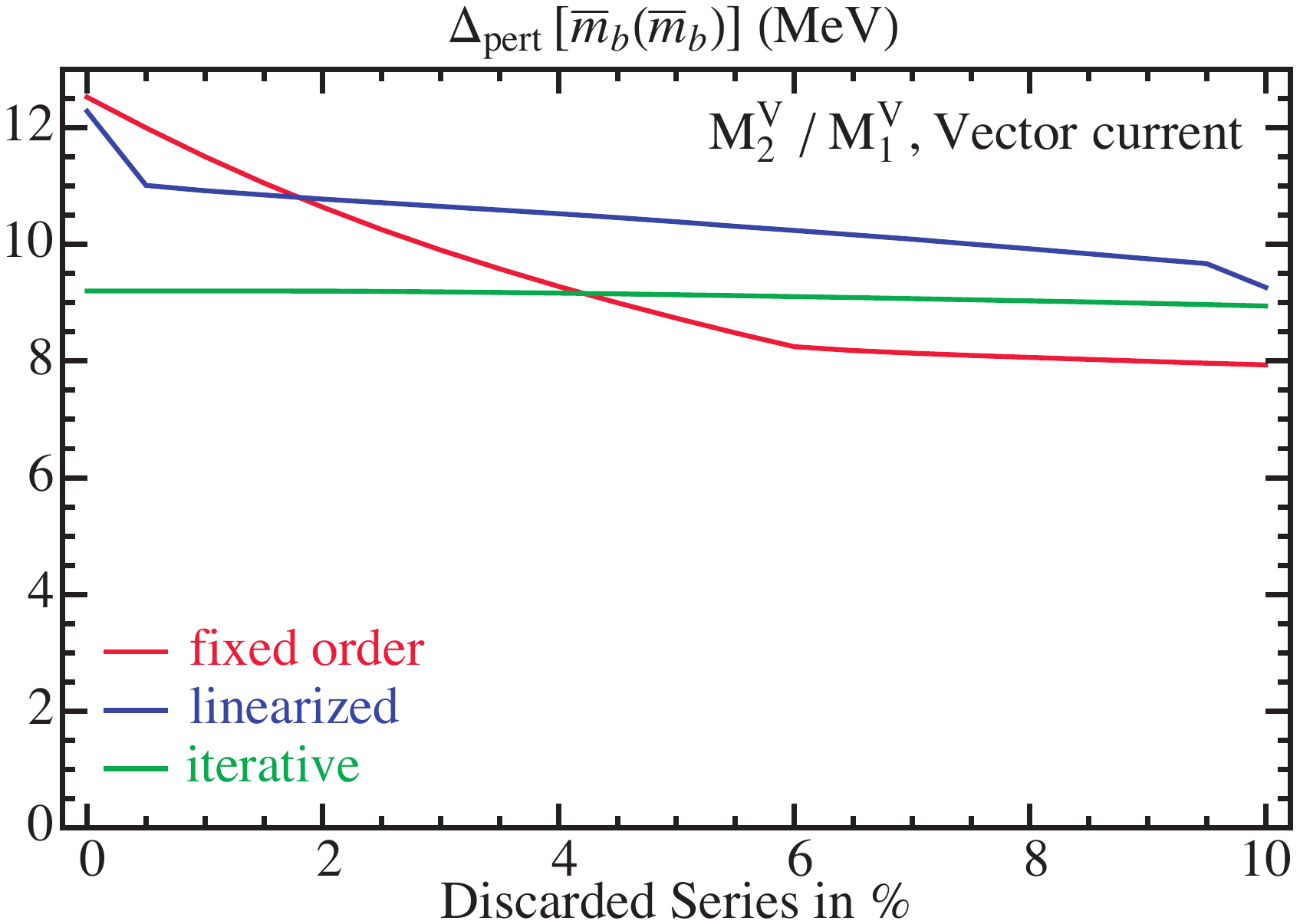}
\label{fig:trimming-bottom-ratio}}
\caption{Half of the scale variation of $\overline{m}_q(\overline{m}_q)$ at 
${\mathcal{O}(\alpha_s^3)}$ as a function of the fraction of the discarded series with highest 
$V_c$ values for the first moment of the vector (a) and pseudoscalar (b) correlators for charm, the 
second moment of the vector correlator for bottom (c); the ratio of the second over the first 
moment for the vector [charm (d) and bottom (f)] and pseudoscalar [charm (e)] correlators.}
\label{fig:trimming}
\end{figure*}

Fig.~\ref{fig:histograms-vector} shows the $V_c$ distributions for expansions (a)\,--\,(d) for the 
extraction of the charm mass from the vector moment $M_1^V$. We find $\langle V_c\rangle_{\rm 
double}=(0.15,\,0.15,\,0.17,\,0.19)$\footnote{Interestingly, the same analysis for the correlated 
variation with $2\,{\rm GeV} \leq \mu_\alpha=\mu_m \le 4$\,GeV yields \mbox{$\langle 
V_c\rangle_{\rm corr}=(0.14,\,0.16,\,0.19,\,0.19)$}, which is similar to the outcome for the double 
variation. This same observation can be made for the rest of the correlators.} and that the 
distributions are peaked around $\langle V_c\rangle$, indicating a very good overall convergence. 
The scale variation error (defined as half the overall variation) as a function of the fraction of 
the series (with the largest $V_c$ values) that are being discarded is shown in 
Fig.~\ref{fig:trimming-vector}. We see that only around 2\% of the series with the highest $V_c$ 
values by themselves cause the increase of the scale variation from well below $15$\,MeV to up 
to $20$\,MeV. These series are located at the upper-left and lower-right corners of 
Figs.~\ref{fig:mccontour1} and \ref{fig:mbcontour1}, and Fig.~6 of Ref.~\cite{Dehnadi:2011gc}, 
corresponding to values of $\mu_m$ and $\mu_\alpha$ far from each other. Given that these series 
have very large $V_c$ values and do not reflect the overall good convergence behavior of the bulk 
of the series, it is justified to remove them from the analysis.

The $V_c$ distributions for the pseudoscalar first moment $M_1^P$ are shown in 
Fig.~\ref{fig:histograms-pseudo}, again showing a clear peak. However, with $\langle 
V_c\rangle_{\rm double}=(0.24,\,0.24,\,0.25,\,0.21)$ [for correlated variation $\langle 
V_c\rangle_{\rm corr}=(0.22,\,0.23,\,0.22,\,0.15)$ with $2$\,GeV as the lower bound], the average 
$V_c$ values are significantly larger than for the vector correlator, indicating that the overall 
perturbative convergence for the pseudoscalar moment is still excellent but worse than for the 
vector moment. This means that the vector correlator method is superior, and we expect that the 
perturbative uncertainty in the charm mass from the pseudoscalar is larger. This expectation is 
indeed confirmed as we discussed in Sec.~\ref{sec:previous-results}, see also 
Sec.~\ref{sec:results}. Fig.~\ref{fig:trimming-pseudo} shows that the effect of discarding the 
series with the worst convergence is very similar to that of the vector correlator.

For our determination of the bottom mass we use the second moment $M_2^V$ (see Sec.~\ref{sec:exp} 
for a discussion on why we discard the first moment), and employ uncorrelated scale variations in 
the range $\overline m_b(\overline m_b) \leq \mu_m , \mu_\alpha \leq 15$\,GeV. 
Fig.~\ref{fig:histograms-bottom} shows the corresponding histograms, and we find that the 
convergence test yields $\langle V_c\rangle_{\rm double}=(0.13,\,0.11,\,0.12,\,0.15)$ for 
expansions (a)\,--\,(d) [for the correlated variation with scales set equal and $5\,{\rm GeV} \leq 
\mu_\alpha=\mu_m \le 15$\,GeV we find $\langle V_c\rangle_{\rm corr}=(0.13,\,0.09,\,0.13,\,0.15)$]. 
As expected, the averages for the bottom are much smaller than for the charm. We further find that 
discarding series with the highest $V_c$ values only has minor effects on the perturbative error 
estimate for fractions up to $5\%$, see Fig~\ref{fig:trimming-bottom}. This is a confirmation that 
the series for bottom moments overall are more stable, which is again expected from the fact that 
perturbation theory should work better for the bottom than for the lighter charm.

The behavior of the ratios of moments is very similar as that for regular moments, as can be seen 
in Figs.~\ref{fig:trimming} and \ref{fig:histograms}, panels (d)\,--\,(f). We find the following 
average values for $V_c$ for methods (a)\,--\,(d): ratios of charm vector moments $\langle 
V_c\rangle_{\rm double}=(0.19,\,0.18,\,0.19)$ [$\langle V_c\rangle_{\rm  corr} = 
(0.16,\,0.16,\,0.23)$]; ratios of charm pseudoscalar moments $\langle V_c\rangle_{\rm 
double}=(0.25,\,0.23,\,0.18)$ [$\langle V_c\rangle_{\rm corr}=(0.25,\,0.20,\,0.16)$]; ratios of 
bottom vector moments $\langle V_c\rangle_{\rm double}=(0.13,\,0.12,\,0.13)$ [$\langle 
V_c\rangle_{\rm corr}=(0.13,\,0.11,\,0.14)$]. Therefore we conclude that the perturbative 
convergence of the ratios of moments is in general terms a bit worse than that of regular moments, 
except for the linearized iterative method of the pseudoscalar ratios.

\begin{figure*}[t!]
\subfigure[]
{
\includegraphics[width=0.31\textwidth]{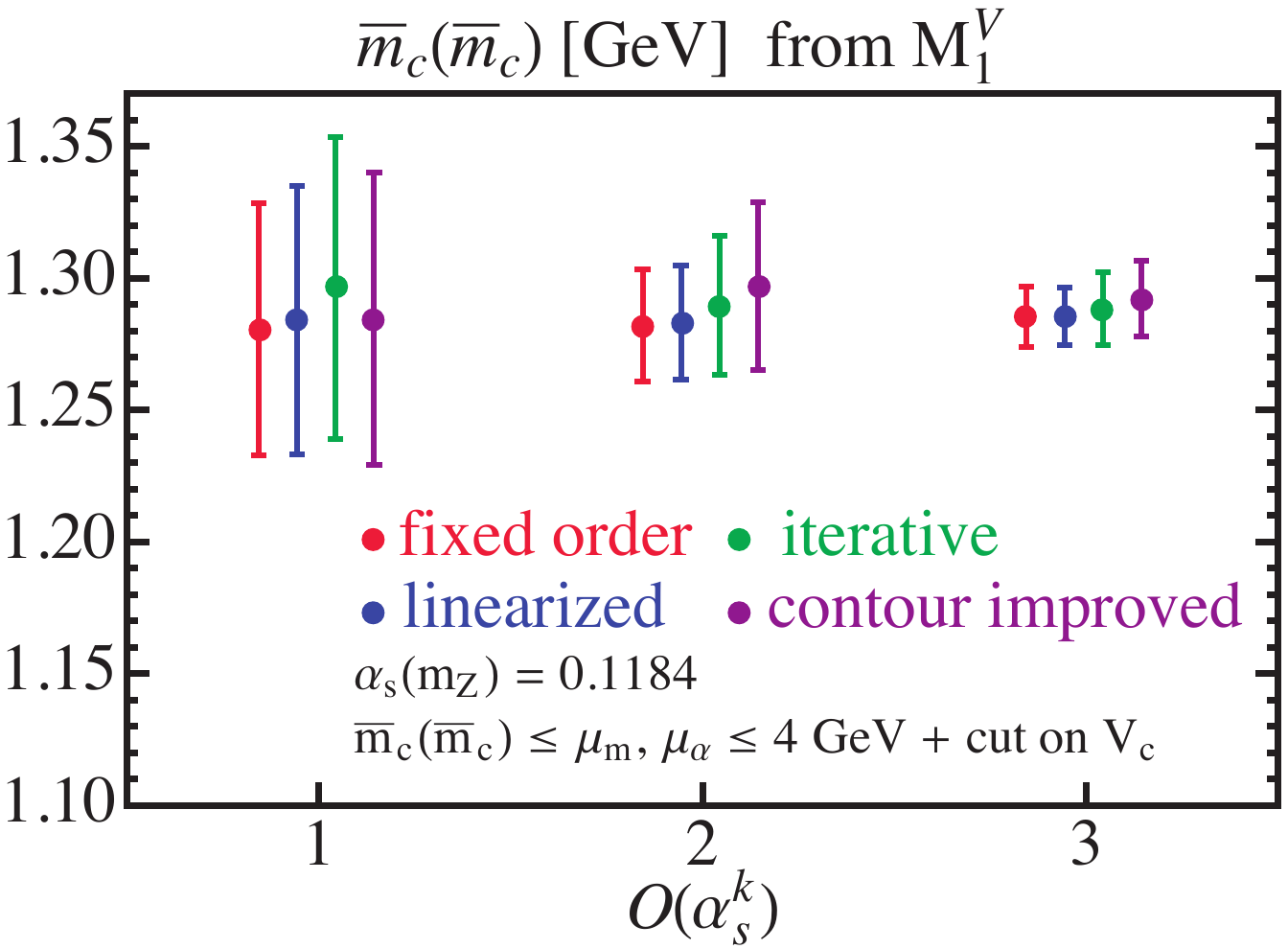}
\label{fig:order-plot-mc-vec}}
\subfigure[]
{
\includegraphics[width=0.31\textwidth]{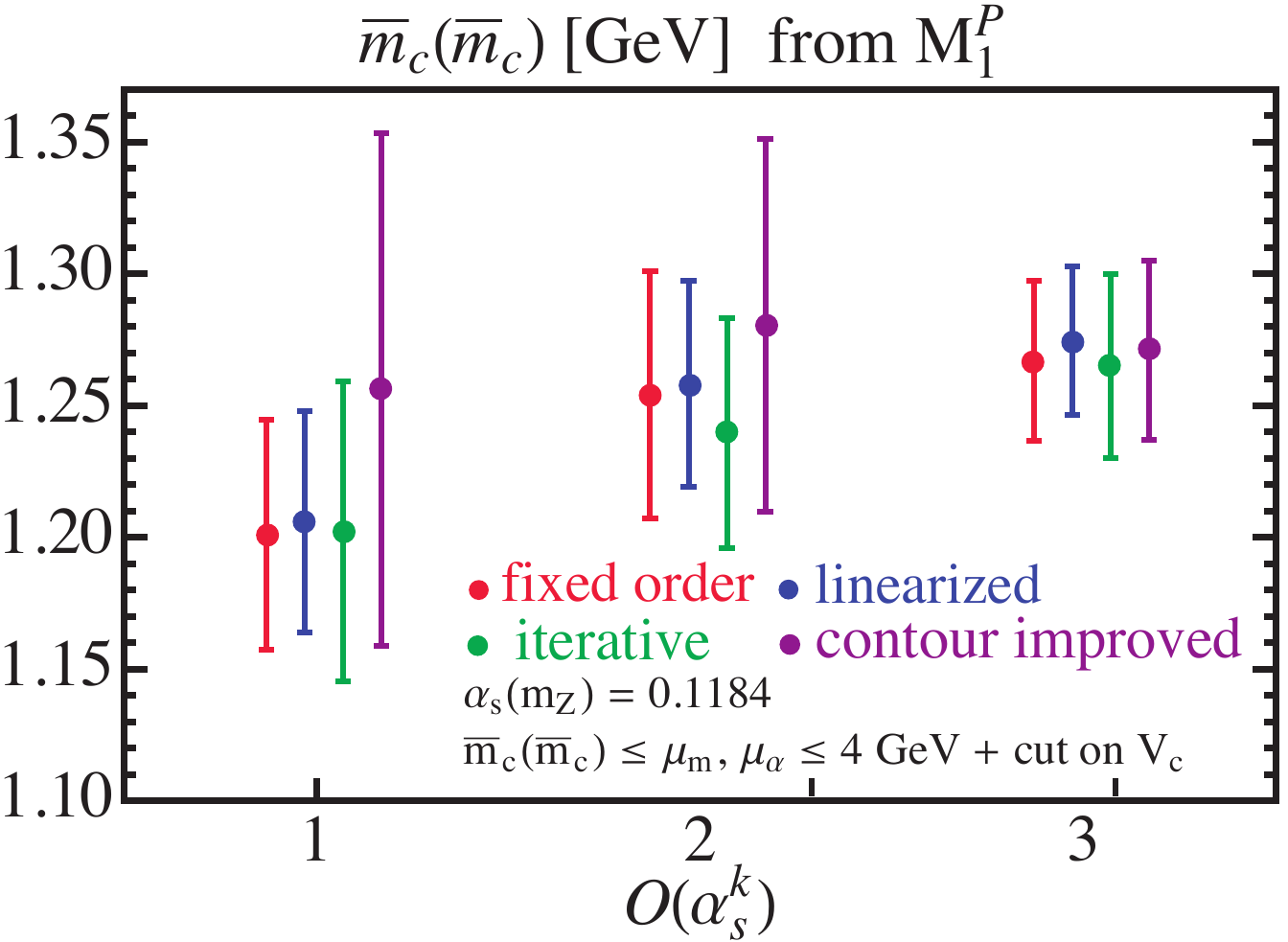}
\label{fig:order-plot-mc-pseudo}}
\subfigure[]
{
\includegraphics[width=0.31\textwidth]{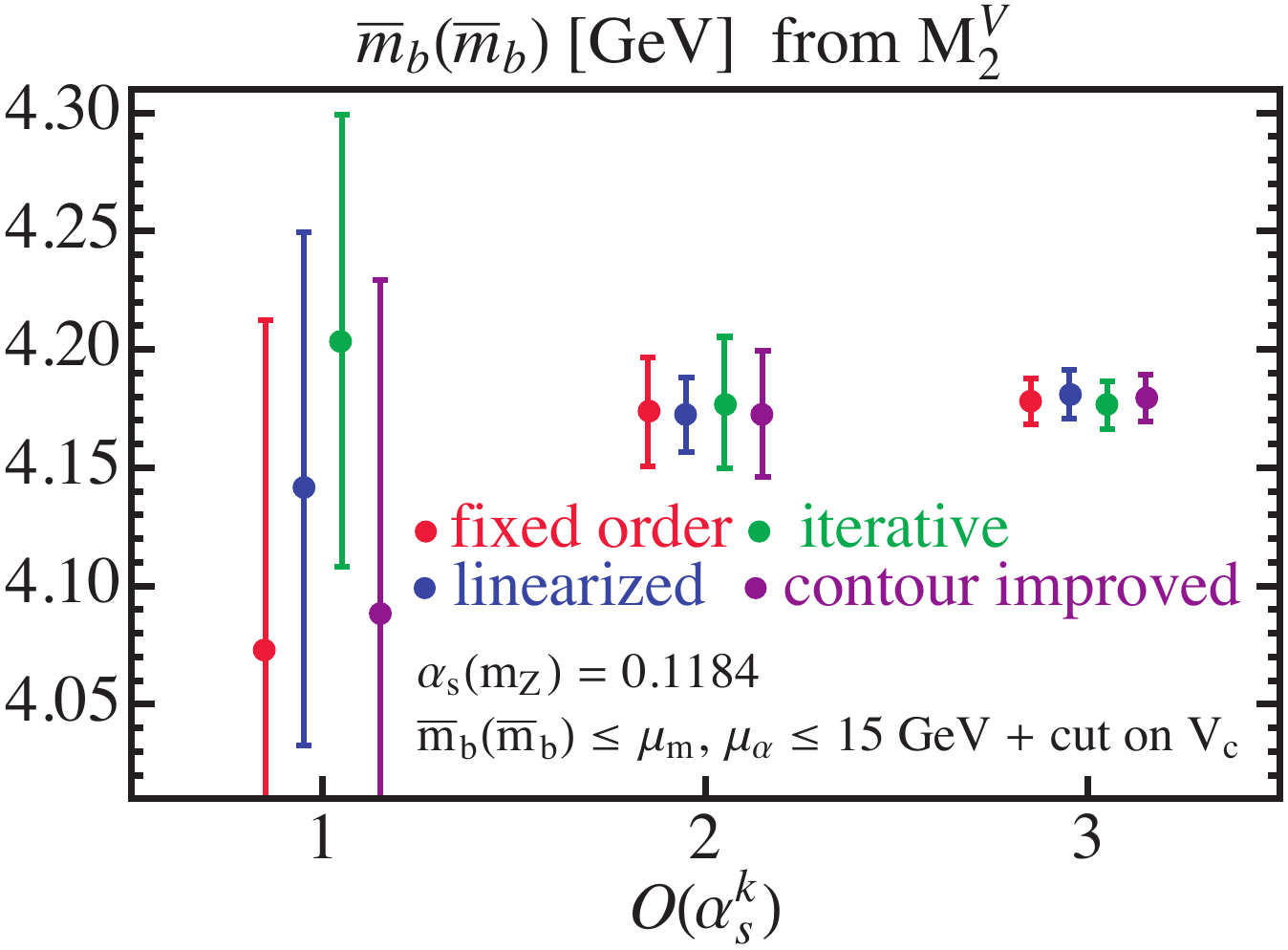}
\label{fig:order-plot-mb-vec}}
\subfigure[]
{
\includegraphics[width=0.31\textwidth]{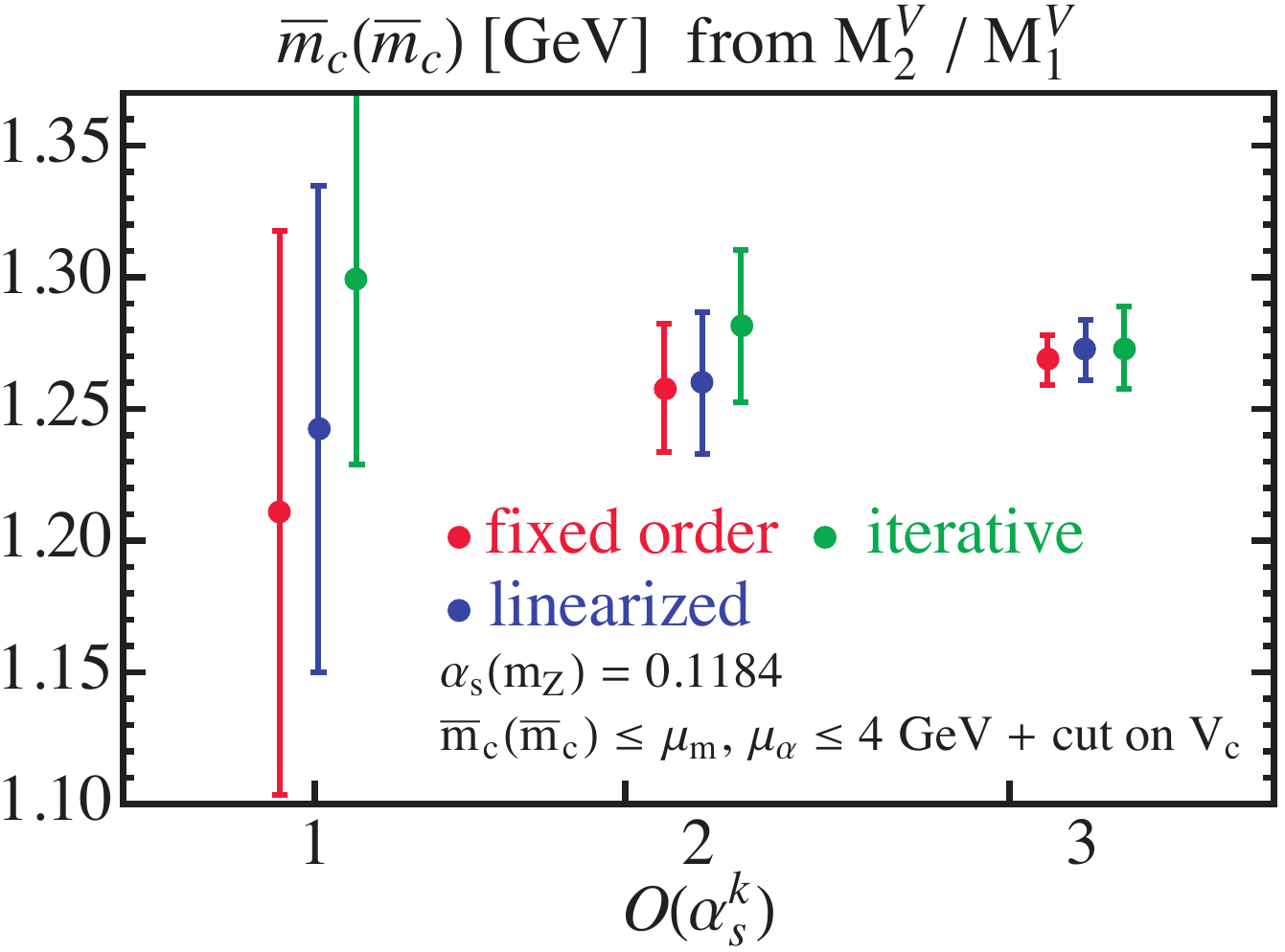}
\label{fig:order-plot-mc-vec-rat}}
\subfigure[]
{
\includegraphics[width=0.31\textwidth]{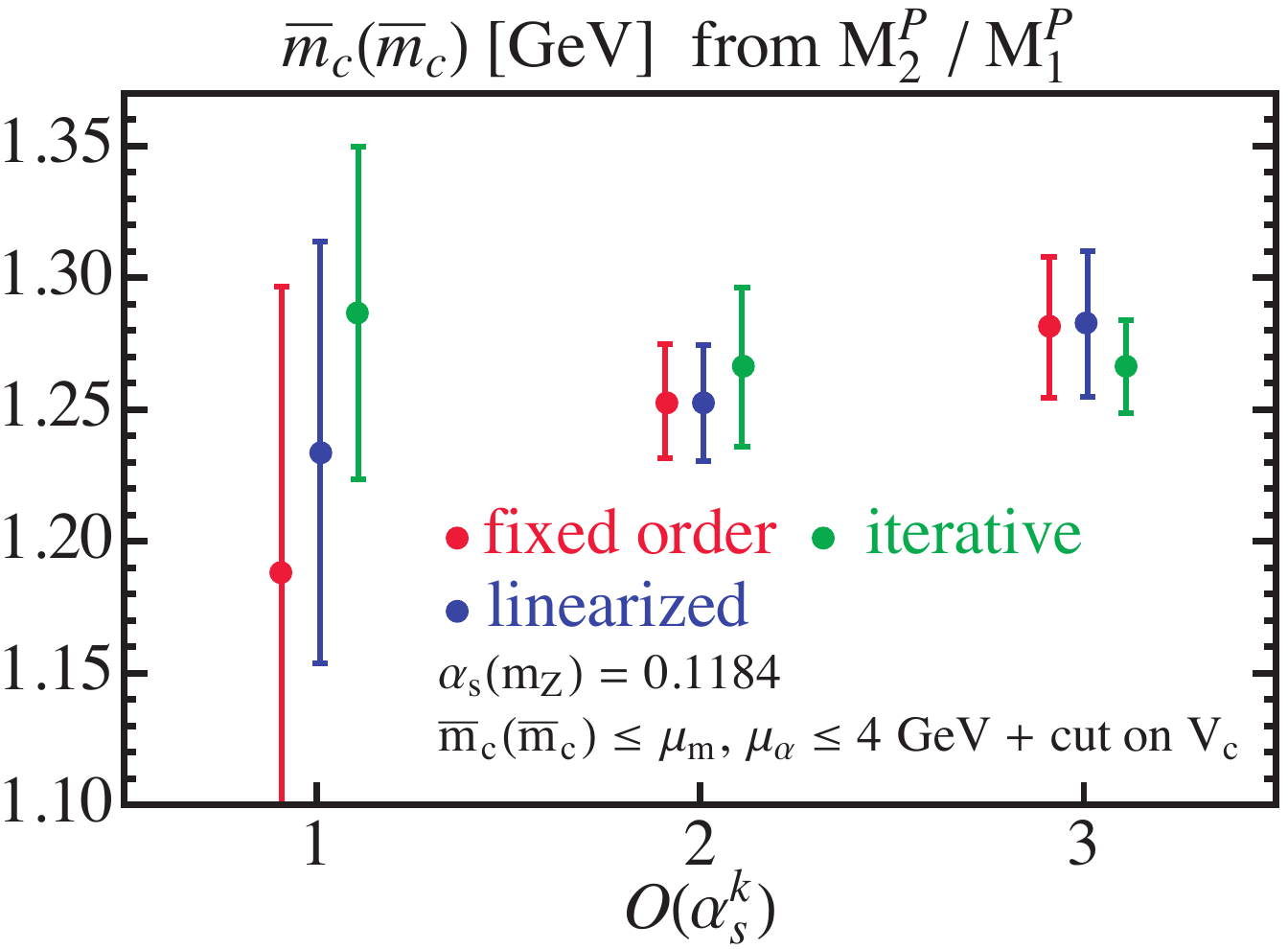}
\label{fig:order-plot-mc-pseudo-rat}}
\subfigure[]
{
\includegraphics[width=0.31\textwidth]{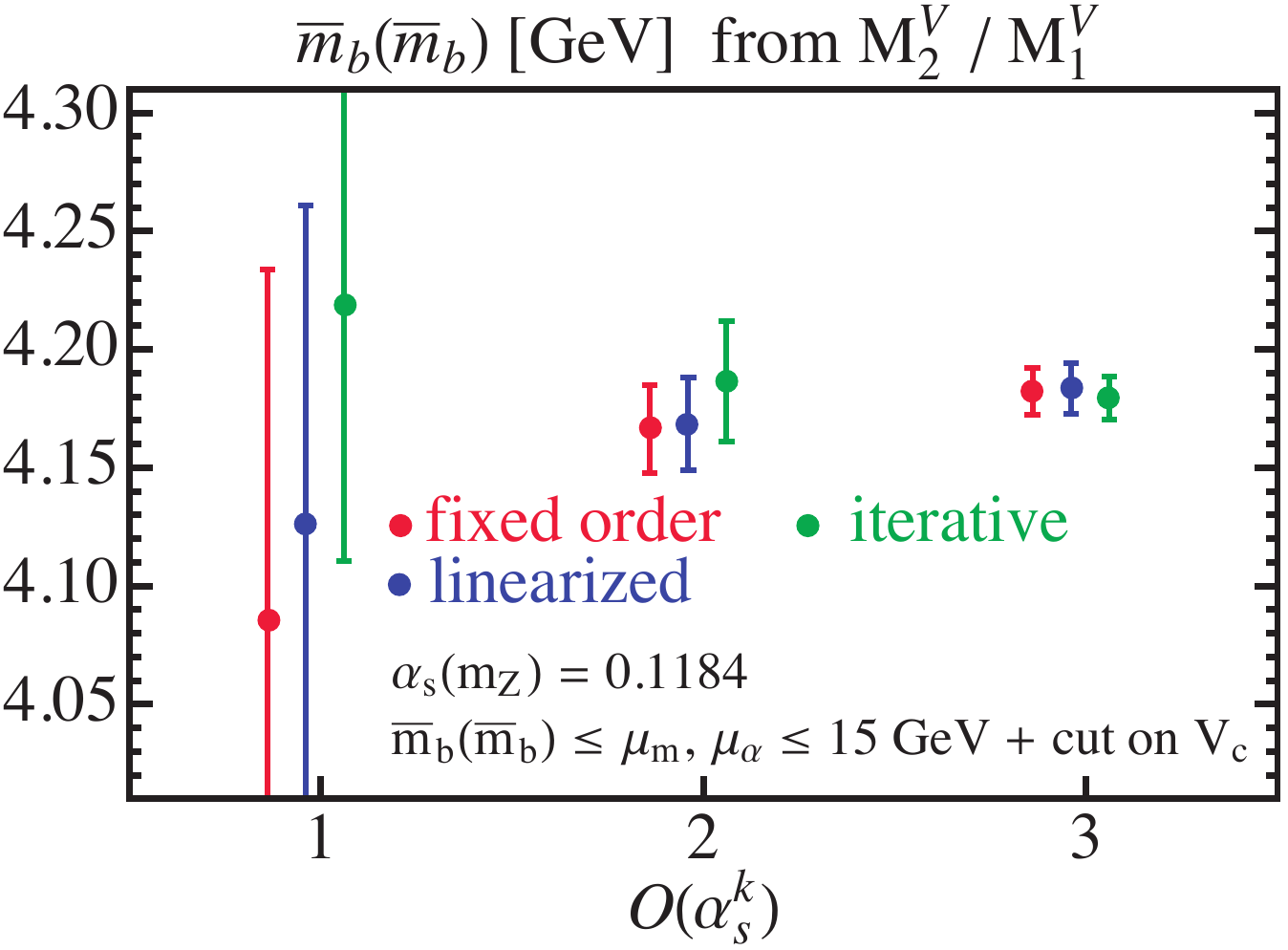}
\label{fig:order-plot-mb-vec-ratio}}
\caption{Charm and bottom mass values from the first [second] moment of the vector (a) for charm 
[(c) for bottom] and pseudoscalar [(b), charm] currents at $\mathcal{O}(\alpha_s^{1,2,3})$; and for 
the ratio of the second over the first moment for the vector [(d) for charm, (g) for bottom] and 
pseudoscalar [(e), charm] correlators for expansions (a)\,--\,(d) [(a)\,--\,(c) for ratios], in 
red, blue, green and purple, respectively.}
\label{fig:order-plots}
\end{figure*}

In our final numerical analyses we discard $3$\% of the series with the worst $V_c$ values. As can 
be seen from Fig.~\ref{fig:histograms}, this only affects series with $V_c$ values much larger than 
the average values for the whole set of series. It is our intention to keep the fraction of 
discarded series as small as possible, since it is our aim to remove only series with convergence 
properties that are obviously much worse than those of the bulk of the series. We call this 
procedure {\it trimming} in the following. As we see in Figs.~\ref{fig:order-plots}, the results 
including the trimming  show a very good order-by-order convergence for the heavy quark mass 
determinations, and at each order every expansion method gives consistent results for the central 
values as well as for the estimate of the perturbative uncertainties. 
Figs.~\ref{fig:order-plot-mc-vec} and \ref{fig:order-plot-mc-pseudo} show the results for 
$\overline{m}_c(\overline{m}_c)$ from the vector and pseudoscalar correlators, respectively, for 
expansions (a)\,--\,(d) at ${\cal O}(\alpha_s^{1,2,3})$ and with 
$\overline{m}_c(\overline{m}_c)\le\mu_m,\mu_\alpha\le 4$\,GeV, using the first moment. 
Figs.~\ref{fig:order-plot-mc-vec-rat} and \ref{fig:order-plot-mc-pseudo-rat} show results for 
methods (a)\,--\,(c), using the ratio of the second over the first moment. Analogously, 
Figs.~\ref{fig:order-plot-mb-vec} and \ref{fig:order-plot-mb-vec-ratio} show the results for 
$\overline{m}_b(\overline{m}_b)$ for the second moment, and the ratio of the second over the 
first moment, respectively, with the uncorrelated variation 
$\overline{m}_b(\overline{m}_b)\le\mu_m,\mu_\alpha\le 15$\,GeV.

\section{Lattice Simulation Data}
\label{sec:lattice-data}

The pseudoscalar current is not realized in nature in a way which is useful to compute the moments 
of the corresponding correlator from experimental data. Results for the moments of the pseudoscalar 
current correlator can, however, be obtained from simulations on the lattice. The strategy of these 
numerical simulations is to tune the lattice parameters (such as bare coupling constant and masses) 
to a selected number of observables (e.g.\ the energy splitting of $\Upsilon$ resonances). Once this 
tuning is performed, the lattice action is fully specified and no further changes are implemented. 
The tuned lattice action can then be used to perform all sorts of predictions, moments of 
correlators among them. Lattice simulations have to face a number of challenges, which usually 
translate into sizeable uncertainties. Among those are the extrapolations to the infinite volume 
and the zero lattice spacing (the latter being much harder), as well as the extrapolation to 
physical quark masses. On top of these systematic uncertainties, there are also statistical ones, 
which are related to the finite sampling used to perform the path integrations. On the other hand 
there are also concerns on the type of lattice action which is being used for the fermions. 
According to Ref.~\cite{Allison:2008xk}, the moments of the pseudoscalar correlator are least 
affected by systematic uncertainties, and so HPQCD has focused on those for their subsequent 
high-precision analyses.

To the best of our knowledge, HPQCD is the only lattice collaboration which has published results 
on QCD correlators. They have used the staggered-quarks lattice action, and MILC configurations for 
gluons. These results have been used to determine the charm mass and the strong coupling 
constant~\cite{Allison:2008xk,McNeile:2010ji,Chakraborty:2014aca} with high accuracy, as well as 
the bottom mass~\cite{McNeile:2010ji,Colquhoun:2014ica}, with smaller precision. We will use the 
simulation results as given in Ref.~\cite{Allison:2008xk}, even if the results quoted in of 
\cite{McNeile:2010ji,Chakraborty:2014aca} are a bit more precise. The reason for this choice is that 
while \cite{Allison:2008xk} makes a straightforward extrapolation to the continuum, which is 
independent of the charm mass and $\alpha_s$ extractions, in 
\cite{McNeile:2010ji,Chakraborty:2014aca} the fit for the quark masses, the strong coupling 
constant and the extrapolation to the continuum is performed all at once, in a single fit. 
Furthermore that fits contains a lot of priors for the parameters one is interested in fitting.
In any case, as we have seen, the charm mass extraction from the pseudoscalar correlator is 
dominated by perturbative uncertainties, as a result of the bad convergence of the series expansion 
for its moments.

Ref.~\cite{Allison:2008xk} provides simulation results for the so-called reduced moments $R_k$, 
which are collected in their Table~II. The index $k$ takes only even values, and starts with the 
value $k = 4$, which is fairly insensitive to the charm mass. Hence the lowest moment we consider 
is $R_6$. Reduced moments are defined as (up to a global power) the full moment divided by the 
tree-level result. By taking this ratio, the authors of Ref.~\cite{Allison:2008xk} claim that large 
cancellations between systematic errors take place. The reduced moments are scaleless, and the 
mass-dimension that one obviously needs to determinate the charm quark mass is regained by dividing 
with the mass of the $\eta_c$ pseudoscalar particle. Thus one can easily translate the reduced 
moments into the more familiar correlator moments $M_n^P$ with the following relation:

\begin{equation}
M_n^P = [C_P(n_f=4)]^{0,0}_{n,0}\bigg(\dfrac{R_{2n+4}}{m_{\eta_c}}\bigg)^{\!2n}\,,
\end{equation}

where the $C_P$ coefficients correspond to the tree-level terms of the standard fixed-order 
expansion of Eq.~(\ref{eq:Mn-theo-FO}). Although the experimental value for $\eta_c$ is 
$2.9836(7)\,$GeV, we use the value $m_{\eta_c} = 2.980\,$GeV given in Ref.~\cite{Allison:2008xk}, 
in order to ease comparison with that analysis. In Ref.~\cite{Chakraborty:2014aca} the value 
$m_{\eta_c} = 2.9863(27)\,$GeV is used. It is claimed that (as for the lattice action) it has no 
QED effects, and the error accounts for $c\bar{c}$ annihilation. Using the quote in 
Ref.~\cite{Chakraborty:2014aca} changes $M_1^P$ by $0.4\%$ and the effect on the charm mass is of 
the order of $2$\,MeV. The uncertainty in the $\eta_c$ mass has no effect on the $M_n^P$ errors. In 
Table~\ref{tab:Latt} we quote the lattice simulation results written as regular moments $M_n^P$.

\begin{table}[tbh!]
\center
\begin{tabular}{|cccc|}
\hline 
$M_1^P$ & $M_2^P$ & $M_3^P$ & $M_4^P$ 
\tabularnewline\hline
$1.402(20)$ & $1.361(40)$ & $1.426(59)$ & $1.558(89)$
\tabularnewline
\hline
\end{tabular}
\caption{Lattice simulation results for the moments of the twice-subtracted pseudoscalar correlator 
$P(q^2)$ for $n_f = 4$. Moments given in units of $10^{-n}\times$\,GeV$^{-2n}$.\label{tab:Latt}}
\end{table}

\section{Computation of the Experimental Moments for the Bottom Correlator}
\label{sec:exp}

In this section we present our computation of the moments for the bottom vector current correlator 
from experimental data. These are made of three distinct contributions: the narrow resonances below 
threshold, the region of broader resonances, explored by BABAR \cite{:2008hx}, and the continuum 
region, where no data has been taken and some modeling is required. The BABAR data has to be 
corrected for initial-state radiation and vacuum polarization effects. In the continuum region 
we use a model which consists of a combination of a linear fit to the BaBar experimental points 
with energy larger than $11.05\,$GeV, and perturbation 
theory as a model for missing experimental data, which are joined smoothly by a cubic interpolation.
We assign a conservative uncertainty guided by the error function of the linear fit to the BABAR 
data in the region with measurements with the highest energy. Our 
determination of the experimental bottom moments differs from Ref.~\cite{Chetyrkin:2009fv} in 
the way we model the uncertainties for the hadronic cross section in the continuum region, plus 
other minor differences in the contributions from the narrow widths and the threshold region, see 
discussion in Sec.~\ref{sec:comp-exp}. We also provide the correlation matrix among different 
moments, which cannot be found in the literature. Therefore, even though there are some 
similarities with the computations outlined in \cite{Chetyrkin:2009fv}, we find it justified to 
discuss our computation of the experimental moments in some detail.

We note that our results for the moments of the bottom vector current correlator have already been 
used in the analyses of high-$n$ moments of Ref.~\cite{Hoang:2012us}, in the context of 
nonrelativistic large-$n$ sum rules.

\subsection{Narrow Resonances}

The contribution of resonances below the open bottom threshold $\sqrt{s}=10.62\,\rm{GeV}$ includes 
$\Upsilon (1S)$ up to $\Upsilon(4S)$. We use the narrow width approximation to compute their 
contribution to the experimental moments, finding

\begin{equation}
M_n^{\rm res} = \frac{9\pi \Gamma_{ee}}{\alpha(M_\Upsilon)^2 M_\Upsilon^{2n+1}}\,.
\end{equation}

The masses and electronic widths of these four resonances are taken from the PDG 
\cite{Agashe:2014kda}, and the values of the effective electromagnetic coupling constant evaluated 
at the $\Upsilon$ masses are taken from Ref.~\cite{Kuhn:2007vp}. This information is collected in 
Table~\ref{tab:NarrowRes}. We have also checked that if one uses a Breit-Wigner instead of the 
narrow width approximation the results change by an amount well within the error due to the 
uncertainty in the electronic width.

In analogy to what we found in our study of the charm moments \cite{Dehnadi:2011gc}, the effect of 
the mass uncertainty in the moments is negligible. Therefore one only needs to consider the 
experimental uncertainty in the electronic widths. There is no information on the correlation 
between the measurements of these widths. The PDG averages of the electronic partial widths for the 
first three resonances is dominated by the CLEO 
measurement~\cite{Rosner:2005eu}.\footnote{Refs.~\cite{Kuhn:2007vp,Chetyrkin:2009fv} assume that 
the error of the electronic width for the first three narrow resonances is $100\%$ correlated, and 
uncorrelated to that of the $\Upsilon(4S)$.} Therefore we take the approach that half of the 
width's 
uncertainty (in quadrature) is uncorrelated (therefore mainly of statistical origin), whereas the 
other half is correlated among the various resonances (therefore coming from common systematics in 
the measurements).

\begin{table}[t!]
 {
\centering
\begin{tabular}{|c|cccc|}
  \hline
  {} & $\Upsilon(1S)$ & $\Upsilon(2S)$ & $\Upsilon(3S)$ & $\Upsilon(4S)$ \\
  \hline
  $M_{\Upsilon}(\rm{GeV})$ & 9.46030(26) & 10.02326(31) & 10.3552(5) & 10.5794(12) \\
 $ \Gamma_{\rm{ee}}(\rm{KeV})$ & 1.340(18) & 0.612(11) & 0.443(8) & 0.272(29) \\
$  \Big(\frac{\alpha_{\rm QED}}{\alpha(M_{\Upsilon})}\Big)^{\!2}$ & 0.932069 & 0.93099 & 0.930811 & 
0.930093 \\
  \hline
\end{tabular}
\caption{Masses and electronic widths of narrow $\Upsilon$ resonances \cite{Agashe:2014kda} and 
effective electromagnetic coupling constant \cite{Kuhn:2007vp}. $\alpha_{\rm QED} = 
1/137.035999084(51)$ represents the fine structure constant. \label{tab:NarrowRes}}
}
\end{table}

\subsection{Threshold Region}

The region between the open bottom threshold and the experimental measurement of the 
\mbox{$R_b$-ratio} at the highest energy, $10.62\,\rm{GeV} \le \sqrt{s}\le 11.2062\,{\rm GeV}$, is 
referred to as the threshold region. The region above the last experimental measurement will be 
collectively denoted as the continuum region. The first experimental data close to the B meson 
threshold were taken by the CLEO \cite{Ammar:1997sk, Besson:1984bd, Huang:2006em} and CUSB 
\cite{Lovelock:1985nb} collaborations. The measurements at each c.m.\ energy have a $6 \% $ 
systematic uncertainty. More recently the BABAR collaboration~\cite{:2008hx} has measured the 
\mbox{$R_b$-ratio} in the energy region between $10.54\,\rm{GeV}$ and $11.20\,\rm{GeV}$, with 
significantly higher statistics and better control of systematic uncertainties (of the order of 
$3\%$). These measurements are taken in small energy bins, densely populating the threshold region. 
The BABAR data supersedes the older data of CLEO and CUSB, and it has already been used in 
Refs.~\cite{Chetyrkin:2009fv, Chetyrkin:2010ic, Bodenstein:2012}, in which the bottom mass was also 
determined.

This BABAR data for the \mbox{$R_b$-ratio} has not been corrected for initial-state radiation and 
vacuum polarization effects. Moreover, the effect of the $\Upsilon(4S)$ resonance has not been 
subtracted,\footnote{The radiative tails of the first three resonances are provided by BABAR, so 
they can be subtracted at the data level, before correcting for vacuum polarization effects.} so 
we have performed the subtraction ourselves, using the Breit-Wigner approximation and using for the 
total width the PDG value $\Gamma_{4S} = 20.5$\,MeV:

\begin{equation}
\label{eq:Y4S}
R^{\rm{BW}}(s)= \frac{9\, M_{4S}^2\, \Gamma^{4S}_{\rm{ee}}}{\alpha(M_{4S})^2}
\frac{\Gamma_{4S}}{(s-M_{4S}^2)^2+\Gamma_{4S}^2 M_{4S}^2}\, .
\end{equation}

For the subtraction of the $\Upsilon(4S)$ resonance and the correction for the initial state 
radiation we take an approach similar to Ref.~\cite{Chetyrkin:2009fv}.

\subsubsection{Subtraction of the $\Upsilon(4S)$ Radiative Tail}
\label{sub:Subtraction}

Before subtracting the radiative tail of the $\Upsilon(4S)$ resonance one has to account for vacuum 
polarization effects. BaBar experimental data has been normalized to the theoretical Born 
level dimuon cross section (using the fine structure constant rather than the running effective 
electromagnetic coupling), instead of normalizing to the number of events with muons in the final 
state. Therefore one has to multiply the BaBar data with $[\alpha(s)/\alpha_{\rm em}]^2$, which we 
take as constant 
with value $0.93$.

The contribution to be subtracted from the BABAR data (already corrected for vacuum 
polarization effects) is the ISR-distorted tail of the $\Upsilon(4S)$, which reaches to energies 
above its mass. The cross section $R$ and the ISR-distorted cross-section $\hat R$ are related by a 
convolution relation

\begin{equation}
\label{convolution}
\hat{R}(s)= \int_{z_0}^{1} \dfrac{{\rm d} z}{z}\, G(z,s) \ R(s \, z) \, ,
\end{equation}

which can be used to determine the ISR effects on the $\Upsilon(4S)$ resonance given in 
Eq.~(\ref{eq:Y4S}). Here the lower integration bound is $z_0=(10\,\rm{GeV})^2/s$. This value is not 
fully fixed by theoretical arguments, and it is chosen such that it excludes the narrow resonances, 
but keeps the major part of the $\Upsilon(4S)$ line shape. The radiator function $G$ is given 
as~\cite{Jadach:1988gb,Chetyrkin:1996tz}

\begin{align}
G(z,s) & =  (1-z)^{\beta(s) \,-\, 1}\, \tilde{G}(z,s)\,,\\
\widetilde{G}(z,s) & = \beta(s)\, e^{\delta_{\rm{yfs}(s)}} \, F(s) \big[\,\delta_{C}^{V+S}(s) + 
\delta_C^H(s,z) \,\big]\,,\nonumber
\end{align}

where the specific form of $\beta$, $F$ and the two $\delta$'s can be found in Eq.~(7) of 
\cite{Chetyrkin:2009fv}. Note that the function $G(z,s)$ is divergent as $z \rightarrow 1 
$, but since $0 < \beta -1 < -1 $, it is integrable. The divergent behavior is absent in 
$\widetilde{G}$, which in the limit $z\rightarrow 1$ reduces to

\begin{equation}
\widetilde{G}(1,s) \,=\, \beta(s)\,e^{\delta_{\rm{yfs}}(s)} \, F(s)\,\delta_{C}^{V+S}(s) \,.
\end{equation}

After subtracting the radiative tail of the $\Upsilon(4S)$ we find that to a good approximation the 
cross section vanishes for energies below $10.62\,\rm{GeV}$. Therefore we add an additional point 
to our BABAR dataset: $R_b(10.62\,\rm{GeV})=0$ and take $R_b=0$ for energies below $10.62$\,GeV. 
Since the subtracted cross section does not exactly vanish between $10.5408$\,GeV and $10.62$\,GeV, 
we take the (small) contribution of the subtracted cross section in that region to the moments as 
an additional source of systematic correlated uncertainty.

\subsubsection{Deconvolution of Initial-State Radiation}

After subtraction of the radiative tails and correcting for vacuum polarization effects, the BABAR 
threshold data are corrected for ISR. The inversion of the convolution in Eq.~(\ref{convolution}), 
can be carried out in an iterative way~\cite{Chetyrkin:2009fv}. Defining $\delta G(z,s)\,=\, 
G(z,s)\, -\, \delta(1-z)$ one can use a successive series of approximations

\begin{equation}\label{eq:ISR-inversion}
R^j(s)=R^0(s) -\int_{z_0}^1 \, \dfrac{{\rm d} z}{z}\,\delta
G(z,s)\,R^{j-1}(s\,z),
\end{equation}

where we denoted the $j$-th approximation of $R(s)$ as $R^{j}(s)$ and use as starting 
point \mbox{$R^0(s)=\hat{R}(s)$}, the BABAR data after correcting for vacuum polarization effects 
and subtracting the radiative tails. In Eq.~(\ref{eq:ISR-inversion}) we take 
$z_0=(10.62\,\rm{GeV})^2/s$, using as a starting point the energy value for which the cross section 
vanishes after the subtraction of the radiative tails. To isolate the singularity at the higher 
endpoint one can perform a subtraction at $z = 1$, resulting in:

\begin{eqnarray}
R^j(s) &=&  R^0(s) + R^{j-1}(s) -
\int_{z_{0}}^{1} \dfrac{{\rm d} z}{z}\, \, \big(1-z \big)^{\beta(s)\,-\,1} \, \Big[
\widetilde{G}(z,s)\,
R^{j-1}(s\, z) - z\,\widetilde{G}(1,s)\, R^{j-1}(s) \Big] \nonumber \\
 & - & \frac{1}{\beta(s)}\, \widetilde{G}(1,s)\, R^{j-1}(s)
 \big(1-z_0 \big)^{\beta(s)}\,.
\end{eqnarray}

We use the trapezoidal rule to evaluate the integration on the discrete set of experimental data 
measurements labeled by the index $i$. Changing the integration variable from $z$ to energy we find

\begin{align}
\label{eq:master-iterative}
&R^j_i =  R^0_i + R^{j-1}_i+\widetilde{G}(1,E_i^2)R_i^{j-1}\Bigg(
1- \frac{E_1^2}{E_i^2}\Bigg)^{\!\beta(E_i^2)}\Bigg(
\frac{E_1(E_2-E_1)}{E_i^2-E_1^2} -\frac{1}{\beta (E_i^2)}\Bigg) \\
& -\sum_{k=2}^{i-1}\Bigg(1- \frac{E_k^2}{E_i^2}
\Bigg)^{\!\beta(E_i^2)\,-\,1}\! E_k \Bigg[ \widetilde{G}\Bigg(
\frac{E_k^2}{E_i^2},E_i^2 \Bigg)\frac{R_k^{j-1}}{E_k^2}-
\widetilde{G}(1,E_i^2) \frac{R_i^{j-1}}{E_i^2} \Bigg](E_{k+1}-E_{k-1})\,,
\nonumber
\end{align}

where we have used $R_1^j=R(10.62\,\rm{GeV})\equiv 0$ for all iterations. After applying the 
procedure as many times as necessary to obtain a stable solution, one obtains the ISR-corrected 
cross section. Among the experimental measurements one finds two data points taken at very similar 
values of the energy: $10.86$\,GeV and $10.8605$\,GeV. It turns out that the fact that they lie very 
close makes the iterative procedure unstable. Therefore we drop the latter point from our analysis. 

\begin{figure}
\center
\includegraphics[width=0.90\textwidth]{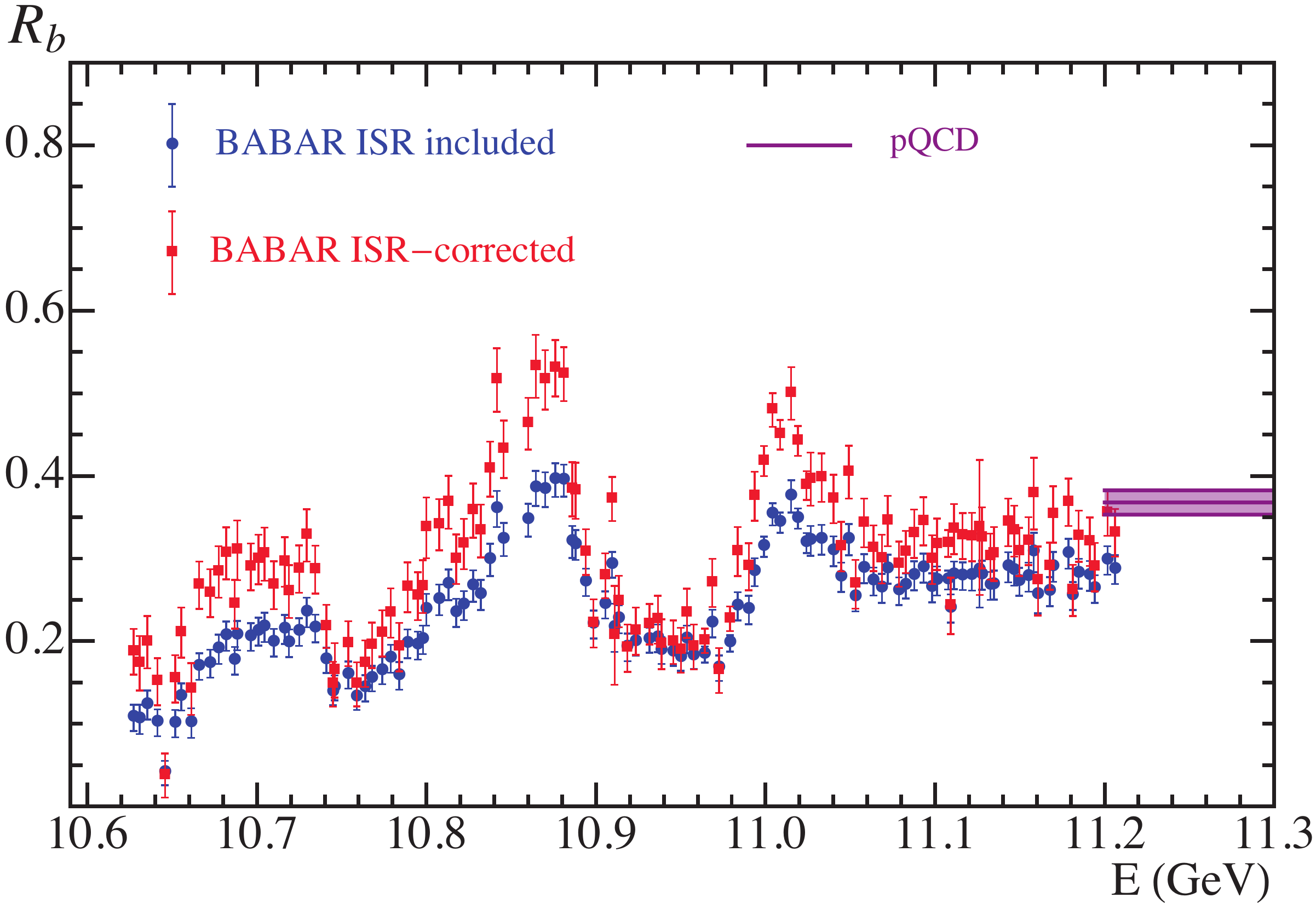}
\caption{BABAR experimental data before (blue) and after (red) the ISR correction is applied. The 
purple bar on the right refers to the pQCD prediction for the continuum region. We have removed one 
data point at $E = 10.8605$\,GeV. \label{fig:BABAR-data}}
\end{figure}

In Fig.~\ref{fig:BABAR-data} we show the BABAR data after the subtraction of all radiative 
tails, before (red) and after (blue) ISR and vacuum polarization corrections.

\subsubsection{Determination of the Unfolding Error Matrix}
\label{sec:unfold}

The BABAR collaboration splits the experimental uncertainties into statistical, systematic 
uncorrelated, and systematic correlated. We add the two former in quadrature to obtain the total 
uncorrelated uncertainty $\epsilon^{\rm uncor}$ and rename the latter as the total correlated 
uncertainty $\epsilon^{\rm{cor}}$. The removal of the radiative tails of the $\Upsilon$ mesons 
has no effect on these uncertainties. Therefore, the correlation matrix for the BABAR data after 
the subtraction of the radiative tails, before it is corrected for ISR effects, can be written as

\begin{equation}\label{eq:M00}
M^{0\,0}_{ij}=(\epsilon_i^{\rm{uncor}})^2\,\delta_{ij}+\epsilon^{\rm{cor}}_i\epsilon^{\rm{cor}}_j\,.
\end{equation}

One needs to compute a new correlation matrix after each iteration. In this way we determine the 
unfolding error matrix.

The master formula in Eq.~(\ref{eq:master-iterative}) can be cast in a matrix form as follows:

\begin{equation}
R^j_i = R^0_i + \sum_{k=2}^{i} G_{ik} R_k^{j-1}\,,
\end{equation}

where $R^j_i$ is to be thought as the $i$-th component of the column vector $R^j$, and $G_{ik}$ 
represents the $(i,k)$-component of a matrix $G$. Here $R_i^j$ depends only on the initial value 
$R_i^0$ and the result of the previous iteration $R_i^{j-1}$. The $G_{ik}$ do not depend on 
$R_k^j$ 
or the iteration step $j$. Therefore, for the error propagation one uses

\begin{equation}
\frac{\partial R^j_i}{\partial R^0_k} = \delta_{ik}\,, \qquad
\frac{\partial R^j_i}{\partial R^{j-1}_k} = G_{ik}\,,
\end{equation}

both of them $j$-independent.
We will denote with $M^{j\,j}$ the correlation matrix among the entries of the vector $R^j$ for a 
given iteration $j$. We also find it convenient to introduce the correlation matrix among $R^j$ and 
$R^0$, referred to as $M^{j\,0}$. Finally we use the notation $M^{0\,j}\equiv(M^{j\,0})^T$. We find 
for the correlation matrix after $j$ iterations:

\begin{align}
M^{j\,j} & = M^{0\,0}+M^{0\,j-1} \ G^{T}+G \ M^{(j-1)\,0} + G \ M^{(j-1)\,(j-1)}\, G^T ,\\
M^{j\,0} & = M^{0\,0} + G \,M^{(j-1)\,0} ,\nonumber
\end{align}

where the elements of the matrix $M^{0\,0}$ are given in Eq.~(\ref{eq:M00}).
We find that after five iterations the result has converged already to a level well below 
the experimental uncertainties. Our unfolded BaBar data agrees well with that worked out in 
Ref.~\cite{Chetyrkin:2010ic}.

\subsubsection{Contribution of the Threshold Region}

After having corrected BABAR data for ISR and vacuum polarization effects, we use the trapezoidal 
rule for integrating the threshold region between $10.62\,\rm{GeV}$ and $11.20\,\rm{GeV}$:

\begin{equation}\label{eq:trapezoid}
M^{\rm{thr}}_n = \dfrac{1}{2n}\bigg[
\sum_{i=2}^{N-1}R_i\bigg(\frac{1}{E_{i-1}^{2n}} - \frac{1}{E_{i+1}^{2n}}\bigg) +
R_N\bigg(\frac{1}{E_{N-1}} - \frac{1}{E_N}\bigg)\bigg]\,,
\end{equation}

where $R_i$ has been already ISR corrected and $N$ is the number of data points. We have added the 
boundary condition point $R_1 = R(10.62) = 0$. From Eq.~(\ref{eq:trapezoid}) one can compute the 
correlation matrix among $M^{\rm{thr}}_n$ for various $n$ values, using the unfolding matrix among 
the $R_i$ computed in Sec.~\ref{sec:unfold}.

\subsection{Continuum Region}
\label{sec:continuum}

\begin{figure}
\center
\includegraphics[width=0.7\textwidth]{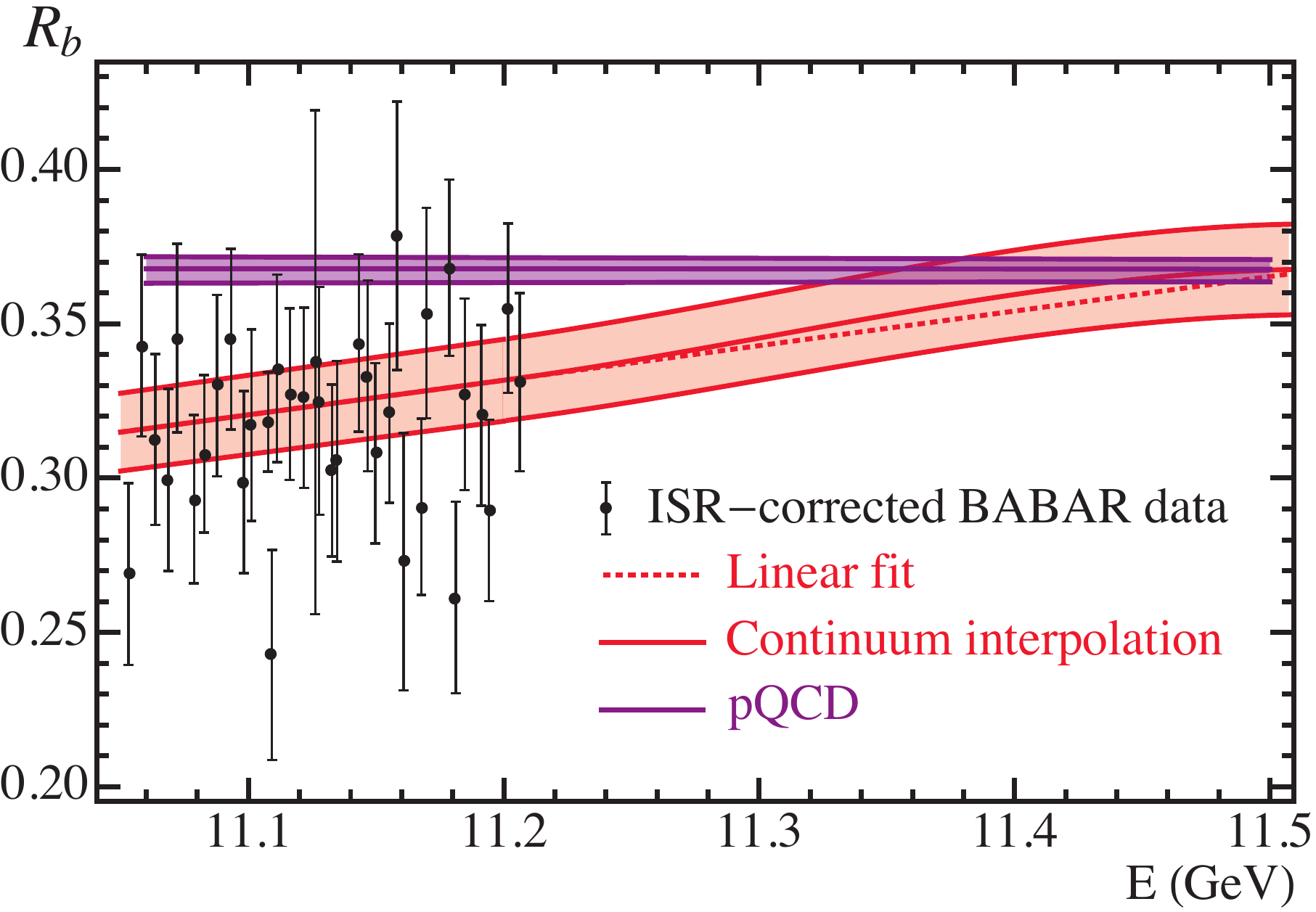}
\caption{Comparison of ISR-corrected BABAR data in the continuum region (black dots with error 
bars) with pQCD (purple band). The red band shows our reconstruction of the continuum, which 
includes a linear fit to the BaBar data, patched to the pQCD prediction in a smooth way using a 
cubic polynomial in the energy. \label{fig:BABAR-pQCD}}
\end{figure}

For the determination of the experimental moments from the region above $11.2$\,GeV we use pQCD 
(which has essentially negligible errors) supplemented by a modeling uncertainty. Comparing pQCD 
(purple line in Fig.~\ref{fig:BABAR-pQCD}) to a linear fit to the BaBar data in the region between 
$11.06\,$GeV and $11.2\,$GeV (red dotted line in Fig.~\ref{fig:BABAR-pQCD}) we find a 
$10\%$ discrepancy concerning the central values. The fit function has a roughly constant $4\%$ 
relative uncertainty. The fit linear function shows a growing pattern such that it would meet the
pQCD prediction at around $11.5\,$GeV. This result is very robust, since a quadratic fit yields 
the same meeting point. To model the continuum in the region between $11.2$\,GeV and $11.52\,$GeV 
we patch together the linear fit function to the BaBar data and the result of pQCD above 
$11.52\,$GeV using a cubic function, demanding continuity and smoothness at $11.2$\,GeV and 
$11.52\,$GeV. The result is shown as the central red line in Fig.~\ref{fig:BABAR-pQCD}. Given that 
the relative discrepancy between experiment and pQCD for $R_b$ at the $Z$-pole is about 
$0.3\%$~\cite{ALEPH:2010aa}, we adopt a relative modeling error that decreases linearly from $4\% $ 
at $11.2$\,GeV to $0.3\%$ at $m_Z$, and stays constant for energies larger than $m_Z$. This is shown 
as the red band in Fig.~\ref{fig:BABAR-pQCD}. This uncertainty makes up for $96.9\%$ of the total 
error for the first moment $M_1^V$ (which has an total $2.45\%$ relative error), and $86.15\%$ of 
the second moment $M_2^V$ (which has a total $1.85\%$ relative error). Note that if we would adopt a 
constant $4\%$ error for all energies above $11.2$\,GeV, this continuum uncertainty would make up 
for $97.24\%$ of the total error for the first moment $M_1^V$ (from a total $2.60\%$ relative 
error), and $86.46\%$ of the second moment $M_2^V$ (from a total $1.87\%$ relative error). The 
difference is small because contributions from higher energies are suppressed. Following our 
procedure in Ref.~\cite{Dehnadi:2011gc} we consider this uncertainty as fully correlated for the 
various moments, but without any correlation to the narrow resonances or the threshold region.

The perturbative QCD theoretical expression which we use to determine this contribution includes 
the non-singlet massless quark cross section supplemented with bottom mass corrections up to 
$O(\overline{m}_b^{\,4}/s^2$).\footnote{We note that the double massive fermion bubble 
contribution to $R_{bb}$ in Eq.~(\ref{eq:Rhad}) includes both virtual and real radiation terms in 
the large energy expansion. However, when this formula is used to compare pQCD to the existing BABAR 
data, below the four-bottom-quarks threshold, the real radiation should be excluded. We have checked 
that this inconsistency has an effect below $0.1$\%.} It takes into account only contributions from 
the electromagnetic current coupled to the bottom quark. It reads:~\cite{Bernreuther:1981sp, 
Gorishnii:1986pz, Chetyrkin:1993hi, Chetyrkin:1994ex, Chetyrkin:2000zk}\footnote{The authors of 
Ref.~\cite{Chetyrkin:2010ic} use the pole mass instead of the $\overline{\rm MS}$, and include 
$\alpha_s^4$ and QED corrections. This explains some numerical differences in the analyses.}

\begin{equation}
R^{\rm{th}}_{b\bar b}(s) = N_c\, Q_b^2\, R^{\rm{ns}}(s,\overline{m}^{\,2}_b(\sqrt{s}),n_f =
5,\alpha_s^{(n_f=5)}(\sqrt{s}))\,,
\end{equation}

where

\begin{eqnarray}\label{eq:Rhad}
&& R^{\rm{ns}} (s,
\overline{m}^{\,2}_b(\mu),n_f=5,\alpha_s^{(n_f=5)}(\mu),\mu) \nonumber \\ &&
= 1 + \frac{\alpha_s}{\pi}+ \Big(\frac{\alpha_s}{\pi} \Big)^2
(1.40923-1.91667\, L_s)+\Big(\frac{\alpha_s}{\pi}\Big)^3(-\,12.7671-7.81872\, 
L_s+3.67361\,L_s^2\,)\nonumber
\\ && +\,\frac{\overline{m}^{\,2}_b(\mu)}{s} \Big[ 12
\frac{\alpha_s}{\pi}+\Big(\frac{\alpha_s}{\pi}\Big)^2 (104.833 - 47\, L_s) 
+\Big(\frac{\alpha_s}{\pi}\Big)^3
(541.753-724.861\, L_s + 137.083\, L_s^2\,) \Big] \nonumber \\ &&
+\,\frac{\overline{m}_b^{\!4}(\mu)}{s^2} \Big[-6 +
\Big(\frac{\alpha_s}{\pi}\Big)(-\,22+24\, L_s)+
\Big(\frac{\alpha_s}{\pi}\Big)^2 (139.014 - 4.83333\, L_m + 214.5\,L_s - 71\, L_s^2\,)\nonumber
\\ && +\Big(\frac{\alpha_s}{\pi}\Big)^3 \
(3545.81 - 158.311\, L_m + 9.66667 \,L_m^2 - 538.813\, L_s + 37.8611 \, L_m\, L_s
\nonumber  \\ && -\,1037.79\, L_s^2+185.389 \,L_s^3\,) \Big]\,,
\end{eqnarray}

with

\begin{equation}
L_s\equiv \ln \Big(\frac{s}{\mu^2}\Big),  \quad L_m\equiv
\ln\Big(\frac{\overline{m}^{\,2}_b(\mu)}{s}\Big)\,,\quad \alpha_s = \alpha_s^{(n_f=5)}(\mu)\,.
\end{equation}

We use the initial conditions $\overline{m}_b(\overline{m}_b)=4.2 \,\rm{GeV}$ and 
$\alpha_s(m_Z)=0.118$.

Therefore, for the continuum region we use the following expression,

\begin{align}
 M^{\rm pQCD}_n = &\int_{s_0}^{s_1}
{\rm d} s\, \frac{R^{\rm{cubic}}_{bb} (s)}{s^{n+1}} \bigg[1 + \gamma'
\dfrac{0.04(m_Z^2 - s) + 0.003(s-s_0)}{m_Z^2-s_0}\bigg] \\
& +\int_{s_1}^{m_Z^2}
{\rm d} s\, \frac{R^{\rm{th}}_{bb} (s)}{s^{n+1}} \bigg[1 + \gamma'
\dfrac{0.04(m_Z^2 - s) + 0.003(s-s_0)}{m_Z^2-s_0}\bigg]\nonumber\\
& +  (1 + 0.003\,\gamma')\int_{m_Z^2}^{\infty}
{\rm d} s\, \frac{R^{\rm{th}}_{bb} (s)}{s^{n+1}}\, ,\qquad
\gamma'  = 0 \pm 1\,,\nonumber
\end{align}

with $s_0 = (11.2062\,{\rm GeV})^2$, $s_1 = (11.52\,{\rm GeV})^2$ and $R^{\rm{cubic}}_{bb}$ is a 
cubic function that smoothly interpolates between the linear fit to BaBar data and pQCD. Here 
$\gamma'$ is the auxiliary variable used to parametrize our uncertainty, which we consider as 
$100\%$ correlated among the various moments. The related entries of the correlation matrix are 
trivially computed as

\begin{equation}
C_{nn'}^{\rm{pQCD}} = \frac{\partial M^{\rm{pQCD}}_n}{\partial \gamma'}
\ \frac{\partial M^{\rm{pQCD}}_{n'}}{\partial \gamma'}\,.
\end{equation}

\subsection{Final Results for the Experimental Moments}

\begin{table}[t!]
\centering
\begin{tabular}{|c|cccc|}
\hline 
n & Resonances  & $10.62-11.2062$ & $11.2062-\infty$ & Total\tabularnewline
\hline
$1$ & $1.394(12|22)$ & $0.270(2|9)$ & $2.862(0|108)$ & $4.526(12|111)$ \\
$2$ & $1.459(12|22)$ & $0.226(1|8)$  & $1.148(0|45)$  & $2.834(12|51)$ \\
$3$ & $1.538(12|22)$ & $0.190(1|7)$  & $0.611(0|24)$  & $2.338(12|34)$ \\
$4$ & $1.630(13|22)$ & $0.159(1|6)$  & $0.365(0|15)$  & $2.154(13|27)$\\
\hline
\end{tabular}
\caption{
Results for our computations of the experimental moments. The second column collects the 
contribution from the first four $\Upsilon$ resonances (using the narrow width approximation). The 
third to fifth columns show the contributions from the threshold (using ISR-corrected BABAR data) 
and continuum (using an interpolation between a linear fit to the BaBar data with highest energy 
and pQCD as a model for the lack of data) regions, and the total moment determinations, 
respectively. The two numbers quoted in parentheses correspond to the uncorrelated and correlated 
experimental uncertainties, respectively. All numbers are given in units of $10^{-(2n+1)} \, 
\rm{GeV}^{-2n}$.\label{tab:moments-results}}
\end{table}

The full result for the experimental moments is obtained by summing up all the portions described 
before,

\begin{equation}
M^{\rm exp}_n=M^{\rm res}_n+M^{\rm thr}_n+M^{\rm pQCD}_n\,.
\end{equation}

We determine two correlation matrices among the first four moments. One of them comes from the 
various uncorrelated uncertainties, whereas the other encodes the systematic uncertainties. We 
denote them as the correlated and uncorrelated correlation matrices, respectively. These are 
computed by summing up the respective individual matrices from each region, and in the same way as 
we did for our charm analysis \cite{Dehnadi:2011gc}, we assume there is no region-to-region 
correlation. We find:

\begin{equation}
\label{eq:corr-mat}\!\!\!\!
C^{\rm exp}_{\rm uc} =
\left(
\begin{array}{cccc}
0.0002 & 0.0002 & 0.0002 & 0.0002 \\
0.0002 & 0.0002 & 0.0002 & 0.0002 \\
0.0002 & 0.0002 & 0.0002 & 0.0002 \\
0.0002 & 0.0002 & 0.0002 & 0.0002 \\
\end{array}\right)\!, \,
C^{\rm exp}_{\rm cor} =
\left(
\begin{array}{cccc}
0.0122 & 0.0055 & 0.0032 & 0.0021 \\
0.0055 & 0.0026 & 0.0017 & 0.0012 \\
0.0032 & 0.0017 & 0.0011 & 0.0009 \\
0.0021 & 0.0012 & 0.0009 & 0.0007 \\
\end{array}\right)\! ,
\end{equation}

where the $(n,m)$ entry of each matrix is given in units of $10^{-2(n + m + 1)} \, 
\rm{GeV}^{-2(n+m)}$ , and the total correlation matrix is the sum of $C^{\rm exp}_{\rm uc}$ and 
$C^{\rm exp}_{\rm cor}$. The contribution of each region to the final experimental moments and the 
corresponding uncertainties are presented in Table~\ref{tab:moments-results}.

\section{Comparison to other Determinations of the Experimental Moments}
\label{sec:comp-exp}

In this section we compare our result for the experimental moments for the bottom vector current 
correlator with previous determinations. These are collected in 
Table~\ref{tab:moments-comparison}.\footnote{In the case of Ref.~\cite{Corcella:2002uu}, we 
reconstruct the experimental moments from their Table~3, where the moments are split in several 
different contributions. For the reconstructed uncertainty, we take one half of the error of the 
narrow resonances correlated to each other, and the other half as uncorrelated. The errors from 
patches where theory input is used are taken as fully correlated to one another. The total 
narrow-resonance error, and the total ``theory-patch'' error are added in quadrature to get the 
final uncertainty.} The most relevant comparison is between the second and third columns, where the 
most recent data on the narrow resonances and the BABAR continuum data are used. For the 
contributions from the narrow resonances we have a perfect agreement with \cite{Chetyrkin:2009fv}, 
although slightly larger errors. For the threshold region our results are slightly smaller, and our 
uncertainties are almost identical; however, this is not a one-to-one comparison, since our 
integration region is slightly smaller. Indeed if we consider their energy range we agree with 
their numbers almost perfectly. The main difference between these two determinations is the 
estimate of the uncertainties coming from the continuum region, where the pQCD prediction for the 
\mbox{$R_b$-ratio} is employed. Whereas we adopt the more conservative approach described in 
Sec.~\ref{sec:continuum}, Ref.~\cite{Chetyrkin:2009fv} employs only the perturbative uncertainties 
related to the purple band in Fig.~\ref{fig:BABAR-pQCD}. In Ref.~\cite{Chetyrkin:2010ic} the same 
collaboration presents a more critical analysis of their errors. In particular they observe that 
the last experimental measurement of BABAR, after being corrected for ISR, disagrees with the pQCD 
prediction at the $20\%$ level (way outside the corresponding uncertainties).\footnote{From our 
own computation of the ISR-corrected $R_b$-ratio, we only observe a $10\%$ deviation between 
the last data point and the pQCD prediction, see Fig.~\ref{fig:BABAR-pQCD}.} To resolve this 
discrepancy they assume two possible scenarios: a) pQCD starts being reliable at energies above 
$13$\,GeV (therefore the authors interpolate between the last experimental point and pQCD at 
$13$\,GeV); b) BABAR systematic errors have been underestimated (therefore the central values of the 
experimental measurements are rescaled by a factor of $1.21$). Ref.~\cite{Chetyrkin:2010ic} quotes 
the values of the experimental moments and the resulting values for the bottom mass for these two 
scenarios. Since the effect of these differences of the two bottom masses obtained from $M_2^V$ is 
only slightly larger than the size of the other uncertainties (that is, the uncertainties of the 
theoretical moments plus the other experimental errors) added quadratically, it is argued that this 
issue can be ignored. We disagree with this argument, since the issue constitutes an independent 
source of uncertainty not covered by the other errors and, in particular, being unrelated to 
uncertainties in the theoretical moments. Therefore this shift must be taken as an additional source 
of error on the experimental moments (and indeed would then dominate the corresponding total error). 
The additional error (to be added in quadrature to the one in round brackets) is quoted in square 
brackets in the third column of Table~\ref{tab:moments-comparison}. It amounts to an additional 
error of $30,18,11$ and $7$\,MeV for $\overline{m}_b(\overline{m}_b)$ extracted from moments 
$M_1^V$ to $M_4^V$, respectively.

Refs.~\cite{Kuhn:2007vp,Boughezal:2006px,Corcella:2002uu} have used the older CLEO and CUSB 
experimental measurements, resulting in relatively large uncertainties. In Ref.~\cite{Kuhn:2007vp} 
the CLEO measurements are divided by a factor of $1.28$, and an error of $10\%$ is assigned. It is 
argued that this procedure is necessary to reconcile old and new CLEO measurements, as well as to 
improve the agreement with pQCD predictions. Ref.~\cite{Corcella:2002uu} uses values for the 
$\Upsilon$-states electronic partial widths given by the PDG 2002, which have larger uncertainties. 
This makes their determination of the experimental moments rather imprecise.

Concerning the continuum region where no measurements exist, while some previous analyses have 
taken a less conservative approach than ours, in Ref.~\cite{Beneke:2014pta} a much more 
conservative approach is adopted. In this region they consider the \mbox{$R_b$-ratio} as constant 
with a $66\%$ uncertainty. In Ref.~\cite{Corcella:2002uu} also a more conservative approach is 
adopted. Between $11.1$ and  $12$\,GeV $\mathcal{O}(\alpha_s^2)$ pQCD errors are used, which are 
larger than $10$\%; for energies above $12$\,GeV a global $10\%$ correlated error is assigned.

\begin{table}[h!]
{
\begin{tabular}{|c|cccc|}
\hline 
$n$ & This work & Chetyrkin et al.\ '09~\cite{Chetyrkin:2009fv} & Kuhn et al.\ 
'07~\cite{Kuhn:2007vp} &
Corcella et al.\ '03~\cite{Corcella:2002uu}\\ \hline
$1$ & $4.526(12|111)$ & $4.592(31)[67]$ & $4.601(43)$ & $4.46(17)$ \\
$2$ & $2.834(12|51)$  & $2.872(28)[51]$ & $2.881(37)$ & $2.76(15)$ \\
$3$ & $2.338(12|34)$  & $2.362(26)[40]$ & $2.370(34)$ & $2.26(13)$ \\
$4$ & $2.154(13|27)$  & $2.170(26)[35]$ & $2.178(32)$ & $2.08(12)$ \\
\hline
\end{tabular}
\caption{Comparison of our results for the experimental moments of the bottom vector 
correlator (2nd column) to previous determinations 
(3rd to 5th columns). The 2nd and 3rd columns use BABAR data from Ref.~\cite{:2008hx}, while 4th 
and 5th use older data from Refs.~\cite{Besson:1984bd,Ammar:1997sk}. The 3rd and 4th columns use 
perturbative uncertainties in the continuum region, while 2nd and 5th use a more conservative 
estimate based on the agreement of data and pQCD. In the 3rd column, we quote in square brackets 
our own estimate of an additional systematic error from the considerations made in 
Ref.~\cite{Chetyrkin:2010ic}. All numbers are given in units of $10^{-(2n+1)} \,\rm{GeV}^{-2n}$. 
\label{tab:moments-comparison}}
}
\end{table}

\begin{figure*}[t!]
\subfigure[]
{
\includegraphics[width=0.48\textwidth]{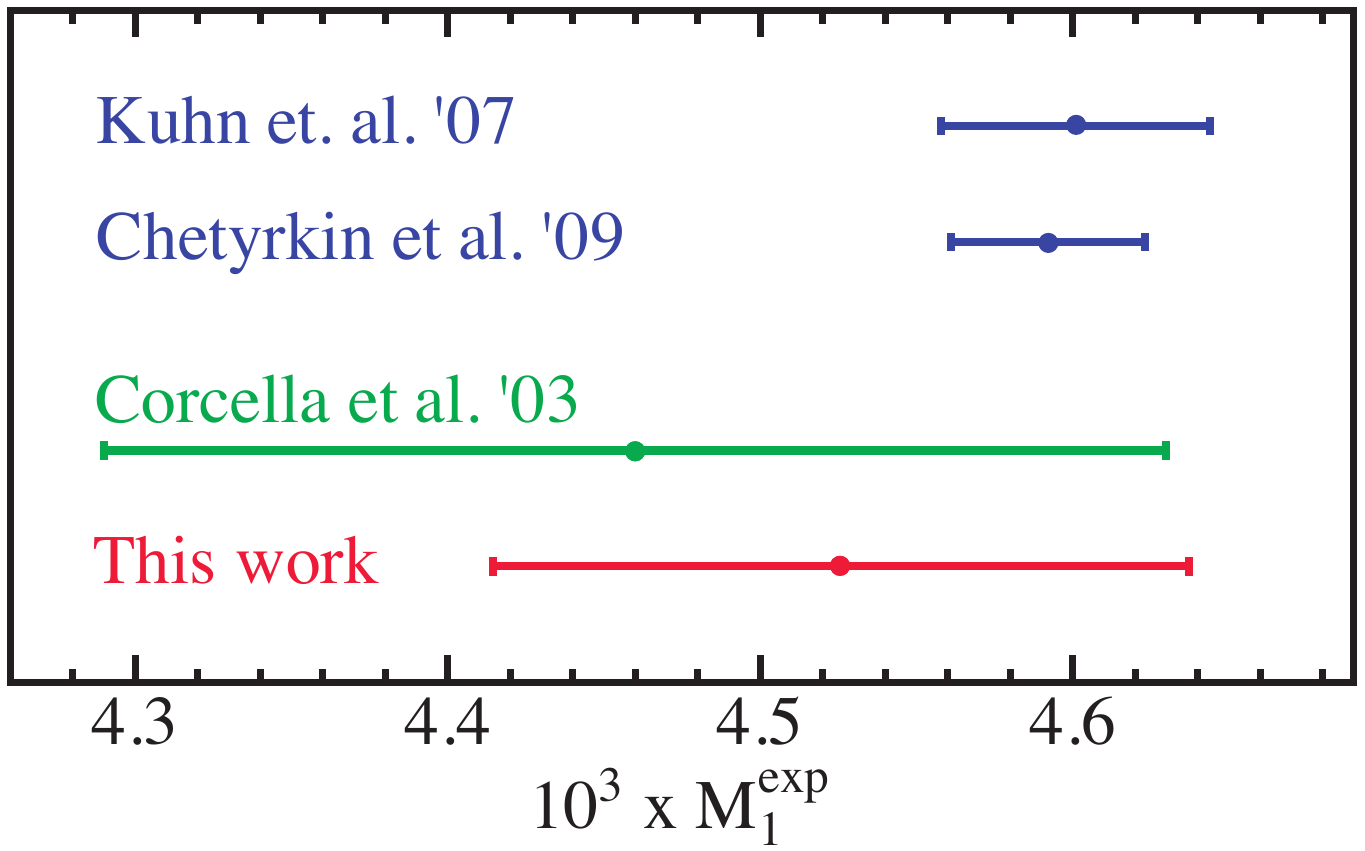}}
\subfigure[]
{
\includegraphics[width=0.48\textwidth]{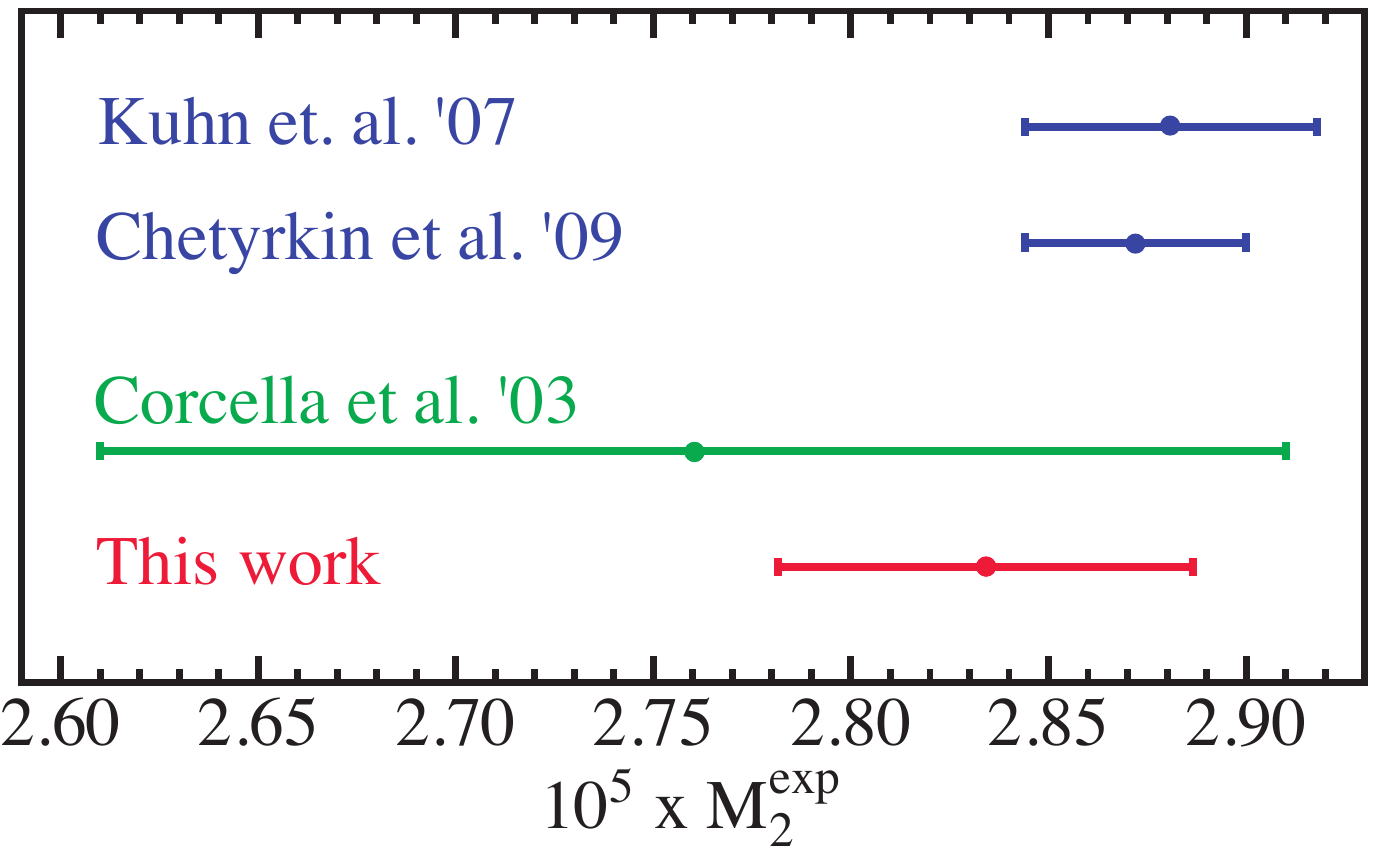}
}
\subfigure[]
{
\includegraphics[width=0.48\textwidth]{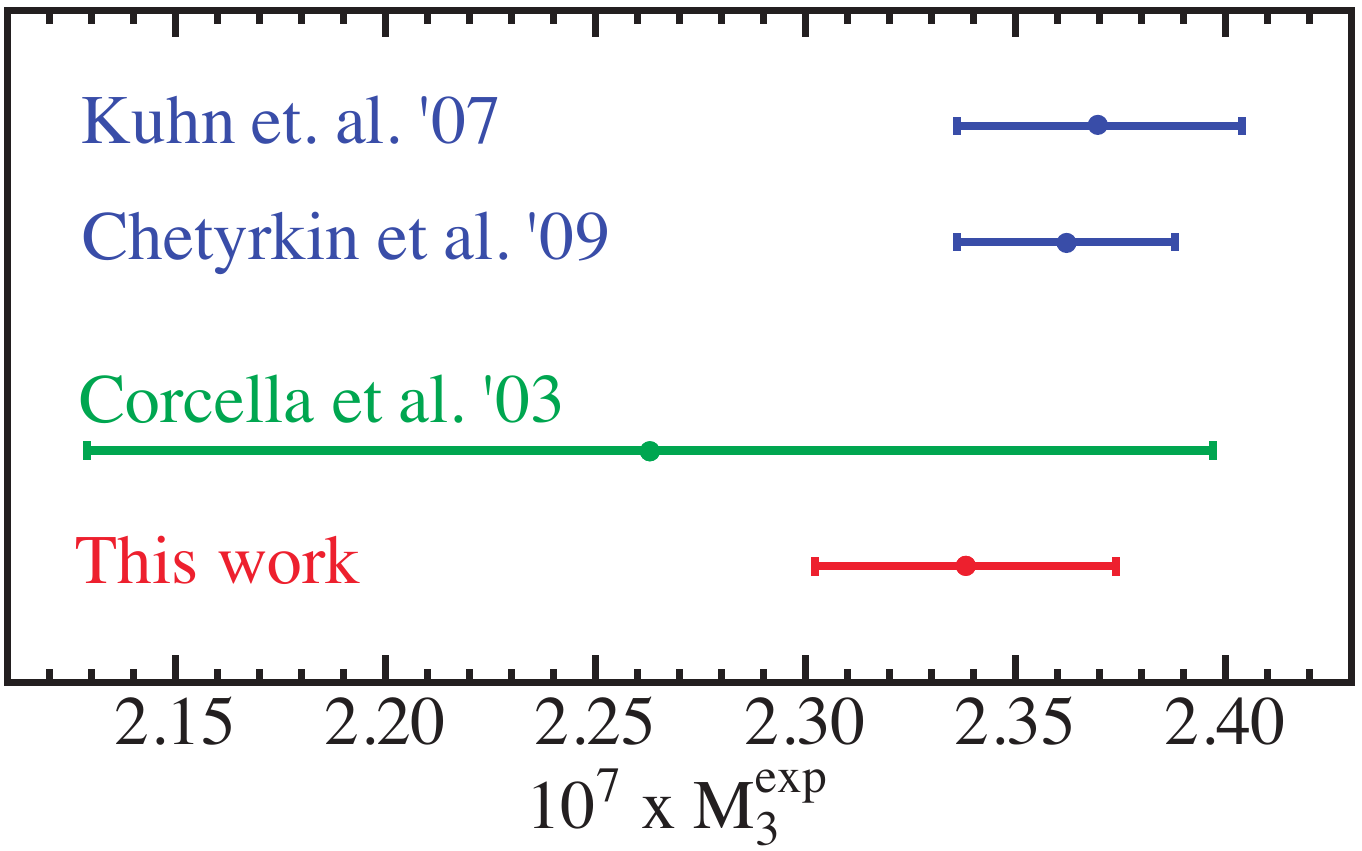}}
\subfigure[]
{
\includegraphics[width=0.48\textwidth]{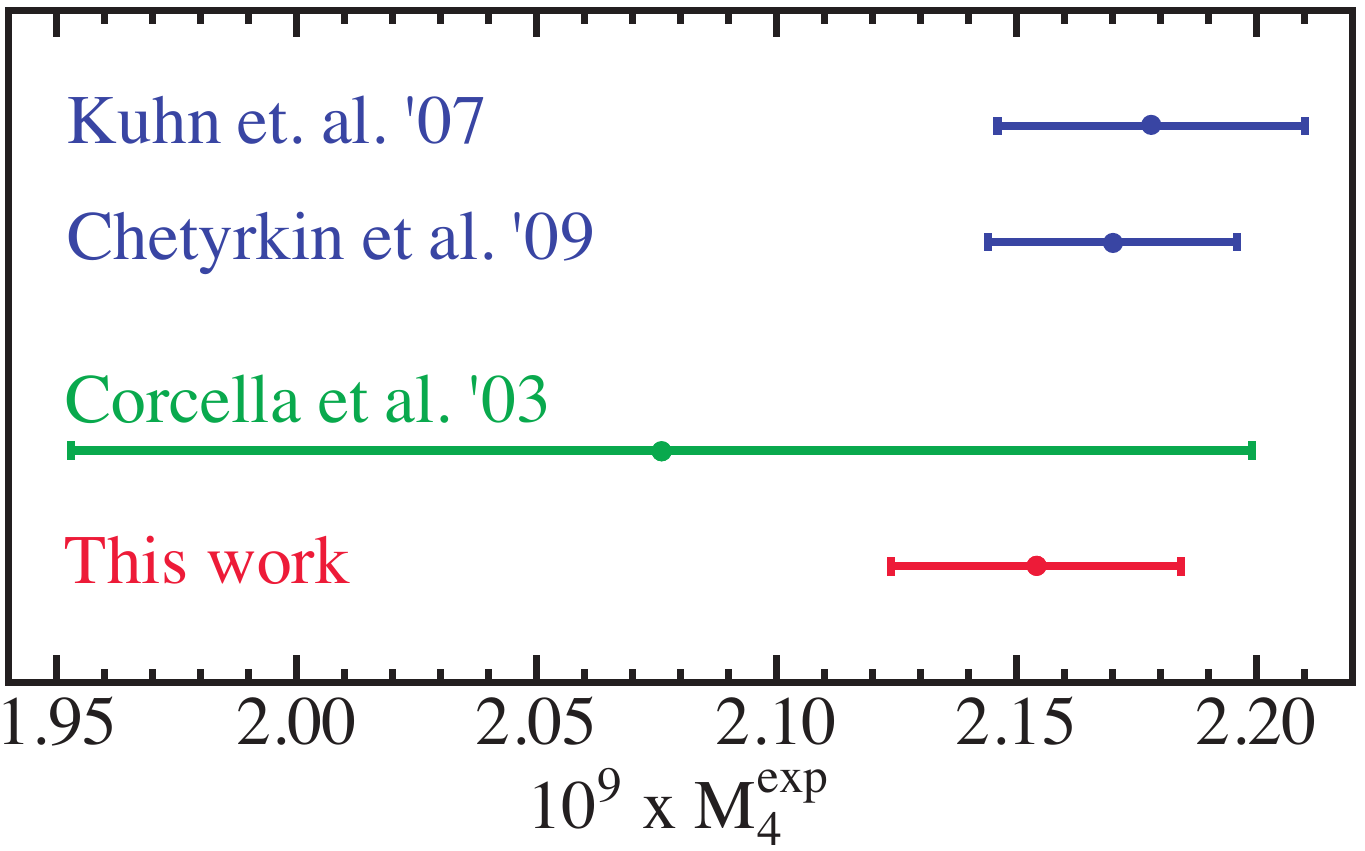}
}
\caption{Comparison of various determinations of the experimental moments for the bottom vector 
correlator. Results in blue correspond to analyses of the same collaboration. The green result and 
the determination at the top do not use the new BABAR results.}
\label{fig:comparison-experimental}
\end{figure*}

\section{Computation of the Experimental Values for the Ratios of Moments}
\label{sec:exp-ratio}

Once the experimental values for the moments of the vector and pseudoscalar correlators have 
been computed, it is in principle a straightforward exercise to calculate ratios of them. The 
central value is obtained by simply taking the ratio of the corresponding central values. To obtain 
the uncertainties (or more generally, the correlation matrix among the different ratios of 
moments), one needs to have access to the complete correlation matrix among the moments. Our 
computation in Ref.~\cite{Dehnadi:2011gc} [see Eqs.~(3.21) and (3.22)] for the charm experimental 
moments, and the procedure presented in Sec.~\ref{sec:exp} [see Eq.~(\ref{eq:corr-mat})] to 
determine the bottom experimental moments, yield the two desired correlation matrices, for 
statistical and systematical correlations. For the pseudoscalar moments the information on 
correlations is not provided in Ref.~\cite{Allison:2008xk}. Therefore we make the simplest possible 
assumption, which is that the moments are fully uncorrelated. This will most certainly 
overestimate the uncertainties for the ratios of moments, but given that we are anyway dominated by 
perturbative uncertainties, our approach appears justified. We collect our results for the 
computation of the ratios of experimental moments in Table.~\ref{tab:exp-ratios}. Readers 
interested in the full correlation matrix among them can send a request to the authors.

\begin{table}[t!]
\centering
\begin{tabular}{|c|ccc|}
\hline 
$n$ & Vector $n_f = 4$ & Vector $n_f = 5$ & Pseudoscalar $n_f = 4$\tabularnewline\hline
$1$ & $6.969(32|59)$ & $6.262(10|53)$ & $0.971(32)$ \tabularnewline
$2$ & $8.807(23|26)$ & $8.251(09|48)$ & $1.048(53)$ \tabularnewline
$3$ & $9.547(14|13)$ & $9.212(08|35)$ & $1.092(77)$ \tabularnewline

\hline
\end{tabular}
\caption{Ratios of experimental moments for the vector correlator with $4$ and $5$ flavors (second 
and third column, respectively), and for the pseudoscalar correlator with $4$ flavors (fourth 
column). For the vector current, the first error in parenthesis corresponds to the statistical 
uncertainty, whereas the second corresponds to the systematic one. For the pseudoscalar correlator 
we only quote the lattice error. Moments given in units of $10^{-2}$\,GeV$^{-2}$, 
$10^{-3}$\,GeV$^{-2}$ and, $10^{-1}$\,GeV$^{-2}$ for the second, third, and fourth column, 
respectively.\label{tab:exp-ratios}}
\end{table}

\section{Results}
\label{sec:results}

In this section we present the final results for our analyses at $\mathcal{O}(\alpha_s^3)$. We take 
method (c) (linearized iterative expansion) as our default expansion. For the estimate of the 
perturbative uncertainty, we perform double scale variation in the ranges $\overline m_c(\overline 
m_c)\le\mu_\alpha,\mu_m\le4$\,GeV for charm (either correlator), and $\overline m_b(\overline 
m_b)\le\mu_\alpha,\mu_m\le15$\,GeV for bottom, and we discard 3\% of the series with the worst 
convergence (that is, with highest values of the $V_c$ convergence parameter). For the charm mass 
determinations (either vector or pseudoscalar correlator) we use the first moment as our default, 
given that it is theoretically more reliable than the higher moments. For the analysis of the 
bottom mass from regular moments, we use $M_2^V$ as our default, since it is less afflicted by 
systematic experimental errors than the first moment, and is nevertheless theoretically sound. For 
the charm and the bottom mass analyses we also examine the ratio of the second over the first moment 
as a cross check and validation of the results from regular moments. The results for the 
experimental moments are collected in: the last column of Table~9 in Ref.~\cite{Dehnadi:2011gc} 
(charm vector correlator regular moments); the last column of Table~\ref{tab:moments-results} 
(bottom vector correlator regular moments); Table~\ref{tab:Latt} (lattice regular moments); and 
Table~\ref{tab:exp-ratios} (all ratios of moments).

\begin{figure*}[t!]
\subfigure[]
{
\includegraphics[width=0.31\textwidth]{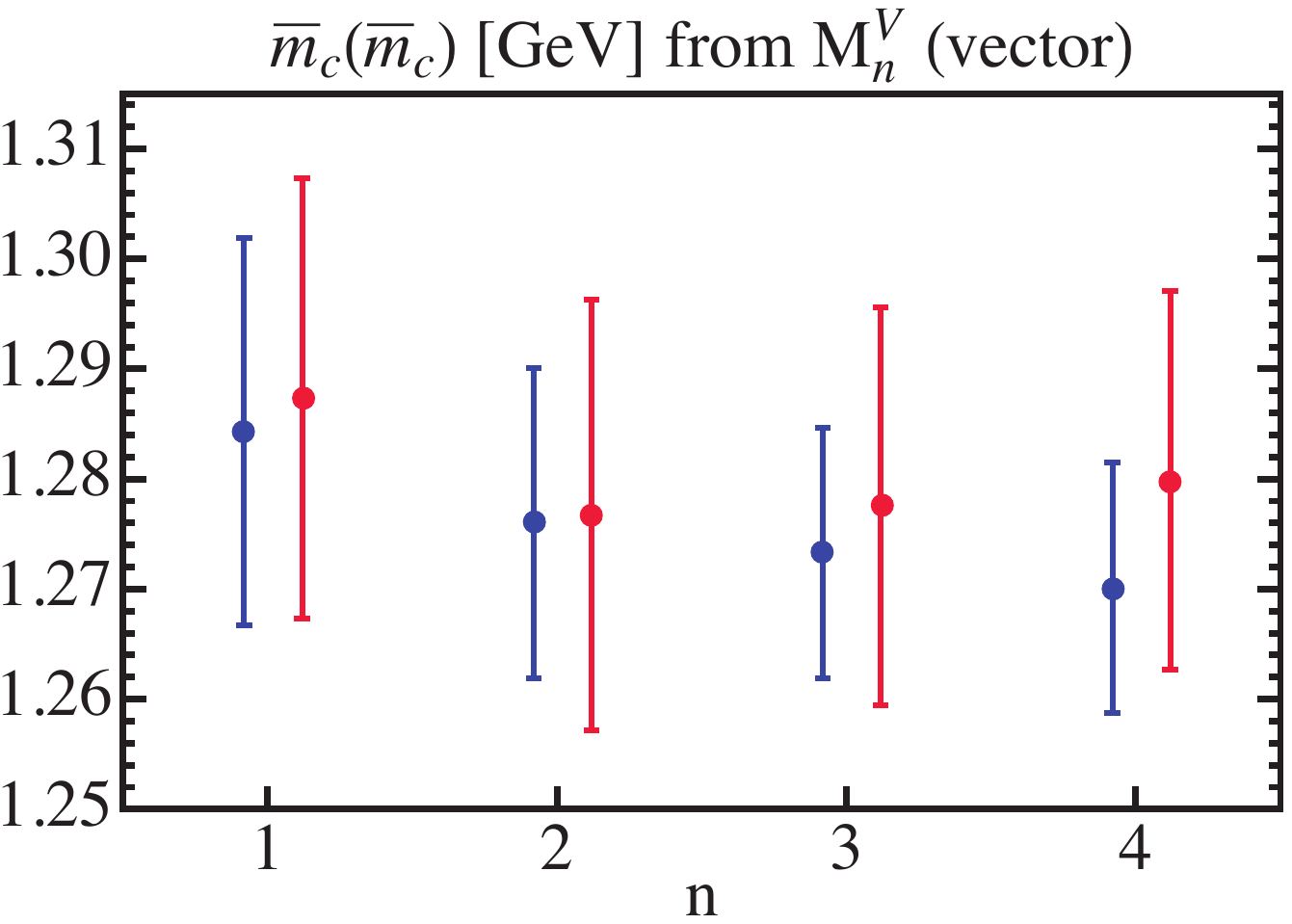}}
\subfigure[]
{
\includegraphics[width=0.31\textwidth]{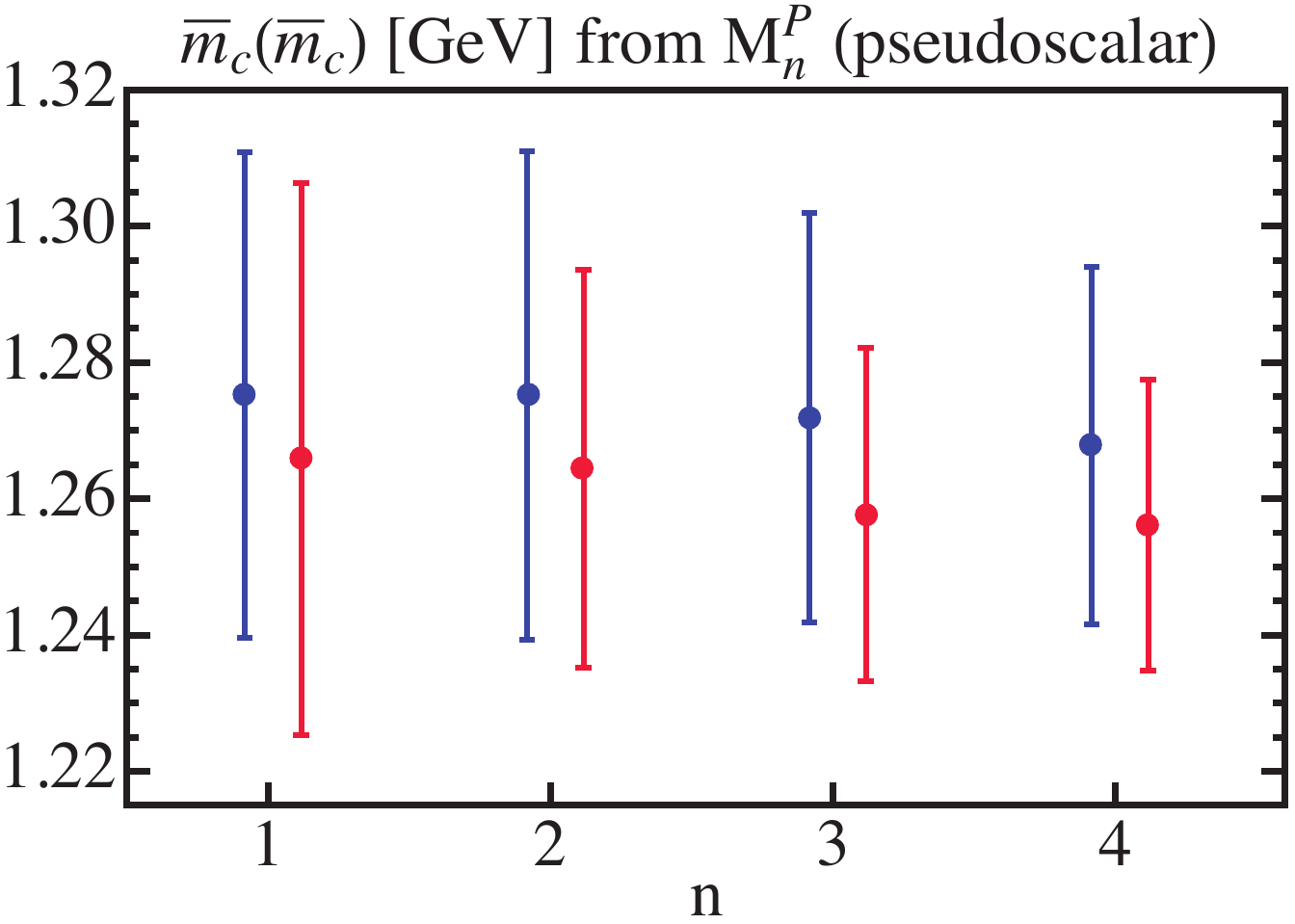}
}
\subfigure[]
{
\includegraphics[width=0.31\textwidth]{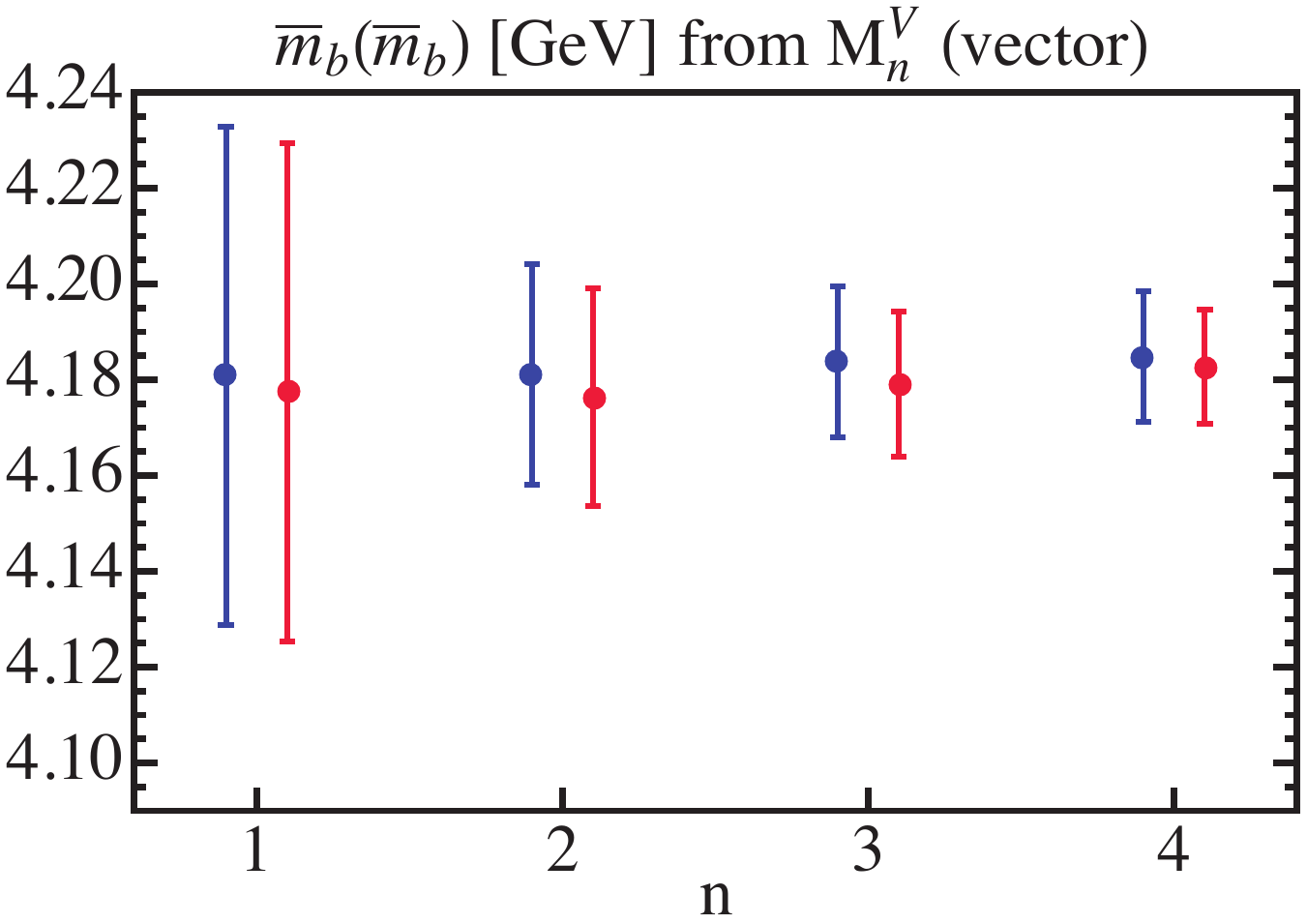}
}
\subfigure[]
{
\includegraphics[width=0.31\textwidth]{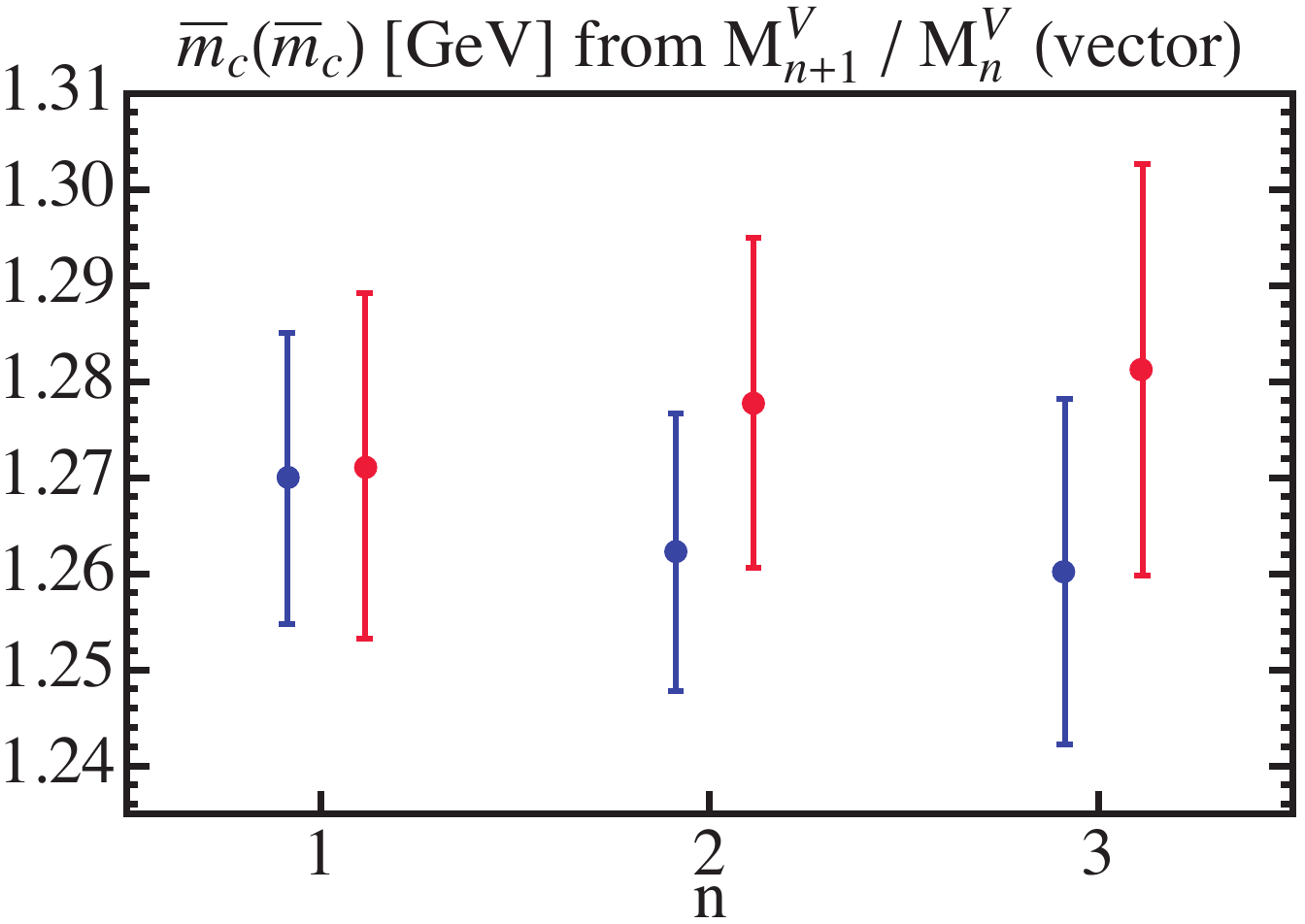}
}
\subfigure[]
{
\includegraphics[width=0.31\textwidth]{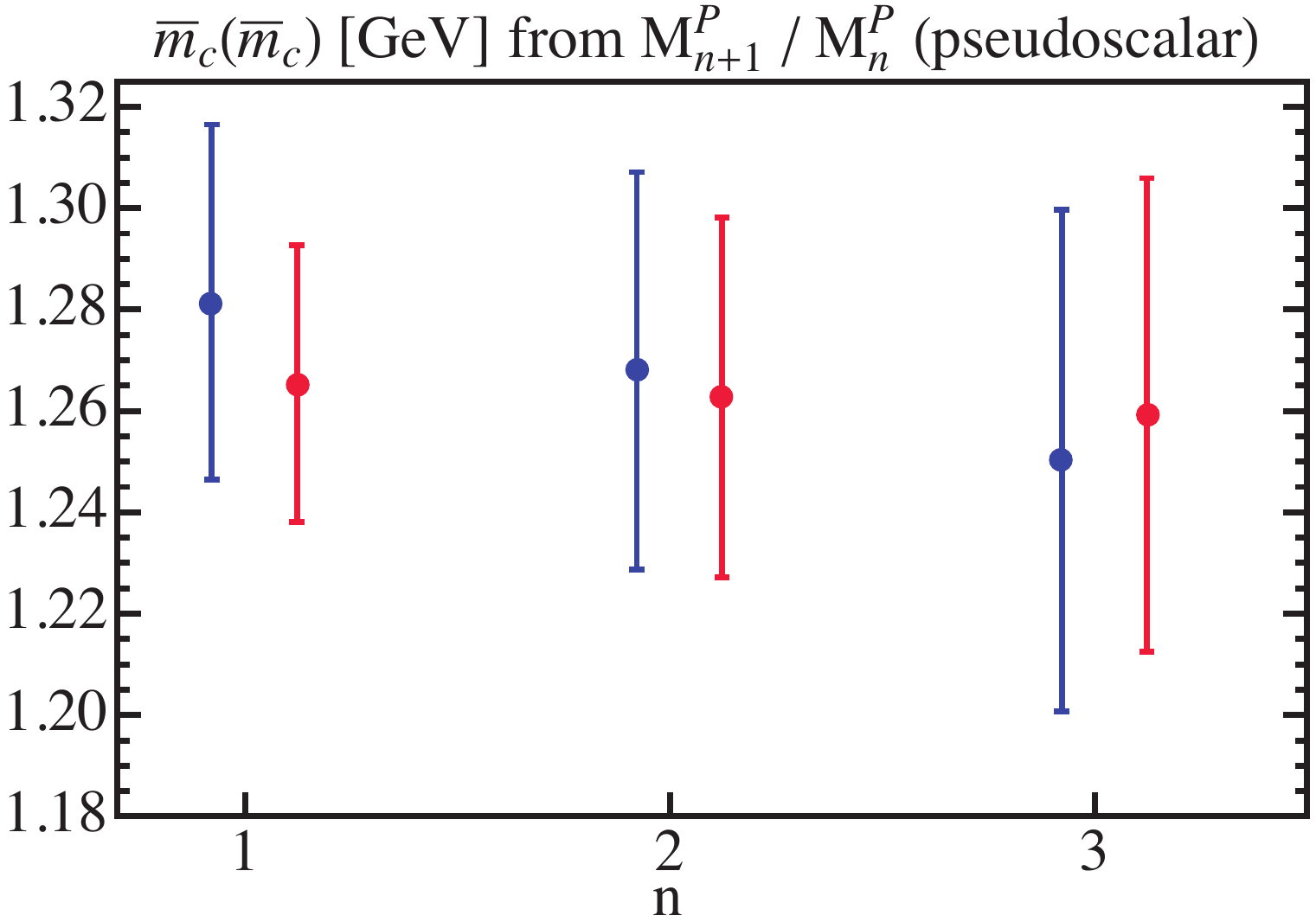}
}
\subfigure[]
{
\includegraphics[width=0.31\textwidth]{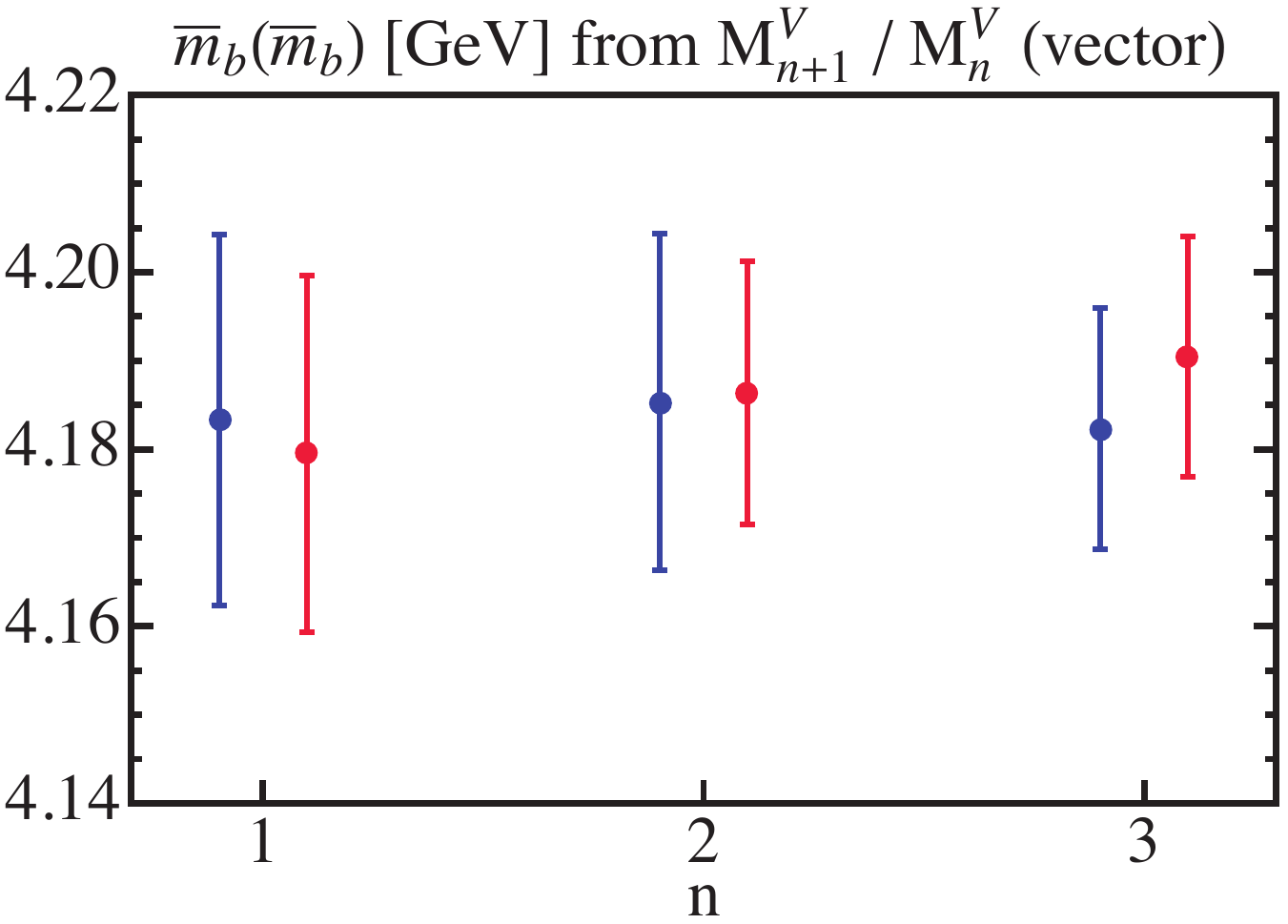}
}
\caption{Charm and bottom quark mass determinations for different moments (upper row) or ratios of 
moments (lower row), for the linearized (in blue) and iterative (in red) methods. Panels (a), (b) 
[(e), (f)] show the results for the charm mass from moments [ratios of moments] of the vector and 
pseudoscalar correlator, respectively. Panels (c) and (g) show the results for the bottom mass from 
the vector correlator, for moments and ratios of moments, respectively.}
\label{fig:higher-moms}
\end{figure*}

We also analyzed higher (and also lower for the case of bottom) moments and ratios of moments for 
all correlator and quark species. Since, as already discussed, fixed-order and contour-improved 
higher moments are particularly afflicted by their nonlinear dependence on the quark mass, we 
only consider the linearized and iterative methods for this analysis. In any case, since higher 
moments have a larger sensitivity to infrared effects and are therefore theoretically less sound, 
the analysis involving higher moments mainly aims at providing cross checks. The results are 
collected in a graphical form in Fig.~\ref{fig:higher-moms}, and the numerical results can be 
obtained from the authors upon request.

Our final determinations include nonperturbative effects through the gluon condensate including its 
Wilson coefficients at order $\mathcal{O}(\alpha_s)$. Furthermore, we assign as a conservative 
estimate of the nonperturbative uncertainty twice the shift caused by including the gluon 
condensate. In any case, this error is very small, particularly for the bottom mass determination. 
One source of uncertainty which we have not discussed so far is that coming from the strong coupling 
constant. Although the world average $\alpha_s(m_Z) = 0.1185 \pm 0.006$ has a very small error, see 
Ref.~\cite{Agashe:2014kda}, one cannot ignore the fact that it is fairly dominated by lattice 
determinations, e.g.~\cite{Chakraborty:2014aca}. Furthermore, there are other precise determinations 
with lower central values and in disagreement with the world average from 
event-shapes~\cite{Abbate:2010xh, Abbate:2012jh, Gehrmann:2012sc,Hoang:2015hka} and 
DIS~\cite{Alekhin:2012ig}. A review on recent $\alpha_s$ determinations can be found in 
Refs.~\cite{Bethke:2011tr,Bethke:2012jm,Pich:2013sqa}. Therefore, in analogy with 
Ref.~\cite{Dehnadi:2011gc}, we perform our analyses for several values of $\alpha_s(m_Z)$ between 
$0.113$ and $0.119$, and provide the central values and perturbative errors as (approximate) linear 
functions of $\alpha_s(m_Z)$. The other uncertainties are essentially $\alpha_s$-independent, so we 
just provide the average. We also quote quark mass results for $\alpha_s$ taken from the world 
average: 

\begin{equation}
\label{eq:alphaswa}
\alpha_s(m_Z) = 0.1185 \pm 0.0021\,,
\end{equation}

where we adopt an uncertainty $3.5$ times larger than the current 
world average~\cite{Agashe:2014kda}. We note that in Ref.~\cite{Dehnadi:2011gc} we have taken 
$\alpha_s(m_Z) = 0.1184 \pm 0.0021$ as an input which causes only tiny sub-MeV differences in the 
quark masses. We refrain ourselves from presenting the $\alpha_s$ dependence of the higher-moment 
result, which the reader can get from the authors upon request.

For the numerical analyses that we carry out in this article we have created two completely 
independent codes: one using Mathematica~\cite{mathematica} and another using 
Fortran~\cite{gfortran}, which is much faster and suitable for parallelized runs on computer 
clusters. The two codes agree for the extracted quark masses at the level of $1\,$eV.

\subsection{Results for the Charm Mass from the Vector Correlator}

\begin{figure*}[t!]
\subfigure[]
{
\includegraphics[width=0.48\textwidth]{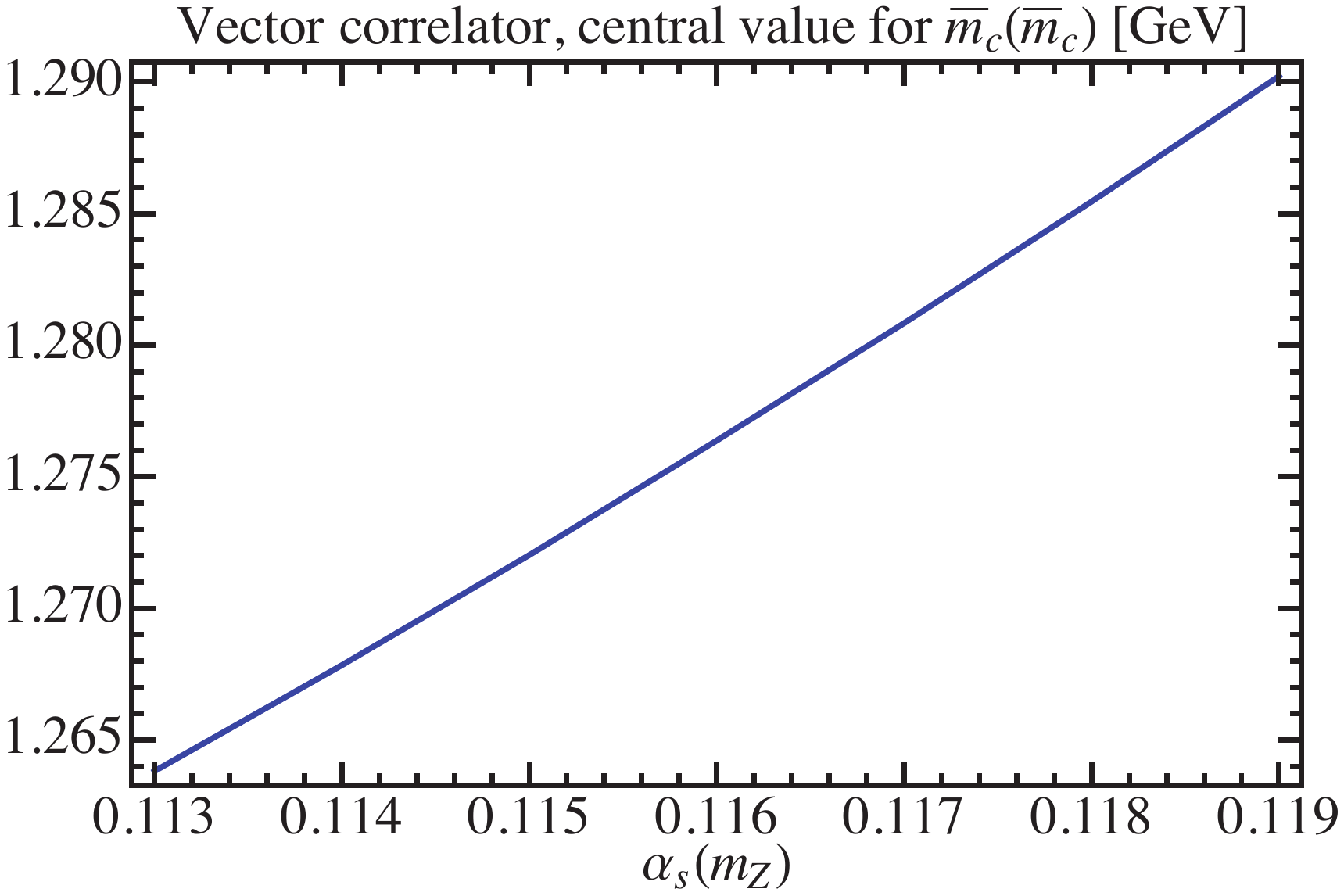}
\label{fig:mc-vector-central}
}
\subfigure[]{
\includegraphics[width=0.48\textwidth]{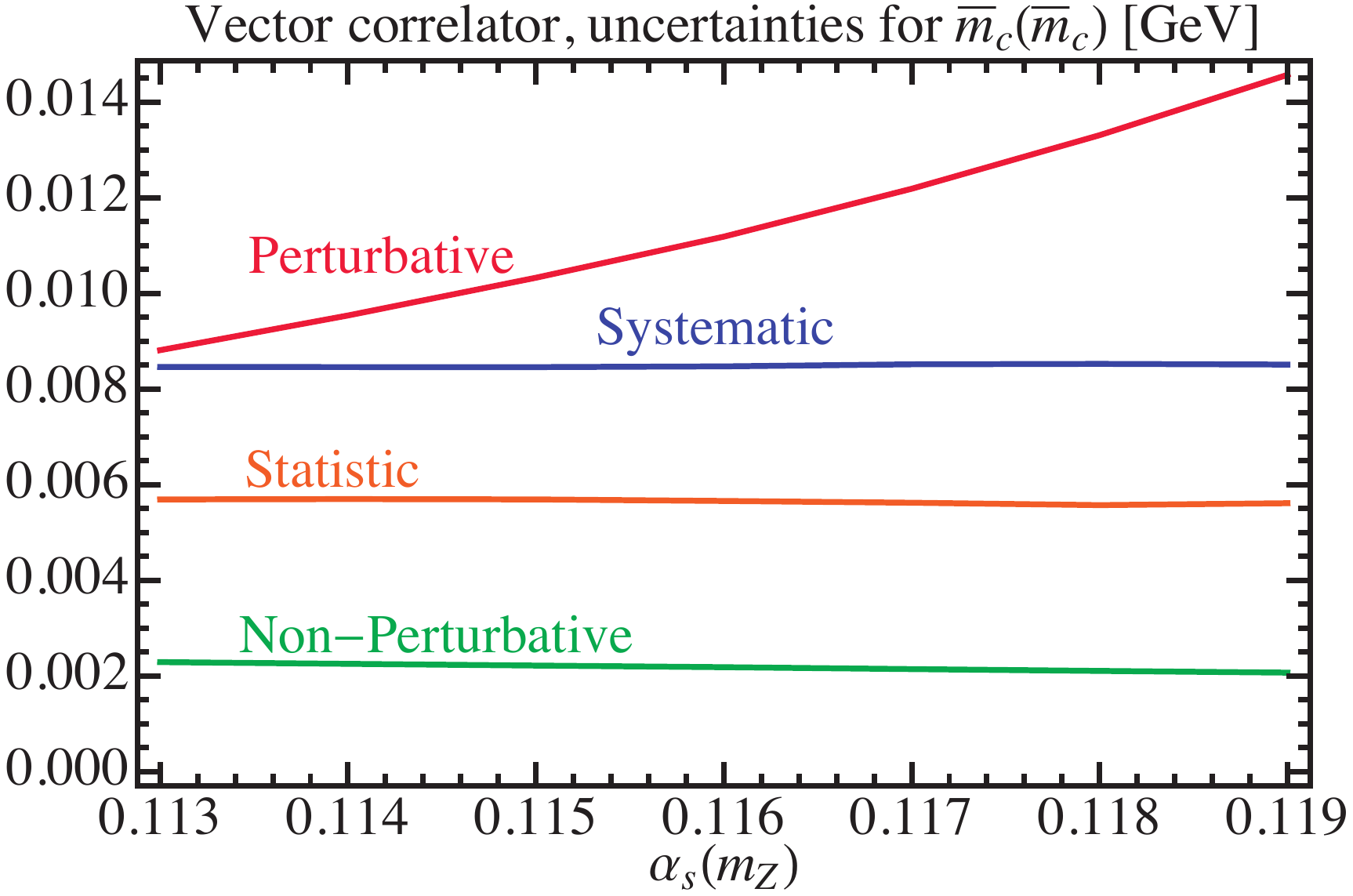}
\label{fig:mc-vector-err}
}
\subfigure[]
{
\includegraphics[width=0.48\textwidth]{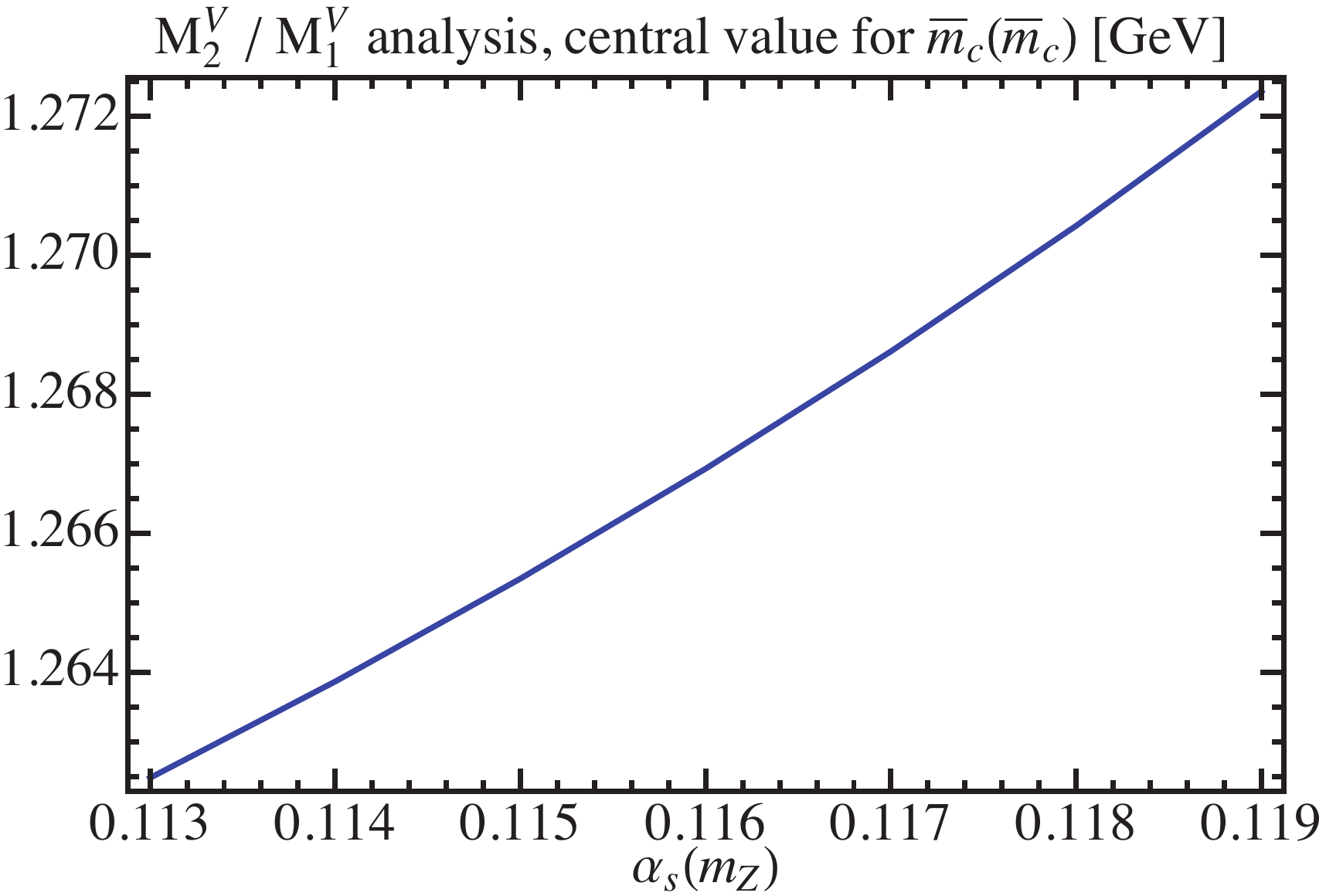}
\label{fig:mc-vector-ratio-central}
}
\subfigure[]{
\includegraphics[width=0.48\textwidth]{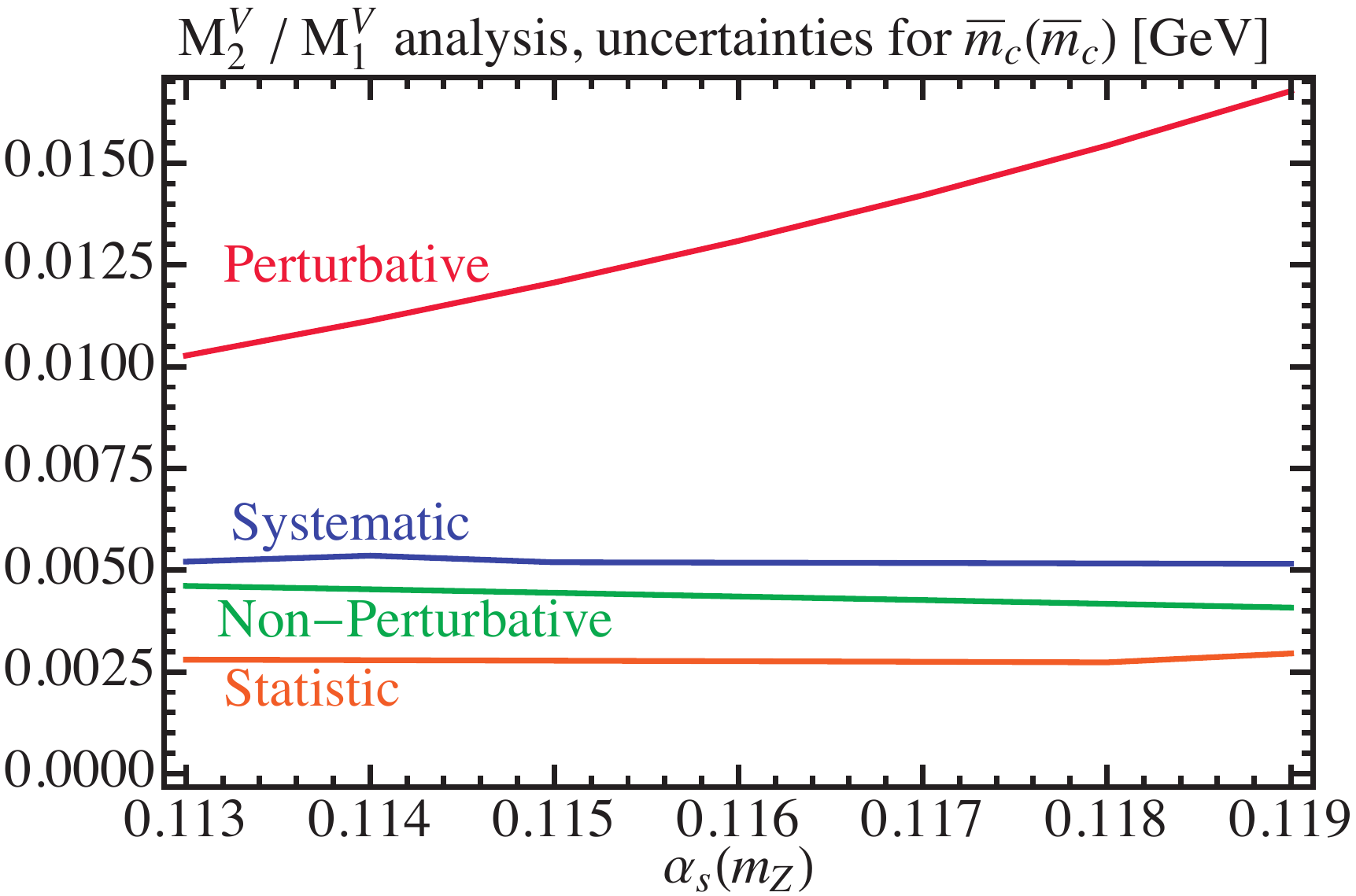}
\label{fig:mc-vector-ratio-err}
}
\caption{
Dependence on $\alpha_s(m_Z)$ of the central values of ${\overline m_c}({\overline m_c})$ and the 
corresponding perturbative (red), statistical (orange), systematical (blue) and nonperturbative 
uncertainties (green), for the analysis of the first moment [panels (a) and (b)] and the ratio of 
the second over the first moment, [panels (c) and (d)], corresponding to the vector correlator.}
\label{fig:charm-alphaS}
\end{figure*}

For the analysis using the first moment of the charm vector correlator we use the experimental 
value quoted in Eq.~(4.1) of Ref.~\cite{Dehnadi:2011gc}:
$M_1^{V,\,\rm exp}=(0.2121\,\pm\, 0.0020_{\rm stat}\,\pm\, 0.0030_{\rm syst})\,{\rm GeV}^{-2}$. The 
outcome of this analysis, and one of the main results of this paper, is:

\begin{align}\label{eq:vector-result}
\overline m_c(\overline m_c) = & \,1.288 \, \pm \, (0.006)_{\rm stat}
\, \pm \, (0.009)_{\rm syst} \, \pm \, (0.014)_{\rm pert}\\
&\, \pm \, (0.010)_{\alpha_s} \, \pm \, (0.002)_{\langle GG\rangle}\,{\rm GeV}\,,\nonumber
\end{align}

where the quoted errors are (from left to right) experimental uncorrelated, experimental 
correlated, peturbative, due to the uncertainty in $\alpha_s$ as given in Eq.~(\ref{eq:alphaswa}), 
and nonperturbative. If we adopt the correlated scale variation $2\,{\rm 
GeV}\le\mu_\alpha=\mu_m\le4$\,GeV, we obtain for method (c) $1.297\, \pm \, (0.005)_{\rm pert}$, 
with the other errors essentially unchanged. For method (a) we would get $1.290\, \pm \, 
(0.0007)_{\rm pert}$, with a scale variation even smaller than the nonperturbative uncertainty, and 
$20$ times smaller than our perturbative error estimate with double scale variation [$3$ times for 
method (c)]. The dependence on $\alpha_s(m_Z)$ is shown graphically in 
Figs.~\ref{fig:mc-vector-central} and \ref{fig:mc-vector-err}, and analytically the result reads:

\begin{eqnarray}
\label{eq:mc-vec-alphas}
\overline m_c(\overline m_c)& = &(1.288 + 4.40\times[\alpha_s(m_Z) - 0.1185])
\, \pm \, (0.006)_{\rm stat} \, \pm \, (0.009)_{\rm syst}\\ &&
\, \pm \, (0.014 + 0.95\times[\alpha_s(m_Z) - 0.1185])_{\rm pert} 
\, \pm \, (0.002)_{\langle GG\rangle}\,.\nonumber
\end{eqnarray}

Eqs.~(\ref{eq:vector-result}) and (\ref{eq:mc-vec-alphas}) supersede the results given in 
Eqs.~(4.5) and (4.2) of Ref.~\cite{Dehnadi:2011gc}, respectively.

For the ratio of the second over the first moment of the vector correlator we use as the 
experimental input
$R_1^{V,\,\rm exp} = (6.969\,\pm\, 0.032_{\rm stat}\,\pm\, 0.059_{\rm syst})\times 
10^{-2}$\,GeV$^{-2}$, which yields the following result for the charm mass:

\begin{align}
\overline m_c(\overline m_c) = & \,1.271 \, \pm \, (0.003)_{\rm stat}
\, \pm \, (0.005)_{\rm syst} \, \pm \, (0.016)_{\rm pert}\\
&\, \pm \, (0.004)_{\alpha_s} \, \pm \, (0.004)_{\langle GG\rangle}\,{\rm GeV}\,.\nonumber
\end{align}

With correlated variation $2\,{\rm GeV}\le\mu_\alpha=\mu_m\le4\,{\rm GeV}$ we get $1.258\, \pm \, 
(0.005)_{\rm pert}$ and $1.279\, \pm \, (0.007)_{\rm pert}$ for methods (a) and (c), respectively. 
In this case the scale variations are a factor of $2$ to $3$ smaller than our perturbative error 
estimate.\footnote{Had we taken the fixed-order expansion (a) and correlated scale variation 
$2\,{\rm GeV}\le\mu_\alpha=\mu_m\le4\,{\rm GeV}$ as the estimate for the perturbative uncertainty, 
the result from $R_1^V$ with all errors added quadratically would be $1.258\pm0.013$\,GeV, whereas 
the result from $M_1^V$  would read $1.290\pm0.015$\,GeV. Both results would not be consistent to 
each other.} The $\alpha_s$ dependence, which can be seen in Figs.~\ref{fig:mc-vector-ratio-central} 
and \ref{fig:mc-vector-ratio-err}, has the form:

\begin{align}
\label{eq:mc-rat-alphas}
\overline m_c(\overline m_c)& = (1.271 + 1.64\times[\alpha_s(m_Z) - 0.1185])
\, \pm \, (0.003)_{\rm stat} \, \pm \, (0.005)_{\rm syst}\\
&\, \pm \, (0.016 + 1.081\times[\alpha_s(m_Z) - 0.1185])_{\rm pert} \, \pm \, (0.004)_{\langle 
GG\rangle}\,.\nonumber
\end{align}

We observe that the central value for the ratios of moments is $17$\,MeV smaller than for the first 
moment analysis, but fully compatible within theoretical uncertainties. Furthermore, the dependence 
on $\alpha_s$ of the central value obtained from the regular moment analysis is larger, which 
translates into a corresponding larger error due to the uncertainty in $\alpha_s$. Both 
determinations have very similar perturbative uncertainties for any value of $\alpha_s$. We also see 
that the charm mass from the ratio of moments has smaller experimental uncertainties, as a result of 
cancellations between correlated errors. Moreover, the charm mass result from $R_1^V$ has a 
nonperturbative error twice as large as that from $M_1^V$. The two charm mass results from the first 
moment and the moment ratio are compared graphically in Fig.~\ref{fig:comparison-observables-charm}.

\subsection{Results for the Charm Mass from the Pseudoscalar Correlator}

\begin{figure*}[t!]
\subfigure[]
{
\includegraphics[width=0.48\textwidth]{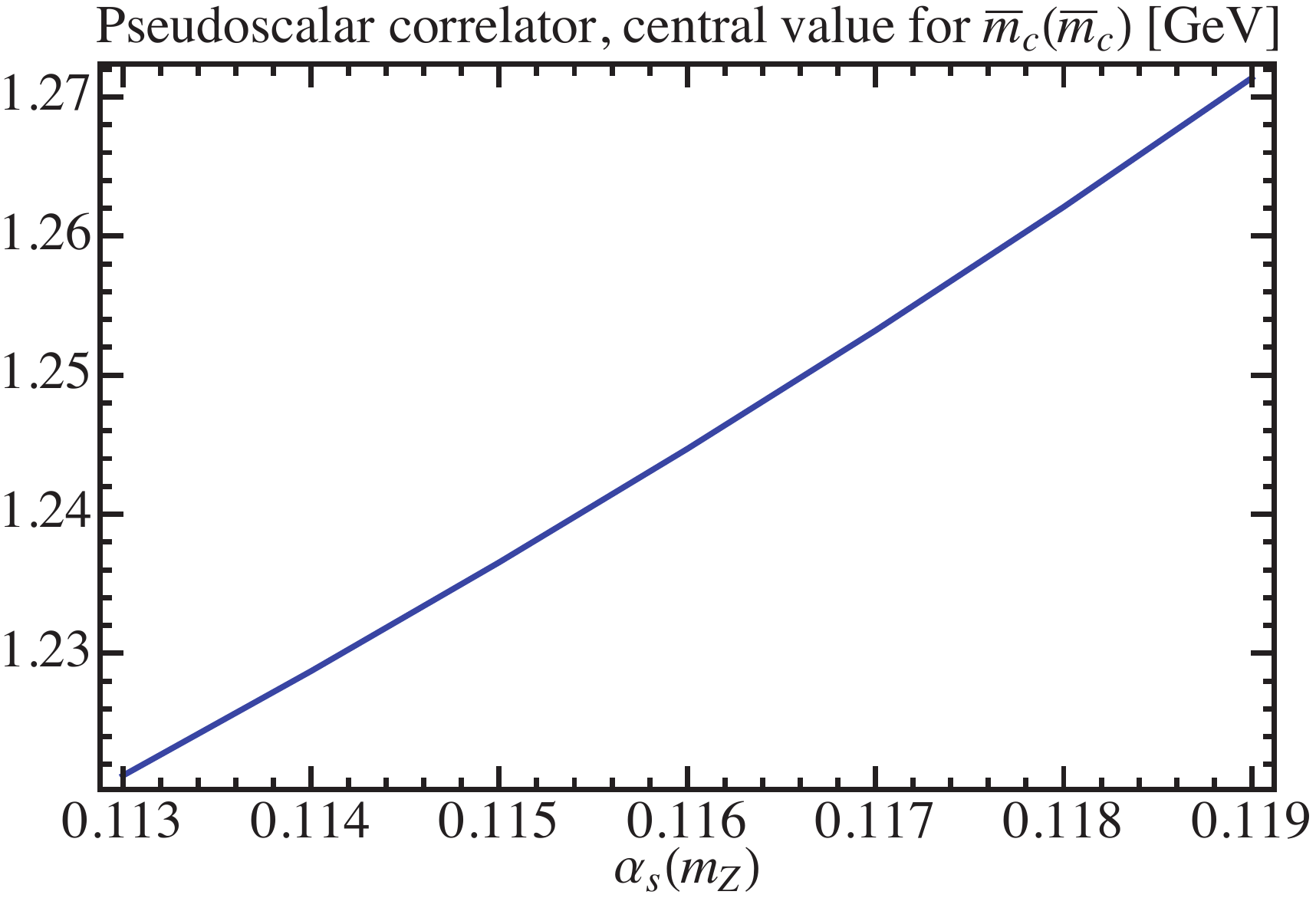}
\label{fig:mc-pseudo-central}
}
\subfigure[]{
\includegraphics[width=0.48\textwidth]{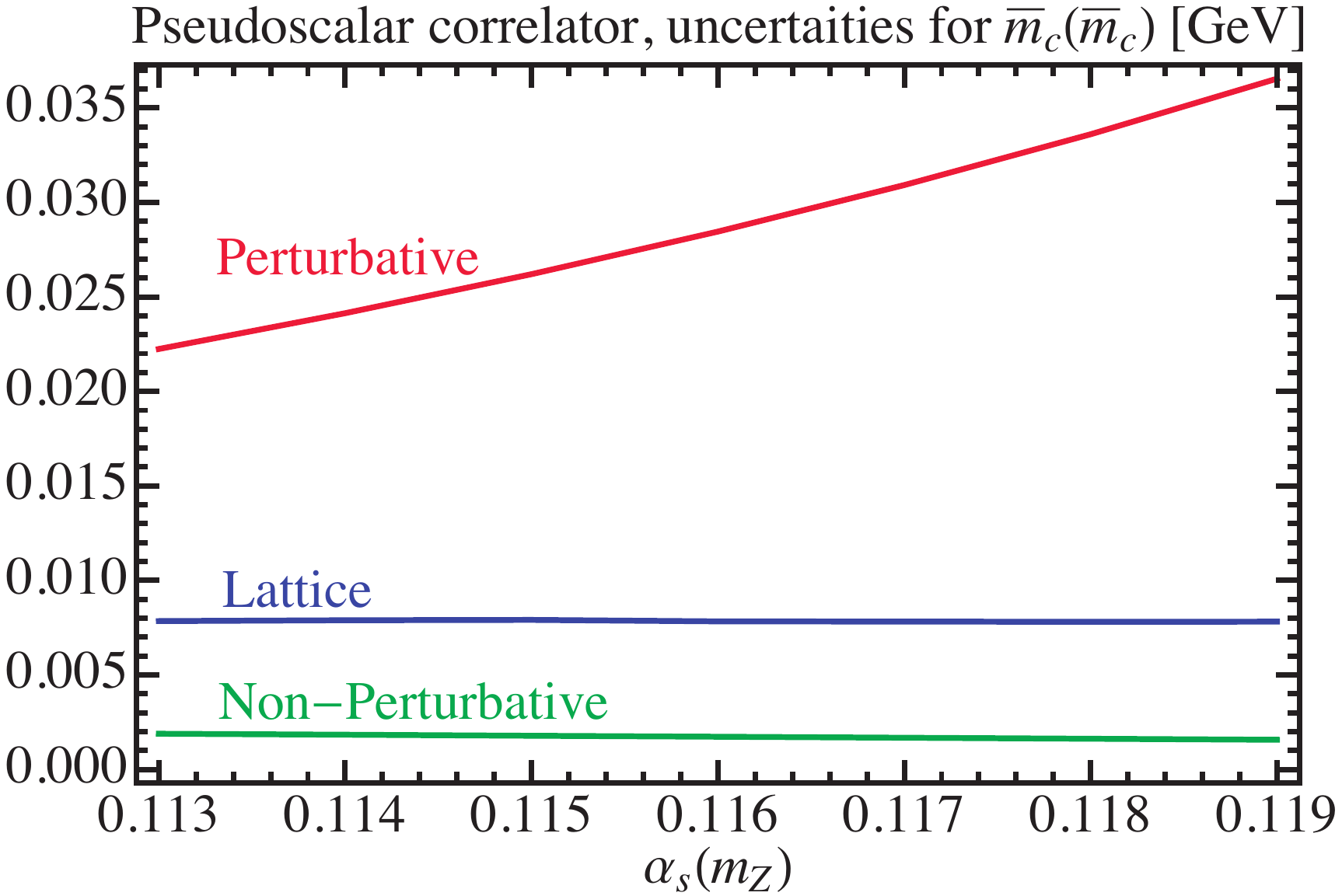}
\label{fig:mc-pseudo-err}
}
\subfigure[]
{
\includegraphics[width=0.48\textwidth]{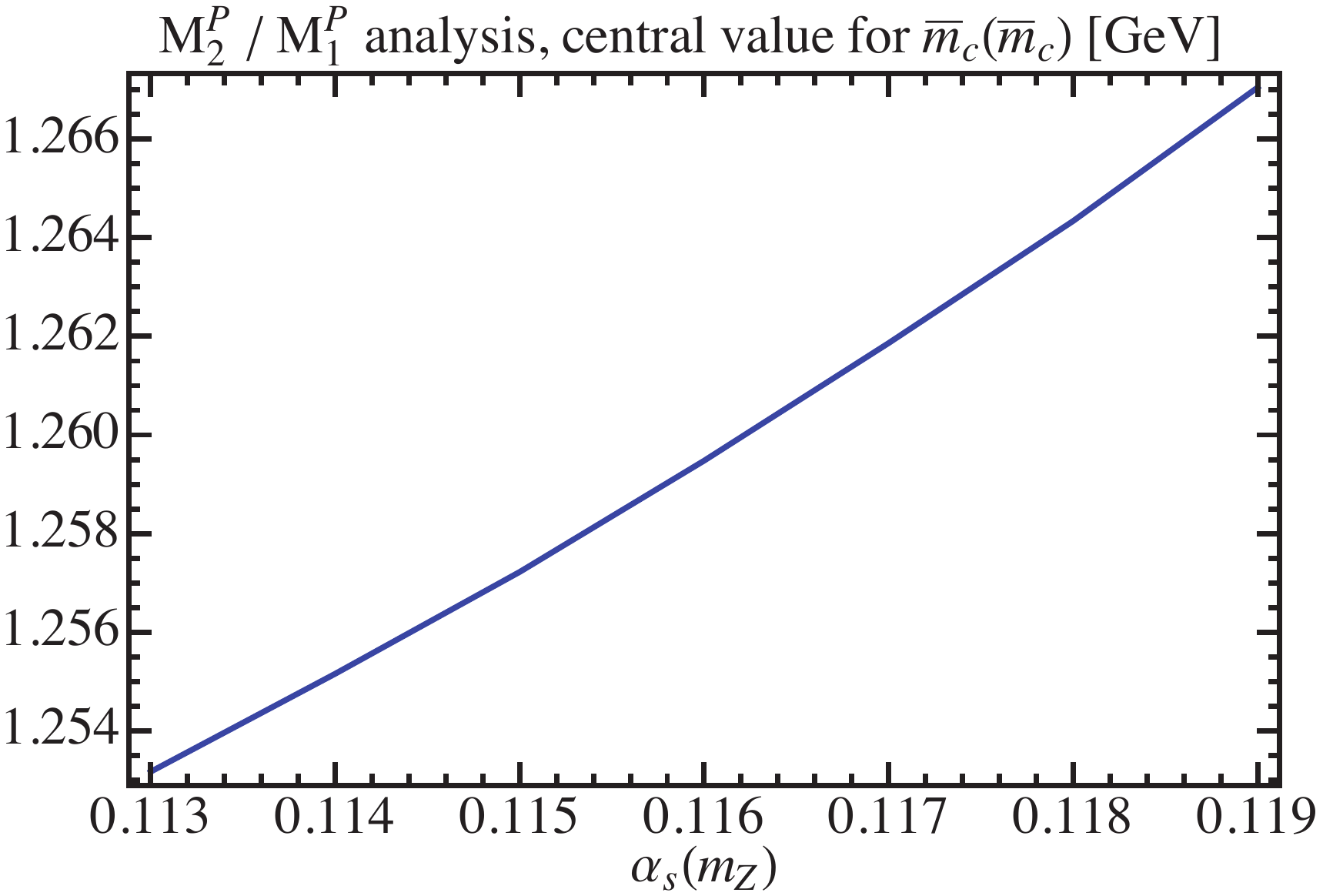}
\label{fig:mc-pseudo-ratio-central}
}
\subfigure[]{
\includegraphics[width=0.48\textwidth]{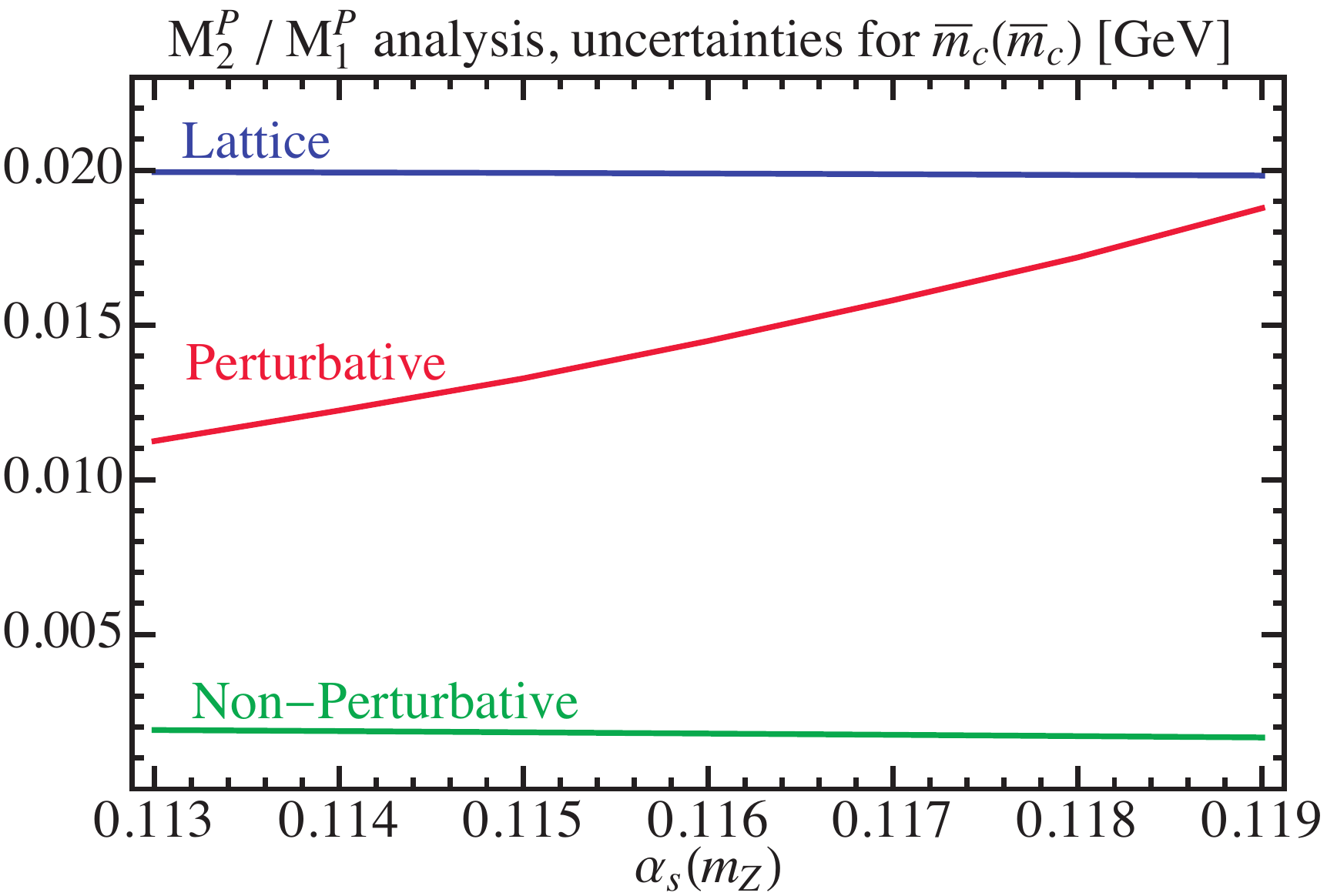}
\label{fig:mc-pseudo-ratio-err}
}
\caption{
Dependence on $\alpha_s(m_Z)$ of the central values of ${\overline m_c}({\overline m_c})$ and the 
corresponding perturbative (red), lattice (blue) and nonperturbative 
uncertainties (green), for the analysis of the first moment [panels (a) and (b)] and the ratio of 
the second over the first moment, [panels (c) and (d)], corresponding to the pseudoscalar 
correlator.}
\label{fig:mc-ratios-alphaS}
\end{figure*}

For the analysis of the first moment of the charm pseudoscalar correlator we employ
$M_1^{P,\,{\rm latt}}\,=\,(0.1402\,\pm\, 0.0020_{\rm latt})\,{\rm GeV}^{-2}$~\cite{Allison:2008xk}, 
which yields the following charm mass determination:

\begin{align}
\overline m_c(\overline m_c) = \,1.267 \, \pm \, (0.008)_{\rm lat}
\, \pm \, (0.035)_{\rm pert} \, \pm \, (0.019)_{\alpha_s}
\, \pm \, (0.002)_{\langle GG\rangle}\,{\rm GeV}\,.
\end{align}

With correlated scale variation $2\,{\rm GeV}\le\mu_\alpha=\mu_m\le4\,{\rm GeV}$ we obtain the 
central values $1.278$ and $1.276$\,GeV, for methods (b) and (c), respectively. In both cases the 
scale variation is $4$\,MeV, 8 times smaller than our perturbative error estimate with double scale 
variation. For the $\alpha_s$ dependence, we find

\begin{align}
\label{eq:mc-lat-alphas}
\overline m_c(\overline m_c)& = (1.267 + 8.36\times[\alpha_s(m_Z) - 0.1185]) \,
\pm \, (0.008)_{\rm lat}\\
&\, \pm \, (0.035 + 2.38\times[\alpha_s(m_Z) - 0.1185])_{\rm pert} \, \pm \, 
(0.002)_{\langle GG\rangle}
\,,\nonumber
\end{align}

which is also displayed in Figs.~\ref{fig:mc-pseudo-central} and \ref{fig:mc-pseudo-err}. As 
expected, the perturbative error is much larger than for the vector correlator, and has a stronger 
dependence on $\alpha_s$. We see that the central value has a much stronger dependence on 
$\alpha_s$ as well, which again translates into a large error due to the uncertainty in the strong 
coupling. The central value is $21$\,MeV lower than Eq.~(\ref{eq:vector-result}), but fully 
compatible within errors [see Fig~\ref{fig:comparison-observables-charm}]. The nonperturbative 
uncertainties are identical to the vector current case.

For the ratio of second over the first moment of the pseudoscalar correlator we use 
$R_1^{P,\,\rm latt} = (0.0971\, \pm\, 0.0032_{\rm latt})\,$GeV$^{-2}$. We find for the charm mass

\begin{align}
\overline m_c(\overline m_c) =  \,1.266\, \pm \, (0.020)_{\rm latt}
\, \pm \, (0.018)_{\rm pert} \, \pm \, (0.006)_{\alpha_s}
\, \pm \, (0.002)_{\langle GG\rangle}\,{\rm GeV}\,.
\end{align}

Using correlated variation $2\,{\rm GeV}\le\mu_\alpha=\mu_m\le4\,{\rm GeV}$ one obtains $1.270\, 
\pm \, (0.007)_{\rm pert}$ and $1.278\, \pm \, (0.003)_{\rm pert}$ for methods (a) and (c), 
respectively. These scale variations are a factor $3$ and $6$ smaller than our perturbative error 
estimate, respectively. The $\alpha_s$ dependence is

\begin{align}
\label{eq:mc-rat-lat-alphas}
\overline m_c(\overline m_c) & = (1.266 + 2.31\times[\alpha_s(m_Z) - 0.1185]) \,
\pm \, (0.020)_{\rm latt}\\
&\, \pm \, (0.018 + 1.25\times[\alpha_s(m_Z) - 0.1185])_{\rm pert} \, \pm \,
(0.002)_{\langle GG\rangle} \,.\nonumber
\end{align}

The central values for both $M_1^P$ and $R_1^P$ are almost identical, but their $\alpha_s$ 
dependence is not: the latter is much smaller (even smaller than for $M_1^V$, but larger than for 
$R_1^V$). Note that the lattice error is larger for the ratio since we made the very conservative 
assumption that they are fully uncorrelated. This is because the correlation matrix for various 
lattice moments is unknown. The perturbative error reduces by a factor of two for any value of 
$\alpha_s$ when using the ratio, but we have checked that this only happens for the iterative 
expansion. On the other hand, the $\alpha_s$ dependence of the perturbative uncertainty is smaller 
for the regular moment determination. The nonperturbative errors are identical.

All charm determinations are illustrated graphically in 
Fig.~\ref{fig:comparison-observables-charm}, where in red we show our preferred determination from 
the first moment of the vector correlator.

\subsection{Results for the Bottom Mass from the Vector Correlator}

\begin{figure*}[t!]
\subfigure[]
{
\includegraphics[width=0.48\textwidth]{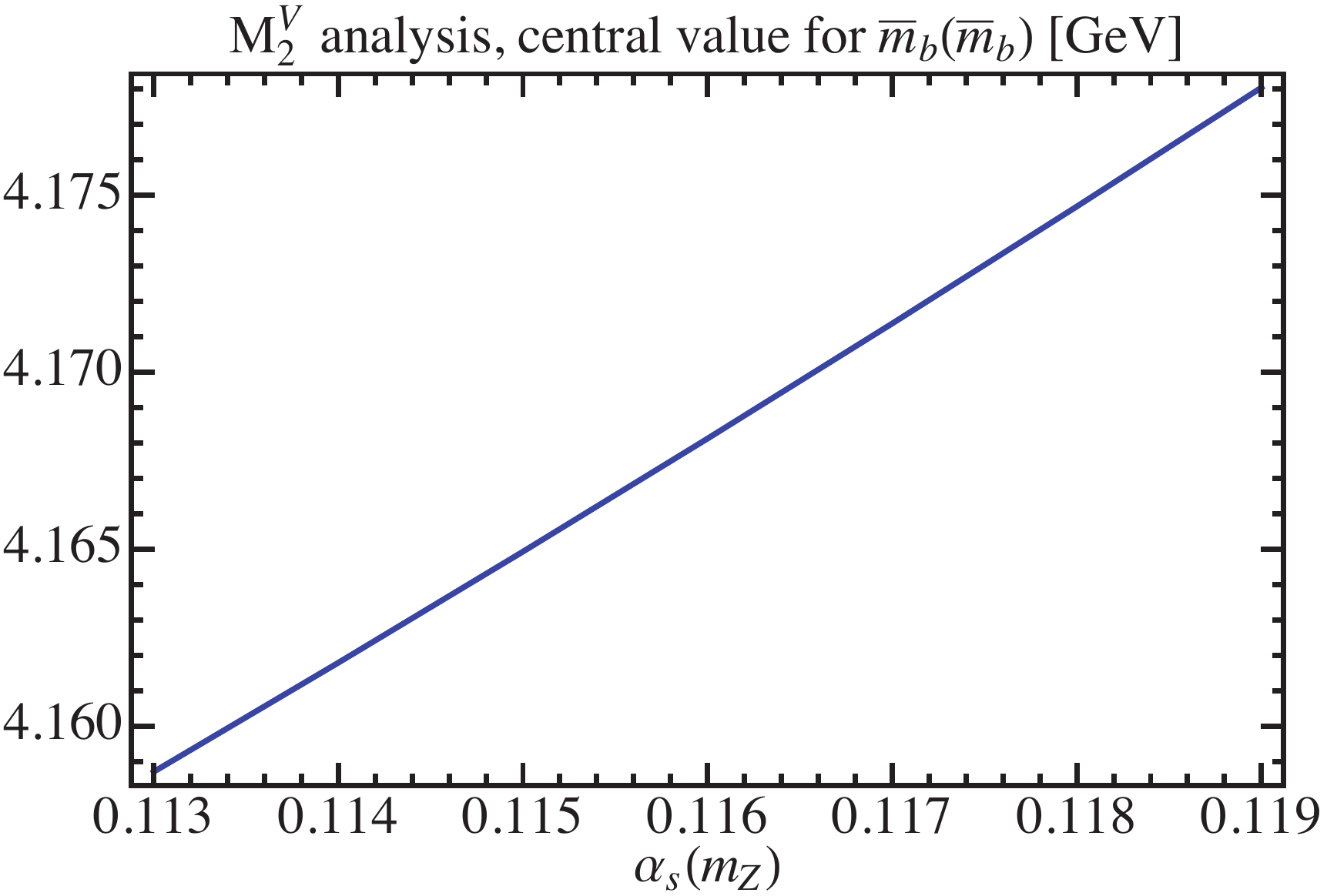}
\label{fig:mb-vector-central}
}
\subfigure[]{
\includegraphics[width=0.48\textwidth]{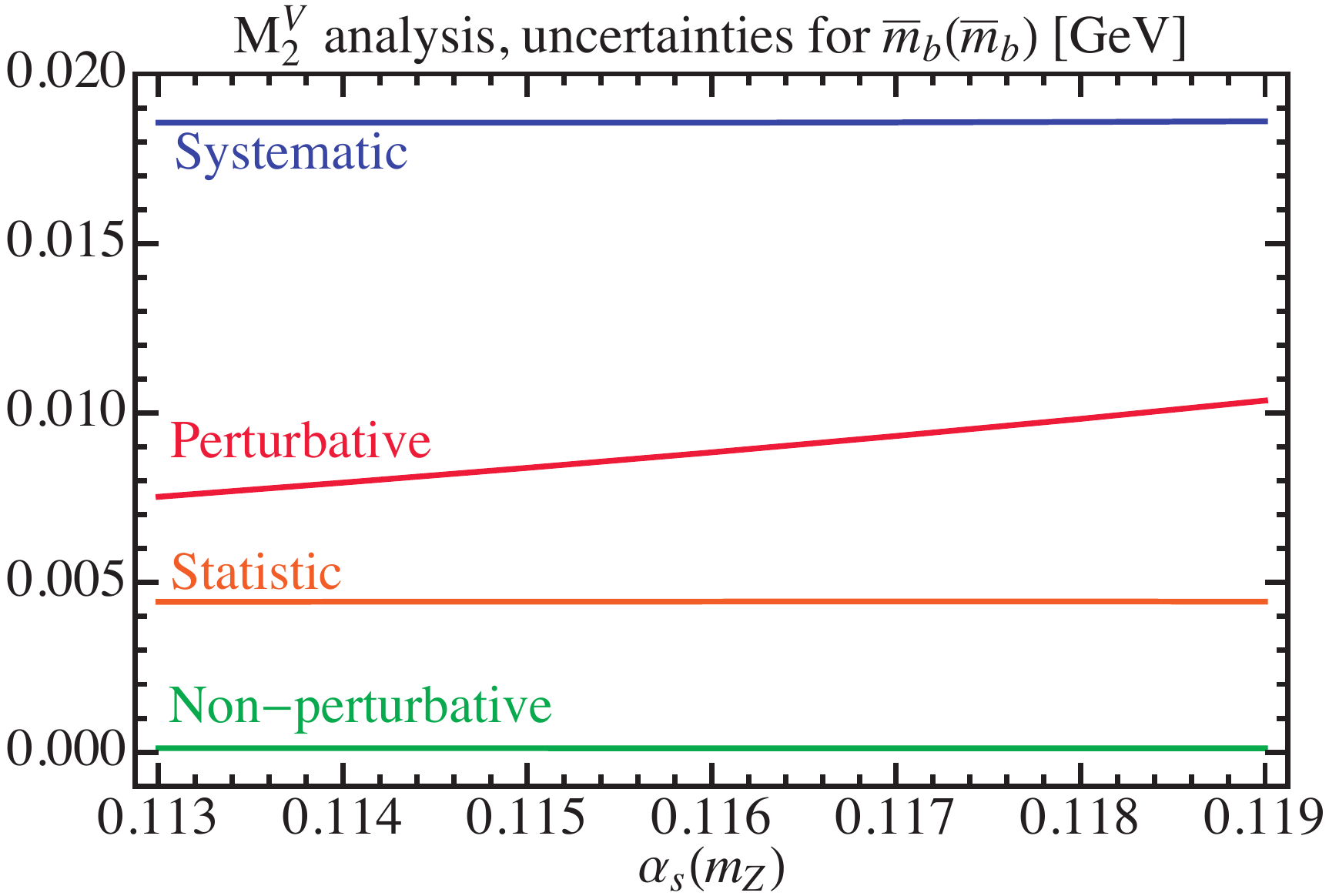}
\label{fig:mb-vector-err}
}
\subfigure[]
{
\includegraphics[width=0.48\textwidth]{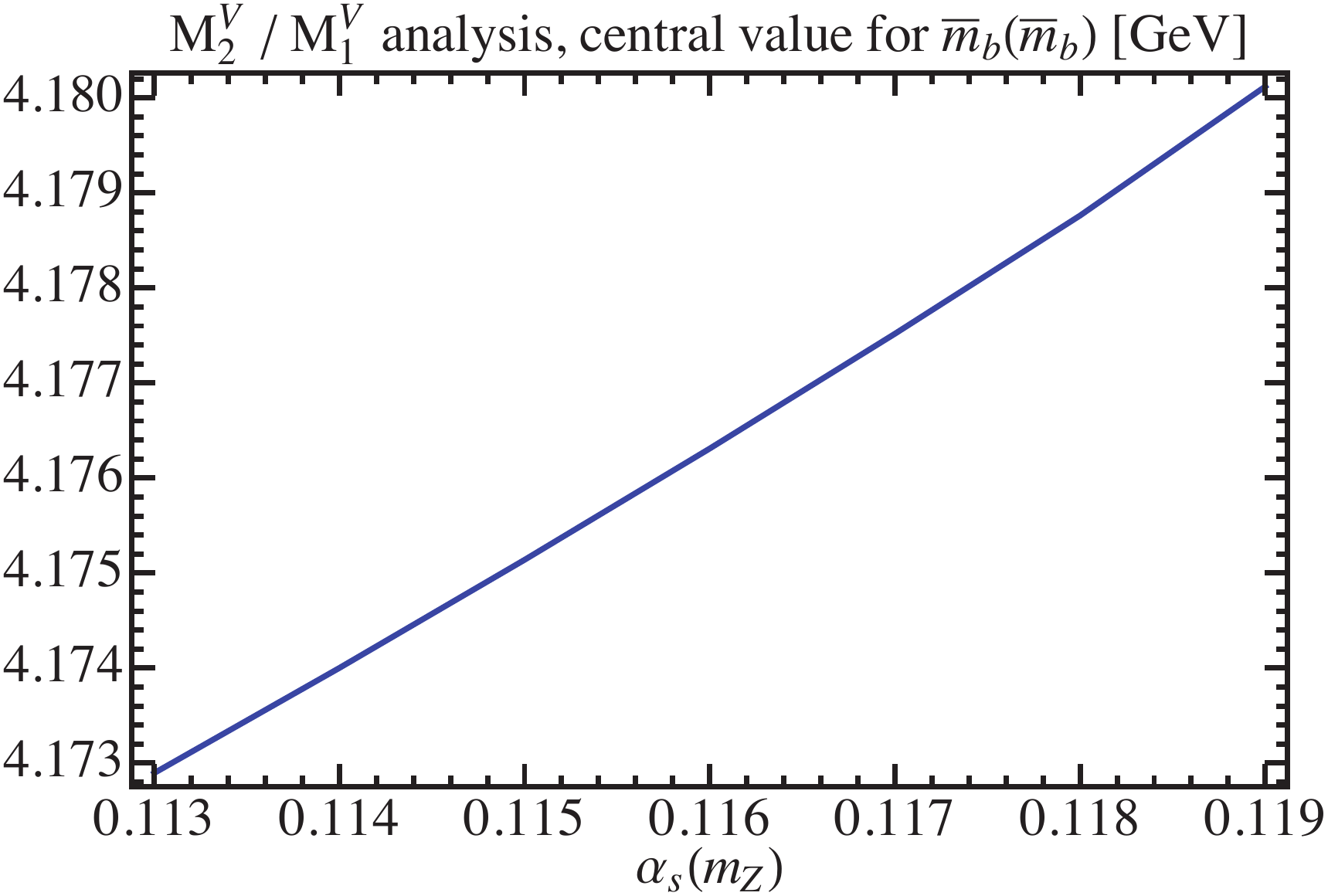}
\label{fig:mb-vector-ratio-central}
}
\subfigure[]{
\includegraphics[width=0.48\textwidth]{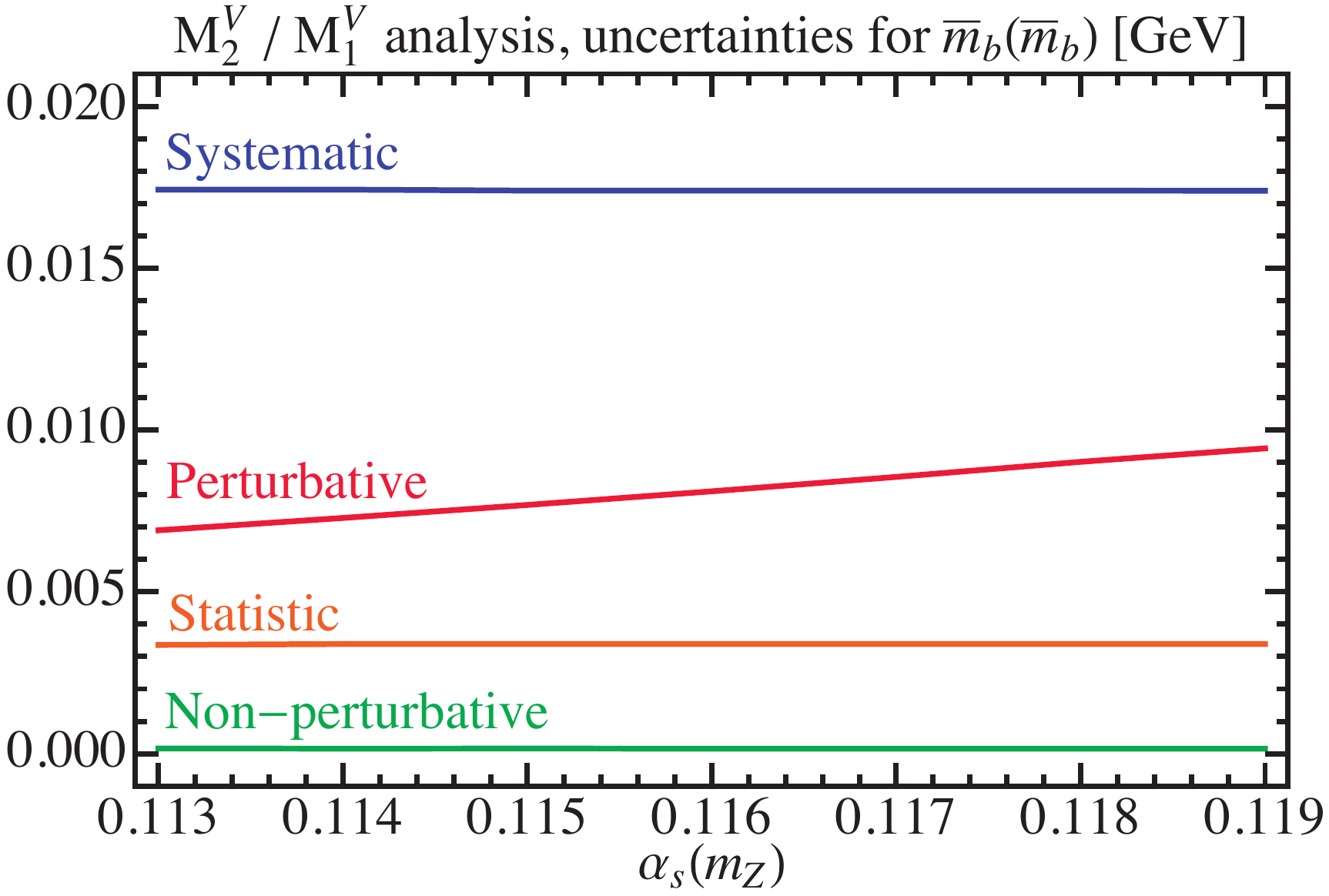}
\label{fig:mb-vector-ratio-err}
}
\caption{
Dependence on $\alpha_s(m_Z)$ of the central values of ${\overline m_b}({\overline m_b})$ and the 
corresponding perturbative (red), statistical (orange), systematical (blue) and nonperturbative 
uncertainties (green), for  for the analysis of the second moment [panels (a) and (b)] and the 
ratio of the second over the the first moment [panels (c) and (d)] of the vector correlator.}
\label{fig:mb-alphaS}
\end{figure*}

For our determination of the bottom quark mass from the second moment of the vector correlator we 
use for the experimental moment
$M_2^{V,\,\rm exp}=(2.834\,\pm\,0.012_{\rm stat}\,\pm\,0.051_{\rm syst})\times 
10^{-5}$\,GeV$^{-4}$, and we obtain

\begin{align}
\overline m_b(\overline m_b) = & \,4.176 \, \pm \, (0.004)_{\rm stat}
\, \pm \, (0.019)_{\rm syst} \, \pm \, (0.010)_{\rm pert}\\
&\, \pm \, (0.007)_{\alpha_s} \, \pm \, (0.0001)_{\langle GG\rangle}\,{\rm GeV}\nonumber\,.
\end{align}

The perturbative error is $30\%$ smaller than for the charm vector correlator analysis, as a 
result of the smaller value of $\alpha_s$ at the scales close to the bottom mass. This is 
consistent with our discussion on the convergence properties of perturbation series for the bottom 
quark carried out in Sec.~\ref{sec:convergence}. The total error is dominated by the experimental 
systematic uncertainty, which in turn mainly comes from the continuum region where one relies on 
modeling in the absence of any experimental measurements. The nonperturbative error is completely 
negligible. This is expected since it is suppressed by two powers of the bottom mass. Using the 
correlated scale variation $5\,{\rm GeV}\le\mu_\alpha=\mu_m\le15\,{\rm GeV}$ for methods (a) and 
(c) we get $4.178$ and $4.182$ for the central values and scale variations which are $20$ and $3$ 
times smaller, respectively. The $\alpha_s$ dependence reads

\begin{align}
\label{eq:mb-alphas}
\overline m_b(\overline m_b)& = (4.176 + 3.22\times[\alpha_s(m_Z) - 0.1185]) \,
\pm \, (0.004)_{\rm stat} \, \pm \, (0.019)_{\rm syst}\\
&\, \pm \, (0.010 + 0.472\times[\alpha_s(m_Z) - 0.1185])_{\rm pert} \,
\pm \, (0.0001)_{\langle GG\rangle} \,.\nonumber
\end{align}

For the analysis based on the ratio of the second moment over the first, we use
$R_2^{V,\,\rm exp} = (6.262\,\pm\,0.010_{\rm stat}\,\pm\,0.053_{\rm syst})\times 
10^{-3}$\,GeV$^{-2}$, and with this value we obtain the bottom mass

\begin{align}
\overline m_b(\overline m_b) = & \,4.179 \, \pm \, (0.003)_{\rm stat}
\, \pm \, (0.017)_{\rm syst} \, \pm \, (0.009)_{\rm pert}\\
&\, \pm \, (0.003)_{\alpha_s} \, \pm \, (0.0002)_{\langle GG\rangle}\,{\rm GeV}\nonumber\,.
\end{align}

With correlated scale variation $5\,{\rm GeV}\le\mu_\alpha=\mu_m\le15\,{\rm GeV}$  we obtain 
$4.175\, \pm \, (0.003)_{\rm pert}$ and $4.182\, \pm \, (0.004)_{\rm pert}$ for methods (a) and (c), 
respectively. In this case the scale variation is smaller by a factor $3$ and $2$, respectively. 
The $\alpha_s$ dependence reads as follows:

\begin{align}
\label{eq:mb-ratio-alphas}
\overline m_b(\overline m_b)& = (4.179 + 1.199\times[\alpha_s(m_Z) - 0.1185]) \, \pm \, 
(0.003)_{\rm stat}  \, \pm \, (0.017)_{\rm syst}\\
&\, \pm \, (0.009 + 0.426\times[\alpha_s(m_Z) - 0.1185])_{\rm pert} 
\, \pm \, (0.0002)_{\langle GG\rangle}\,.\nonumber
\end{align}

Although the central value for the ratio analysis is $3$\,MeV higher, this has no significance 
given the size of the uncertainties. The dependence of the central value on $\alpha_s$ is three 
times smaller for the ratio analysis. The perturbative error and its $\alpha_s$ dependence are 
roughly the same for the ratio and the single moment analysis. Moreover, the two experimental 
errors are very similar. This is because, even though there is some cancellation of correlated 
errors in the ratio, a significant part of the huge systematic error of the first moment remains 
uncanceled.

\begin{figure*}[t!]
\subfigure[]
{
\includegraphics[width=0.48\textwidth]{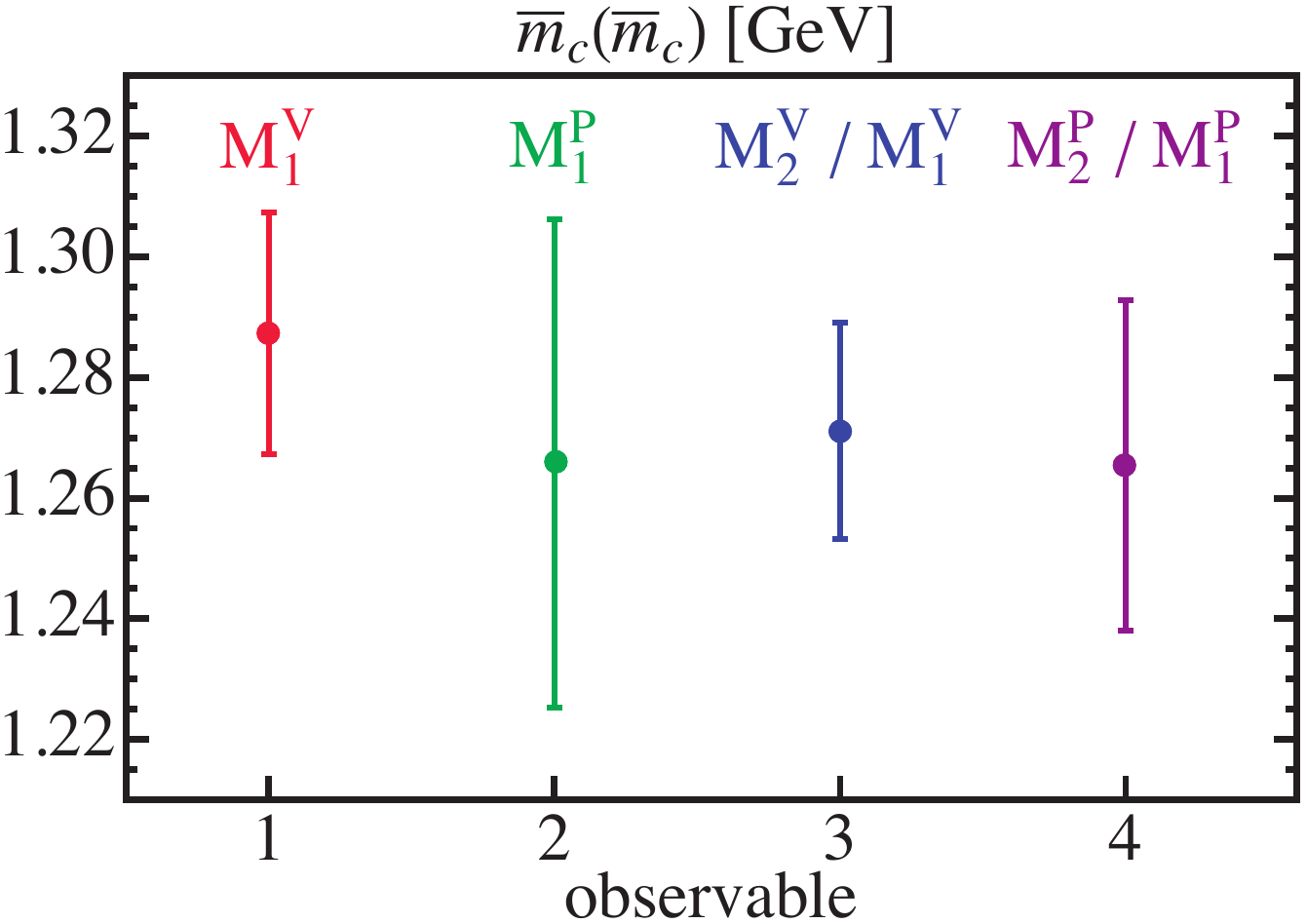}
\label{fig:comparison-observables-charm}}
\subfigure[]
{
\includegraphics[width=0.48\textwidth]{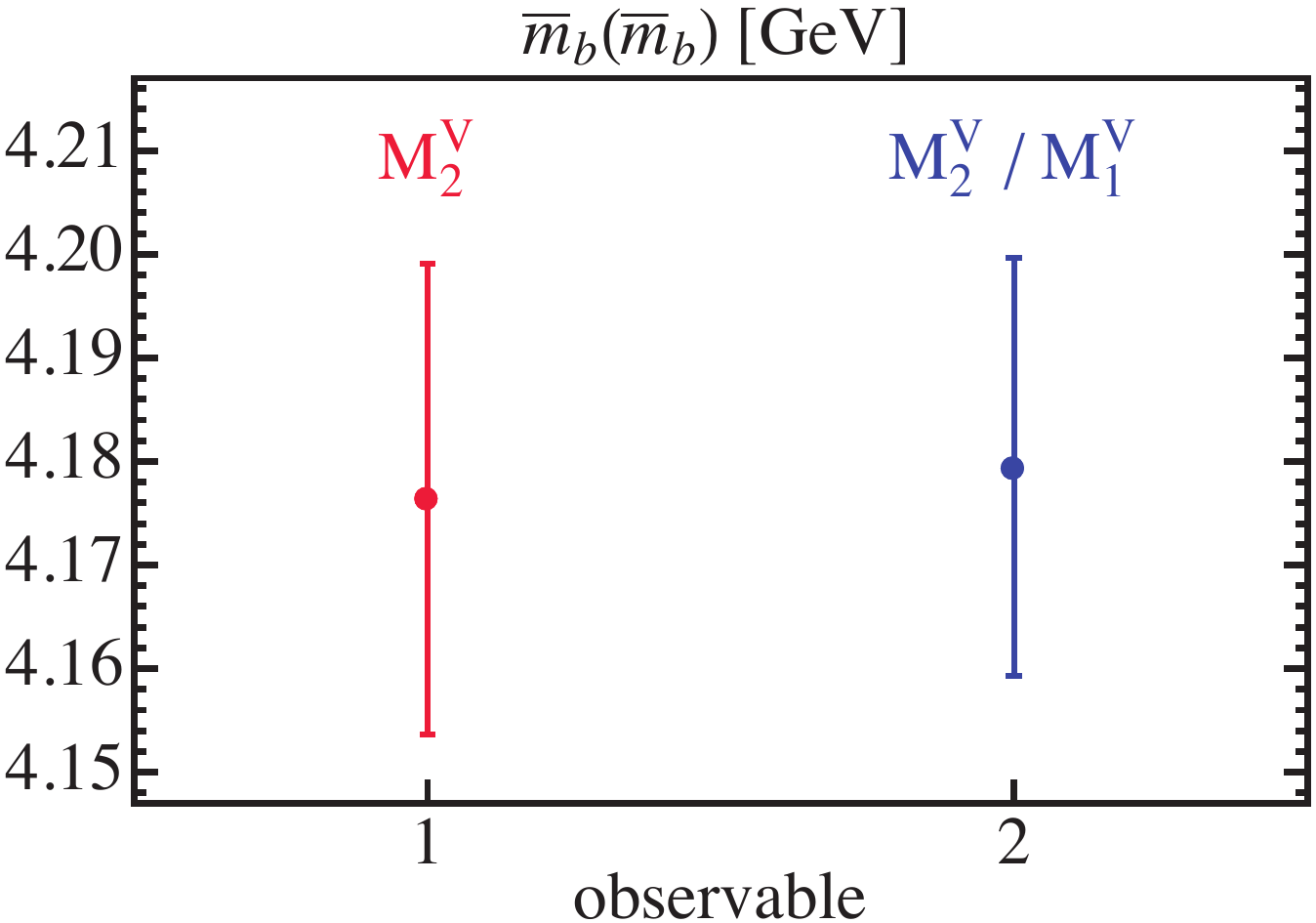}
\label{fig:comparison-observables-bottom}}
\caption{Charm (a) [bottom (b)] mass determinations from the first [second] moment of the vector 
correlator (in red), the first moment of the pseudoscalar correlator (green, charm only), and the 
ratio of the second over the first moment of the vector (blue) and pseudoscalar correlator (purple, 
charm only).}
\label{fig:comparison-observables}
\end{figure*}

A graphical illustration of the two bottom mass determinations is shown in 
Fig.~\ref{fig:comparison-observables-bottom}. Both combined uncertainties and central values are 
rather similar, and we adopt the result from the second moment (in red) as our default result.

\section{Comparison to other Determinations}
\label{sec:comparison-masses}

In this section we make a comparison to previous analyses of our updated charm mass determination 
from the vector correlator, our new results for the charm mass from the pseudoscalar current 
correlator and of our bottom mass determination. We restrict our discussion to determinations which 
use QCD sum rules with infinite as well as finite energy range for the vector or pseudoscalar 
current correlators, and including relativistic and nonrelativistic versions of the sum rules. We 
do not cover charm mass determinations from DIS or bottom mass determinations from jets (which are 
in any case rather imprecise), as well as determinations which are based on the mass of bound 
states (B mesons or quarkonia) or B decays.

In Figs.~\ref{fig:charm-comparions} and \ref{fig:bottom-comparions} we present in a graphical form 
a compilation of recent sum rule determinations of the charm and bottom masses, respectively. We 
have labeled them from top to bottom with numbers from $1$ to $14$. We note that comparing these 
results, one has to keep in mind that different analyses in general employed different values and 
uncertainties for the strong coupling. Only the analyses in Refs.~\cite{Hoang:2004xm, 
Chetyrkin:2009fv, Bodenstein:2010qx, Dehnadi:2011gc, Hoang:2012us} and ours have provided the 
dependence of their results on the value of $\alpha_s(m_Z)$.

\subsection{Charm Mass}

Let us first focus our attention on the charm mass, Fig.~\ref{fig:charm-comparions}. Within each 
color, determinations are ordered according to publication date. In red (determinations $12$ to 
$14$) we show the results of our collaboration: $12$ and $13$ for the vector correlator, the former 
(dashed) corresponding to Ref.~\cite{Dehnadi:2011gc} without trimming procedure, and the latter 
(solid) corresponding to this work, which includes the trimming procedure. Determinations $12$ to 
$14$ are the only analyses using uncorrelated scale variation. Determination $6$ 
(gray)~\cite{Hoang:2004xm} sets $\mu_m = \overline{m}_c(\overline{m}_c)$, and all the other 
analyses have set $\mu_m = \mu_\alpha$. Determinations in blue ($1$~\cite{Chakraborty:2014aca}, 
$2$~\cite{McNeile:2010ji} and $3$~\cite{Allison:2008xk}) were performed by the HPQCD collaboration, 
which employ method (b) for the pseudoscalar correlator moments used for the mass determination in 
their lattice analyses. Only $1$ to $3$ and $14$ use pseudoscalar moments, while all the other 
analyses use the vector correlator. Among those $7$~\cite{Bodenstein:2011ma}, 
$8$~\cite{Chetyrkin:2009fv}, $9$~\cite{Kuhn:2007vp} and $10$~\cite{Boughezal:2006px} use data in the 
threshold region only up to $4.8$\,GeV; analysis $6$ uses two patches of data in the threshold 
region, one from threshold to $4.7$\,GeV, and another between $7.2$ and $11$\,GeV; analyses $12$ 
and $13$ use all available data (see Ref.~\cite{Dehnadi:2011gc} for the complete bibliographic 
information on charm data). The result in $6$ uses only $\mathcal{O}(\alpha_s^2)$ 
perturbative input [all the other analyses utilize $\mathcal{O}(\alpha_s^3)$ computations] and 
older information on the narrow resonances, from the PDG 2006. They also study the fixed-order 
expansion and two methods of contour improvement. In black ($8$ to $10$) we display results using 
fixed order analyses at ${\cal O}(\alpha_s^3)$ from the Karlsruhe ($8$ and $9$) and W\"urzburg 
($10$) collaborations.

Analysis $7$~(orange) corresponds to weighted finite-energy sum rules. They employ a kernel which 
enhances the sensitivity to the charm mass, and at the same time reduces the sensitivity to the 
continuum region. Green color analyses, collected in $4$~\cite{Narison:2010cg, Narison:2011xe} 
and $5$~\cite{Bodenstein:2010qx}, apply other kinds of sum rules.  Analysis $5$ uses a finite 
energy sum rule similar to $7$, but the kernel makes the sensitivity to the charm mass quite small. 
On the other hand, the two determinations of $4$ use shifted moments, ratios of shifted moments, 
and exponential sum rules, and consider only the contributions from the first 6 vector resonances in 
the narrow width approximation, and the pQCD prediction for the continuum. The lower analysis of 
$4$~\cite{Narison:2010cg} includes contributions from condensates up to dimension 6; the 
higher~\cite{Narison:2011xe} includes condensates up to dimension 8. In purple 
($11$~\cite{Signer:2008da}) we show the only analysis which uses large-$n$ moments for the charm 
mass fits, employing NRQCD methods to sum up large logs and threshold enhanced perturbative 
corrections, supplemented with fixed order predictions for the formally power suppressed terms. 
This analysis uses contributions from narrow resonances, plus a crude model for the threshold and 
continuum patches, for which a conservative uncertainty is assigned. We note, however, that this 
analysis might be questioned since perturbative NRQCD is in general not applicable for the 
charmonium states.

Our new vector correlator result agrees well with the world average, having a similar uncertainty. 
Our result is fully compatible with the other determinations shown in 
Fig.~\ref{fig:charm-comparions}. As mentioned already before, we disagree with the small 
perturbative uncertainties related to the scale variations of the vector and/or pseudoscalar 
moments adopted in analyses $1$ to $3$ and $7$ to $10$.

\subsection{Bottom Mass}

\begin{figure*}[tbh!]
\subfigure[]
{\label{fig:charm-comparions}
\includegraphics[width=0.497\textwidth]{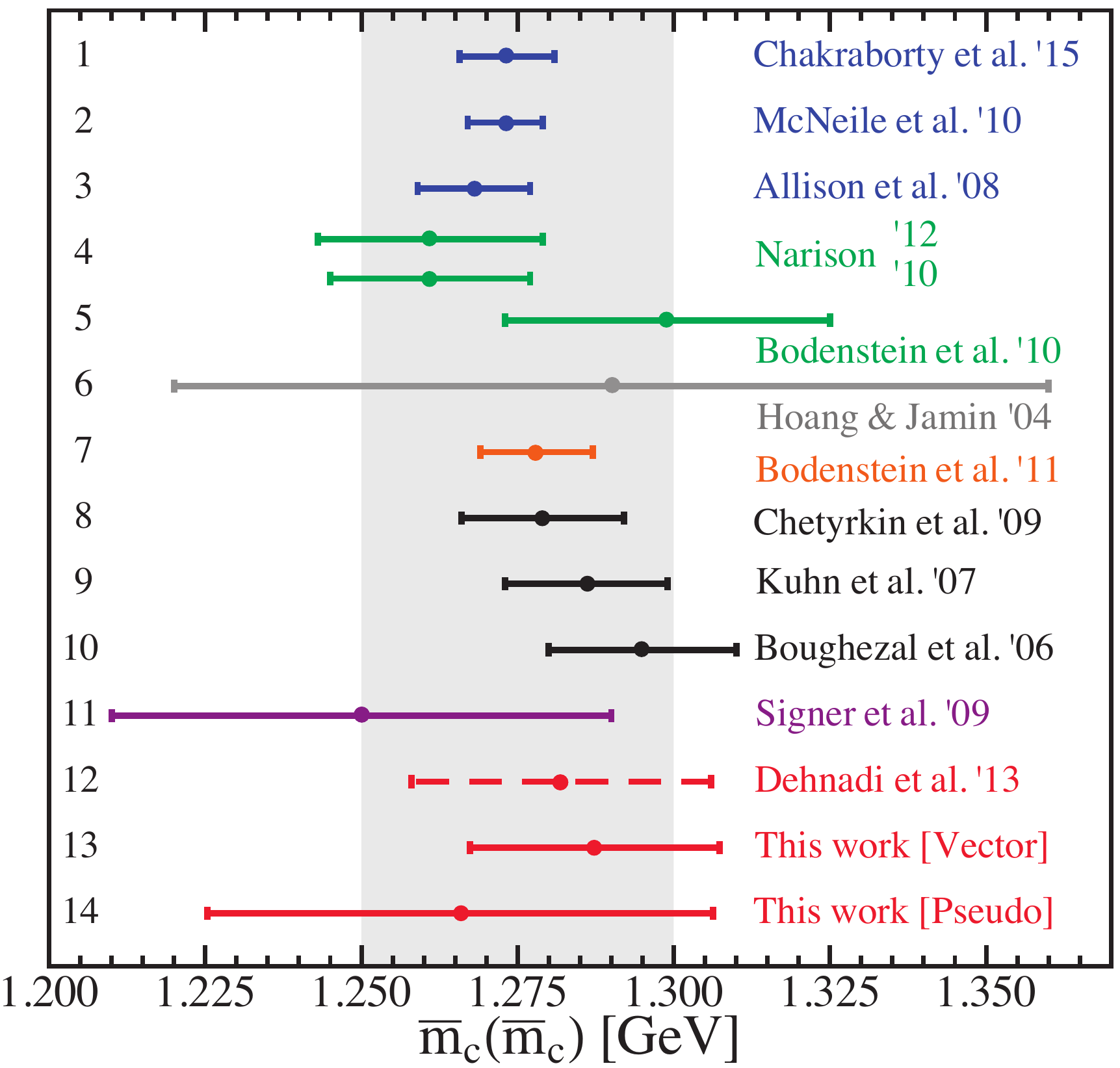}}~~~
\subfigure[]
{\label{fig:bottom-comparions}
\includegraphics[width=0.481\textwidth]{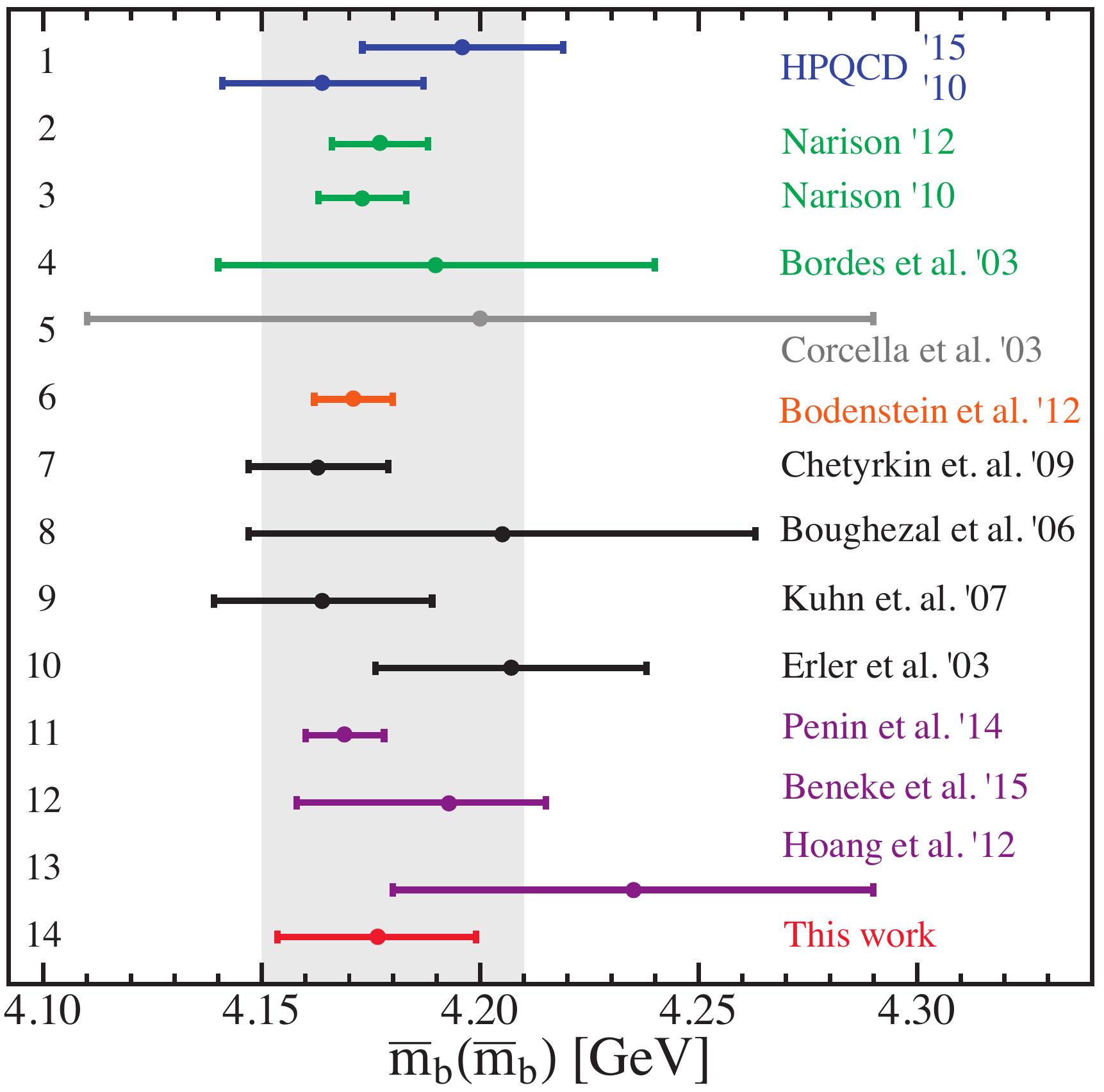}
}
\caption{Comparison of recent determinations of charm (a) and bottom (b) quark masses from sum rule 
analyses. Red results correspond to our determination. Black and gray correspond to 
$\mathcal{O}(\alpha_s^2)$ and $\mathcal{O}(\alpha_s^3)$ analyses, respectively. Purple results use 
nonrelativistic sum rules. Orange use weighted finite energy sum rules. Blue results are based on 
QCD sum rules using lattice simulation results as experimental data. Green labels other kinds of 
sum rule analyses (FESR, $Q^2$-dependent moments, ratios of moments).\label{fig:comparison}}
\end{figure*}

Let us now turn our attention to the bottom mass results, see Fig.~\ref{fig:bottom-comparions}. The 
coloring and chronological conventions are analogous to Fig.~\ref{fig:charm-comparions}, and we try 
to keep a similar ordering. We show three nonrelativistic determinations 
($11$~\cite{Hoang:2012us}, $12$~\cite{Beneke:2014pta} and $13$~\cite{Penin:2014zaa}) in purple; 
$\mathcal{O}(\alpha_s^2)$ fixed-order analyses are shown in gray ($5$~\cite{Corcella:2002uu}), 
black ($10$ \cite{Erler:2002bu}) and green ($4$~\cite{Bordes:2002ng}); finite energy sum rules also 
based on fixed-order appear in orange ($6$~\cite{Bodenstein:2012}) and green ($4$); there are two 
lattice analyses in blue, collected in $1$~\cite{Colquhoun:2014ica, McNeile:2010ji}. Analyses 
$3$, $6$, $7$ and $11$ to $13$ use the new BABAR data, whereas the others use the older CLEO and 
CUSB data. Analyses $4$ and $11$ include only the contributions of the first six vector resonances. 
Analyses $4$ and $5$ use older measurements of the electronic width for the narrow resonances. 
Analyses $3$, $4$, $6$ to $9$ use pQCD in the high-energy spectrum for the experimental moments. 
The theoretical treatment of the bottom mass analyses in red, gray, black, blue and green are in 
complete analogy to their charm mass analyses: $3$ and $4$ for bottom correspond to $4$ and $5$ for 
charm, respectively; $1$ in bottom corresponds to $1$ and $2$ for charm. 

The upper analysis of $1$~\cite{Colquhoun:2014ica} uses a nonrelativistic lattice action to compute
ratios of large-$n$ moments, which are later compared to relativistic continuum perturbation theory.
Because the continuum computation do not sum up Sommerfeld enhanced terms, this procedure is
questionable. The analysis $10$ uses a combination of 
$M_6^V$ with the infinite momentum transfer moment, both in fixed-order, in order to constrain the 
continuum region. They only use experimental information on narrow resonances, and model the rest of 
the spectrum with theory predictions. Finally, they make the following scale choice: $\mu_\alpha = 
\mu_m = \overline{m}_b(\overline{m}_b)$ and estimate the truncation error from an ansatz for the 
$\mathcal{O}(\alpha_s^3)$ term.\footnote{Ref.~\cite{Erler:2002bu} also makes a determination of the 
charm mass. We exclude it from our comparison since it is not used in the PDG average.} Analyses 
$11$ and $13$ use large-$n$ moments and NRQCD methods for their theoretical moments. Analysis $12$ 
uses NRQCD fixed-order perturbation theory at N$^3$LO (which accounts for the summation of the 
Coulomb singularities) and $13$ uses renormalization group improved perturbation theory in the 
framework of vNRQCD at N$^2$LL order. Both analyses employ low-scale short-distance masses to avoid 
ambiguities related to the pole mass renormalon. Analysis $12$ also uses N$^3$LO NRQCD fixed-order 
input, but is incomplete concerning the contributions from the continuum region in the theoretical 
moments. Moreover, they extract the pole mass which is then converted to the ${\overline{\rm MS}}$ 
scheme.

Our result $14$ is in full agreement with the world average, having a slightly smaller uncertainty. 
It also agrees with the other analyses shown, with slightly smaller or comparable uncertainties.
We disagree with the small perturbative uncertainties related to scale variations quoted in $6$ to 
$10$.

\section{Conclusions}
\label{sec:conclusions}

In this work we have determined the $\overline {\rm MS}$ charm and bottom quark masses from 
quarkonium sum rules in the framework of the OPE, using $\mathcal{O}(\alpha_s^3)$ perturbative 
computations, plus nonperturbative effects from the gluon condensate including its Wilson 
coefficient at $\mathcal{O}(\alpha_s)$. For the determination of the perturbative uncertainties we 
independently varied the renormalization scales of the strong coupling and the quark masses, in 
order to account for the variations due to different possible types of $\alpha_s$ expansions, as 
suggested earlier in Ref.~\cite{Dehnadi:2011gc}.

In order to avoid a possible overestimate of the perturbative uncertainties, coming from the double 
scale variation in connection with a low scale of $\alpha_s$ and resulting in badly convergent 
series, we have re-examined the charm mass determination from charmonium sum rules (vector 
correlator) supplementing the analysis with a convergence test. The convergence test is based on 
Cauchy's radical test, which is adapted to the situation in which only a few terms of the series 
are known, and quantifies the convergence rate of each series by the parameter $V_c$.  We find that 
the distribution of the convergence parameter $V_c$ coming from the complete set of series peaks 
around its mean value, and allows to quantify the overall convergence rate of the set of series for 
each moment in a meaningful way. This justifies discarding (or ``trimming'') series with values of 
$V_c$ much larger than the average. For our analysis we discard $3\%$ of the series having the 
highest $V_c$ values, which results in a reduction of the perturbative uncertainty for the 
$\overline{\rm MS}$ charm mass $\overline m_c(\overline m_c)$ from $19$\,MeV 
in~\cite{Dehnadi:2011gc} to $14$\,MeV (which amounts to 26\%), and a small shift of $+\,5$\,MeV in 
the central value. Our new determination of the charm mass from the first moment (which is 
theoretically the cleanest) of the vector correlator reads:

\begin{equation}\label{eq:vector-quadrature}
\overline m_c(\overline m_c) = \,1.288 \, \pm 0.020\,\rm GeV\,, \qquad{\rm [Vector~Correlator]}
\end{equation}

where all sources of uncertainty have been added in quadrature. This result supersedes our 
corresponding earlier result from Ref.~\cite{Dehnadi:2011gc}, which was $1.282\,\pm\, 0.024$\,GeV. 
This makes it clear that the trimming procedure discards series which produce small values of the 
charm mass.

We have applied the same method of theory uncertainty estimate to analyze the HPQCD lattice 
simulation results for the pseudoscalar correlator. Our convergence test signals that the 
pseudoscalar moments have far worse convergence than the corresponding vector ones. This translates 
into an uncertainties of $35$\,MeV due to the truncation of the perturbative series and the error in 
$\alpha_s$ (roughly twice as big as for the vector determination). In contrast, using correlated 
scale variation (e.g.\ setting the scales in the mass and the strong coupling equal) the 
scale variation can be smaller by a factor of $8$. Our new determination from the first moment 
(again being the most reliable theoretical prediction) of the pseudoscalar correlator reads
$\overline m_c(\overline m_c) = 1.267 \pm 0.041$\,GeV, where again all individual errors have been 
added in quadrature. The combined total error is twice as big as for the vector correlator, and 
therefore we consider it as a validation of Eq.~(\ref{eq:vector-quadrature}) in connection with the 
convergence test. The result is in sharp contrast with the analyses carried out by the HPQCD 
collaboration~\cite{Allison:2008xk, McNeile:2010ji, Chakraborty:2014aca}, where perturbative 
uncertainties of $4$\,MeV are claimed. We have checked that, as for the vector correlator, for the 
different possible types of \mbox{$\alpha_s$-expansion} the correlated variation in general leads to 
a bad order-by-order convergence of the charm mass determination.

The second important result of this work is the determination of the bottom quark mass from the 
vector correlator. We have reanalyzed the experimental moments by combining experimental 
measurements of the first four narrow resonances, the threshold region covered by BABAR, and a 
theoretical model for the continuum. This theoretical model is an interpolation between a linear 
fit to the BaBar data points with highest energy and pQCD, to which we assign a $4\%$ systematic 
uncertainty which decreases linearly to reach $0.3\%$ at the $Z$-pole, and stays constant at $0.3\%$ 
for higher energies. Our treatment is motivated by the error function yielded by the fit to 
BaBar data in the energy range between $11.0$ and $11.2$\,GeV and the discrepancy between pQCD and 
experimental measurements at the \mbox{$Z$-pole}. This results into a large error for the first 
moment, and therefore we choose the second moment (which is theoretically as clean as the first one 
for the case of the bottom quark) for our final analysis, giving a total experimental uncertainty 
of $18$\,MeV. Our treatment of the experimental continuum uncertainty is in contrast to 
Ref.~\cite{Chetyrkin:2009fv}, where instead the very small perturbative QCD uncertainties (less 
than $1$\%) are used, claiming an experimental uncertainties of $6$\,MeV. In the light of the 
analysis carried out here, supported by the observations made in Ref.~\cite{Chetyrkin:2010ic}, we 
believe this is not justified. Our convergence test reveals that, as expected for the heavier 
bottom quark, the perturbative series converge faster than for the charm quark. Correspondingly, 
the perturbative and $\alpha_s$ uncertainties are $\sim 30\%$ smaller than those for charm. Taking 
correlated scale variation as used in Refs.~\cite{Chetyrkin:2009fv,Bodenstein:2012} the 
perturbative error estimate can shrink up to a factor of $20$. We also find that correlated 
variation leads to incompatible results for the different types of $\alpha_s$-expansions. Our final 
result for the bottom mass from the second moment, with all errors added in quadrature, reads:

\begin{equation}\label{eq:bottom-quadrature}
\overline m_b(\overline m_b) = \,4.176 \, \pm 0.023\,\rm GeV\,,
\qquad{\rm [Vector~Correlator]}
\end{equation}

where the total error is fairly dominated by the systematic error, which comes from the 
continuum region of the spectrum. Our uncertainty is very similar to the one obtained by the HPQCD 
analysis, but $30\%$ larger than the $16$\,MeV claimed by \cite{Chetyrkin:2009fv}. Our central 
value is $13$\,MeV larger than the latter. This good agreement is a result of two effects 
that push in opposite directions: smaller value of the second experimental moment, and different 
perturbative analysis. Curiously enough, a similar accidental cancellation was observed for the 
charm mass in \cite{Dehnadi:2011gc}.

In order to further validate the results discussed above, we have also analyzed the ratios of 
consecutive moments of each one of the three correlators as alternative observables. In all cases 
the results from the moment ratios agree very well the regular moment analyses.

\bigskip
\acknowledgments
This work was supported in part by the European Community's Marie-Curie Research Network under 
contract PITN-GA-2010-264564 (LHCphenOnet). VM has been partially supported by a Marie Curie
Fellowship under contract PIOF-GA-2009-251174. BD thanks the FWF Doktoratskollegs ``Particles and 
Interactions'' (DK W 1252-N27) for partial support. We thank the {\it Erwin Schr\"odinger 
International Institute for Mathematical Physics} (ESI Vienna), where a part of this work has been 
accomplished, for partial support. We thank Riccardo Faccini for information on the treatment of 
vacuum polarization effects by the BaBar collaboration, and Matthias Steinhauser and Christian 
Sturm for pointing out that they were missing in a previous version of the manuscript.

\appendix
\section{Numerical Values for the Perturbative Coefficients}
\label{app:coefs}

In this appendix we succinctly collect the numerical values for all of the coefficients appearing 
in the perturbative series, such that our analysis can be reproduced. We organize these values in 
tables, each of them corresponding to a different equation.

\renewcommand{\arraystretch}{1.2}\setlength{\LTcapwidth}{\textwidth}

\begin{table}[tbh!]
\center
\begin{tabular}{|c|cccc|}
\hline
 & $n=1$ & $n=2$ & $n=3$ & $n=4$\tabularnewline
\hline
$[a_V(n_f=5)]_{n}^{0,0}$ & $-\,4.011$ & $-\,6.6842$ & $-\,9.7224$ & $-\,13.0879$  \tabularnewline
$[a_V(n_f=5)]_{n}^{1,0}$ & $-\,36.9604$ & $-\,69.9123$ & $-\,112.669$ & $-\,165.326$ \tabularnewline
\hline
$[a_P(n_f=4)]_{n}^{0,0}$ & $8.02101$ & $0$ & $-\,9.72244$ & $-\,20.9406$ \tabularnewline
$[a_P(n_f=4)]_{n}^{1,0}$ & $39.1439$ & $-\,36.3842$ & $-\,152.67$ & $-\,309.925$ \tabularnewline
\hline
\end{tabular}
\caption{Numerical values for the coefficients of Eq.~(\ref{eq:GG}) for the vector correlator
with $n_f = 5$ (first two columns), and for the pseudoscalar correlator with $n_f = 4$ (last two 
columns). (Gluon condensate contribution).\label{tab:gluoncondensate}}
\end{table}

\begin{table}[tbh!]
\center\scriptsize
\begin{tabular}{|c|cccc|cccc|}
\hline 
 & $[C_V]_{0,i}^{0,0}$ & $[C_V]_{0,i}^{1,0}$ & $[C_V]_{0,i}^{0,1}$ & $[C_V]_{0,i}^{0,2}$ & 
$[C_P]_{0,i}^{0,0}$ & $[C_P]_{0,i}^{1,0}$ & $[C_P]_{0,i}^{0,1}$ & $[C_P]_{0,i}^{0,2}$ 
\tabularnewline
\hline
 & \multicolumn{4}{|c|}{$n_f=5$}& \multicolumn{4}{|c|}{$n_f=4$}\tabularnewline
\hline
$i=0$ & $0$ & $0$ & $0$ & $0$ & $1.33333$ & $0$ & $0$ & $0$\tabularnewline
$i=1$ & $1.44444$ & $0$ & $0$ & $0$ & $3.11111$ & $0$ & $0$ & $0$\tabularnewline
$i=2$ & $3.21052$ & $0$ & $-\,2.76852$ & $0$ & $0.115353$ & $0$ & $-\,6.48148$ & $0$\tabularnewline
$i=3$ & $-\,6.28764$ & $5.53704$ & $-\,15.7977$ & $5.30633$ & $-\,1.22241$ & $12.963$ & 
$-\,10.4621$ 
& $13.5031$\tabularnewline
\hline
\end{tabular}
\caption{Numerical values of the coefficients for Eq.~(\ref{eq:Pi0msbar}) for $\widehat\Pi_V(0)$ in 
the $\overline{\mbox{MS}}$ scheme and $n_f = 5$, and $P(q^2=0)$ for $n_f = 4$.\label{tab:cPi0vec}}
\end{table}

\begin{table}[tbh!]
\center
\scriptsize
\begin{tabular}{|c|ccccccccc|}
\hline 
 & $[C_V]_{n,i}^{0,0}$ & $[C_V]_{n,i}^{1,0}$ & $[C_V]_{n,i}^{2,0}$ & $[C_V]_{n,i}^{3,0}$ & 
$[C_V]_{n,i}^{0,1}$ & $[C_V]_{n,i}^{1,1}$
 & $[C_V]_{n,i}^{2,1}$ & $[C_V]_{n,i}^{0,2}$ & $[C_V]_{n,i}^{1,2}$\tabularnewline\hline
\multicolumn{10}{|c|}{$n=1$} \\\hline
$i=0$ & $1.06667$ & $0$ & $0$ & $0$ & $0$ & $0$ & $0$ & $0$ & $0$\tabularnewline
$i=1$ & $2.55473$ & $2.13333$ & $0$ & $0$ & $0$ & $0$ & $0$ & $0$ & $0$\tabularnewline
$i=2$ & $3.15899$ & $8.33909$ & $4.17778$ & $0$ & $-\,4.89657$ & $-\,4.08889$ & $0$ & $0$ & 
$0$\tabularnewline
$i=3$ & $-\,7.76244$ & $18.2235$ & $29.3221$ & $8.12346$ & $-\,18.2834$ & $-\,37.1221$ & $-16.0148$ 
& $9.38509$ & $7.83704$\tabularnewline
\hline
\multicolumn{10}{|c|}{$n=2$} \\\hline
$i=0$ & $0.457143$ & $0$ & $0$ & $0$ & $0$ & $0$ & $0$ & $0$ & $0$\tabularnewline
$i=1$ & $1.10956$ & $1.82857$ & $0$ & $0$ & $0$ & $0$ & $0$ & $0$ & $0$\tabularnewline
$i=2$ & $3.23193$ & $7.20649$ & $5.40952$ & $0$ & $-\,2.12665$ & $-\,3.50476$ & $0$ & $0$ & 
$0$\tabularnewline
$i=3$ & $-\,2.64381$ & $19.0805$ & $35.2229$ & $14.1249$ & $-\,15.0705$ & $-\,32.0439$ & 
$-\,20.7365$ & $4.07609$ & $6.71746$\tabularnewline
\hline
\multicolumn{10}{|c|}{$n=3$} \\\hline
$i=0$ & $0.270899$ & $0$ & $0$ & $0$ & $0$ & $0$ & $0$ & $0$ & $0$\tabularnewline
$i=1$ & $0.519396$ & $1.6254$ & $0$ & $0$ & $0$ & $0$ & $0$ & $0$ & $0$\tabularnewline
$i=2$ & $2.06768$ & $5.57705$ & $6.43386$ & $0$ & $-\,0.995509$ & $-\,3.11534$ & $0$ & $0$ & 
$0$\tabularnewline
$i=3$ & $-\,1.17449$ & $14.8309$ & $36.8953$ & $21.0888$ & $-\,9.18132$ & $-\,25.3067$ & 
$-\,24.6631$ & $1.90806$ & $5.97108$\tabularnewline
\hline
\multicolumn{10}{|c|}{$n=4$} \\\hline
$i=0$ & $0.184704$ & $0$ & $0$ & $0$ & $0$ & $0$ & $0$ & $0$ & $0$\tabularnewline
$i=1$ & $0.203121$ & $1.47763$ & $0$ & $0$ & $0$ & $0$ & $0$ & $0$ & $0$\tabularnewline
$i=2$ & $1.22039$ & $3.86194$ & $7.3266$ & $0$ & $-\,0.389316,$ & $-\,2.83213$ & $0$ & $0$ & 
$0$\tabularnewline
$i=3$ & $-\,1.386$ & $9.10716$ & $34.8581$ & $28.8994$ & $-\,5.16902$ & $-\,18.3751$ & $-\,28.0853$ 
& $0.746189$ & $5.42825$\tabularnewline
\hline
\end{tabular}
\caption{Numerical values of the coefficients for Eq.~(\ref{eq:Mn-theo-FO}) for the vector current 
with $n_f = 5$.
(Standard fixed-order expansion).\label{tab:cfixedorder}}
\end{table}

\begin{table}[tbh!]
\center
\scriptsize
\begin{tabular}{|c|ccccccccc|}
\hline 
 & $[\bar{C}_V]_{n,i}^{0,0}$ & $[\bar{C}_V]_{n,i}^{1,0}$ & $[\bar{C}_V]_{n,i}^{2,0}$ & 
$[\bar{C}_V]_{n,i}^{3,0}$ &
 $[\bar{C}_V]_{n,i}^{0,1}$ & $[\bar{C}_V]_{n,i}^{1,1}$ & $[\bar{C}_V]_{n,i}^{2,1}$ & 
$[\bar{C}_V]_{n,i}^{0,2}$ & $[\bar{C}_V]_{n,i}^{1,2}$\tabularnewline\hline
\multicolumn{10}{|c|}{$n=1$} \\
\hline
$i=0$ & $1.0328$ & $0$ & $0$ & $0$ & $0$ & $0$ & $0$ & $0$ & $0$\tabularnewline
$i=1$ & $1.2368$ & $1.0328$ & $0$ & $0$ & $0$ & $0$ & $0$ & $0$ & $0$\tabularnewline
$i=2$ & $0.788784$ & $2.80034$ & $1.50616$  & $0$ & $-\,2.37054$ & $-\,1.97952$ & $0$ & $0$& 
$0$\tabularnewline
$i=3$ & $-\,4.70257$ & $4.68012$ & $9.59148$ & $2.42659$ & $-\,6.01262$ & $-\,13.2306$ & 
$-\,5.77361$ & $4.54354$ & $3.79409$\tabularnewline\hline
\multicolumn{10}{|c|}{$n=2$} \\\hline
$i=0$ & $0.822267$ & $0$ & $0$ & $0$ & $0$ & $0$ & $0$ & $0$ & $0$\tabularnewline
$i=1$ & $0.498944$ & $0.822267$ & $0$ & $0$ & $0$ & $0$ & $0$ & $0$ & $0$\tabularnewline
$i=2$ & $0.999196$ & $1.74376$ & $1.19914$ & $0$ & $-\,0.956309$ & $-\,1.57601$ & $0$ & $0$ & 
$0$\tabularnewline
$i=3$ & $-3.19148$ & $1.49991$ & $6.92794$ & $1.93195$ & $-\,5.03603$ & $-\,8.67158$ & $-\,4.5967$ 
& 
$1.83292$ & $3.02069$\tabularnewline\hline
\multicolumn{10}{|c|}{$n=3$} \\\hline
$i=0$ & $0.804393$ & $0$ & $0$ & $0$ & $0$ & $0$ & $0$ & $0$ & $0$ \tabularnewline
$i=1$ & $0.257044$ & $0.804393$ & $0$ & $0$ & $0$ & $0$ & $0$ & $0$ & $0$\tabularnewline
$i=2$ & $0.817928$ & $1.4748$ & $1.17307$ & $0$ & $-\,0.492667$ & $-\,1.54175$ & $0$ & $0$ & 
$0$\tabularnewline
$i=3$ & $-1.97558$ & $0.0722677$ & $6.44038$ & $1.88995$ & $-\,3.75658$ & $-\,7.59737$ & 
$-\,4.49678$ & $0.944279$ & $2.95503$\tabularnewline\hline
\multicolumn{10}{|c|}{$n=4$} \\\hline
$i=0$ & $0.809673$ & $0$ & $0$ & $0$ & $0$ & $0$ & $0$ & $0$ & $0$ \tabularnewline
$i=1$ & $ 0.111301$ & $0.809673$ & $0$ & $0$ & $0$ & $0$ & $0$ & $0$ & $0$\tabularnewline
$i=2$ & $0.615164$ & $1.33706$ & $1.18077$ & $0$ & $-\,0.213327$ & $-\,1.55187$ & $0$ & $0$& 
$0$\tabularnewline
$i=3$ & $-1.36612$ & $-\,0.923734$ & $6.26766$ & $1.90236$ & $-\,2.62711$ & $-\,7.08209$ & 
$-\,4.5263$ & $0.408876$ & $2.97442$\tabularnewline
\hline
\end{tabular}
\caption{Numerical values of the coefficients for Eq.~(\ref{eq:Mn-theo-exp}) for the vector current 
with $n_f = 5$.
  (Linearized expansion). \label{tab:ctildefixedorder}}
\end{table}

\begin{table}[tbh!]
\scriptsize
\begin{tabular}{|c|ccccccccc|}
\hline 
 & $[\tilde{C}_V]_{n,i}^{0,0}$ & $[\tilde{C}_V]_{n,i}^{1,0}$ & $[\tilde{C}_V]_{n,i}^{2,0}$ & 
$[\tilde{C}_V]_{n,i}^{3,0}$ &
 $[\tilde{C}_V]_{n,i}^{0,1}$ & $[\tilde{C}_V]_{n,i}^{1,1}$ & $[\tilde{C}_V]_{n,i}^{2,1}$ & 
$[\tilde{C}_V]_{n,i}^{0,2}$ & $[\tilde{C}_V]_{n,i}^{1,2}$\tabularnewline\hline
\multicolumn{10}{|c|}{$n=1$} \\
\hline
$i=0$ & $1$ & $0$ & $0$ & $0$ & $0$ & $0$ & $0$ & $0$ & $0$\tabularnewline
$i=1$ & $1.19753$ & $1$ & $0$ & $0$ & $0$ & $0$ & $0$ & $0$ & $0$\tabularnewline
$i=2$ & $3.1588$ & $4.71142$ & $1.45833$  & $0$ & $-\,2.29527$ & $-1.91667$ & $0$ & $0$ & 
$0$\tabularnewline
$i=3$ & $1.32698$ & $14.7867$ & $13.2036$ & $2.34954$ & $-\,15.0028$ & $-\,20.4771$ & $-\,5.59028$ 
& 
$4.39926$ & $3.67361$\tabularnewline\hline
\multicolumn{10}{|c|}{$n=2$} \\\hline
$i=0$ & $1$ & $0$ & $0$ & $0$ & $0$ & $0$ & $0$ & $0$ & $0$\tabularnewline
$i=1$ & $0.60679$ & $1$ & $0$ & $0$ & $0$ & $0$ & $0$ & $0$ & $0$\tabularnewline
$i=2$ & $2.42875$ & $4.12068$ & $1.45833$ & $0$ & $-\,1.16301$ & $-1.91667$ & $0$ & $0$ & 
$0$\tabularnewline
$i=3$ & $1.7702$ & $11.9808$ & $12.3421$ & $2.34954$ & $-\,10.7766$ & $-\,18.2126$ & $-\,5.59028$ & 
$2.22911$ & $3.67361$\tabularnewline\hline
\multicolumn{10}{|c|}{$n=3$} \\\hline
$i=0$ & $1$ & $0$ & $0$ & $0$ & $0$ & $0$ & $0$ & $0$ & $0$ \tabularnewline
$i=1$ & $0.31955$ & $1$ & $0$ & $0$ & $0$ & $0$ & $0$ & $0$ & $0$\tabularnewline
$i=2$ & $1.65593$ & $3.83344$ & $1.45833$ & $0$ & $-\,0.612471$ & $-\,1.91667$ & $0$ & $0$ & 
$0$\tabularnewline
$i=3$ & $1.53408$ & $10.1987$ & $11.9232$ & $2.34954$ & $-\,7.11997$ & $-\,17.1115$ & $-\,5.59028$ 
& 
$1.1739$ & $3.67361$\tabularnewline\hline
\multicolumn{10}{|c|}{$n=4$} \\\hline
$i=0$ & $1$ & $0$ & $0$ & $0$ & $0$ & $0$ & $0$ & $0$ & $0$ \tabularnewline
$i=1$ & $ 0.137464$ & $1$ & $0$ & $0$ & $0$ & $0$ & $0$ & $0$ & $0$\tabularnewline
$i=2$ & $1.0347$ & $3.65135$ & $1.45833$ & $0$ & $-\,0.263473$ & $-\,1.91667$ & $0$ & $0$ & 
$0$\tabularnewline
$i=3$ & $0.744809$ & $8.93759$ & $11.6576$ & $2.34954$ & $-\,4.29854$ & $-\,16.4135$ & $-\,5.59028$ 
& $0.504989$ & $3.67361$\tabularnewline
\hline
\end{tabular}
\caption{Numerical values of the coefficients for Eq.~(\ref{eq:iterative-general}) for the vector 
current with $n_f = 5$.
(Iterative linearized expansion).\label{tab:vector-it}}
\end{table}

\begin{table}[tbh!]
\center
\scriptsize
\begin{tabular}{|c|ccccccccc|}
\hline 
 & $[C_P]_{n,i}^{0,0}$ & $[C_P]_{n,i}^{1,0}$ & $[C_P]_{n,i}^{2,0}$ & $[C_P]_{n,i}^{3,0}$ & 
$[C_P]_{n,i}^{0,1}$ & $[C_P]_{n,i}^{1,1}$
 & $[C_P]_{n,i}^{2,1}$ & $[C_P]_{n,i}^{0,2}$ & $[C_P]_{n,i}^{1,2}$\tabularnewline\hline
\multicolumn{10}{|c|}{$n=1$} \\\hline
$i=0$ & $0.5333$ & $0$ & $0$ & $0$ & $0$ & $0$ & $0$ & $0$ & $0$\tabularnewline
$i=1$ & $2.0642$ & $1.06667$ & $0$ & $0$ & $0$ & $0$ & $0$ & $0$ & $0$\tabularnewline
$i=2$ & $7.23618$ & $5.89136$ & $2.17778$ & $0$ & $-\,4.30041$ & $-\,2.22222$ & $0$ & $0$ & 
$0$\tabularnewline
$i=3$ & $7.06593$ & $29.1882$ & $19.5609$ & $4.47654$ & $-\,36.7734$ & $-\,27.9695$ & $-\,9.07407$ 
& 
$8.95919$ & $4.62963$\tabularnewline
\hline
\multicolumn{10}{|c|}{$n=2$} \\\hline
$i=0$ & $0.30476$ & $0$ & $0$ & $0$ & $0$ & $0$ & $0$ & $0$ & $0$\tabularnewline
$i=1$ & $1.21171$ & $1.21905$ & $0$ & $0$ & $0$ & $0$ & $0$ & $0$ & $0$\tabularnewline
$i=2$ & $5.9992$  & $6.86166$ & $3.70794$ & $0$ & $-\,2.5244$ & $-\,2.53968$ & $0$ & $0$ & 
$0$\tabularnewline
$i=3$ & $14.5789$ & $36.2468$ & $31.4945$ & $10.0938$ & $-\,28.8842$ & $-\,32.5014$ & $-15.4497$ & 
$5.25916$ & $5.29101$\tabularnewline
\hline
\multicolumn{10}{|c|}{$n=3$} \\\hline
$i=0$ & $0.20318$ & $0$ & $0$ & $0$ & $0$ & $0$ & $0$ & $0$ & $0$\tabularnewline
$i=1$ & $0.71276$ & $1.21905$ & $0$ & $0$ & $0$ & $0$ & $0$ & $0$ & $0$\tabularnewline
$i=2$ & $4.26701$ & $6.29135$ & $4.92698$ & $0$& $-\,1.48491$ & $-\,2.53968$ & $0$ & $0$ & 
$0$\tabularnewline
$i=3$ & $13.3278$ & $34.8305$ & $38.066$ & $16.697$ & $-\,20.066$ & $-\,30.1251$ & $-\,20.5291$ & 
$3.09356$ & $5.29101$\tabularnewline
\hline
\multicolumn{10}{|c|}{$n=4$} \\\hline
$i=0$ & $0.147763$ & $0$ & $0$ & $0$ & $0$ & $0$ & $0$ & $0$ & $0$\tabularnewline
$i=1$ & $0.401317$ & $1.18211$ & $0$ & $0$ & $0$ & $0$ & $0$ & $0$ & $0$\tabularnewline
$i=2$ & $2.91493$  & $5.1643$ & $5.95979$  & $0$ & $-\,0.836077$ & $-\,2.46272$ & $0$ & $0$ & 
$0$\tabularnewline
$i=3$ & $9.9948$ & $29.5129$ & $40.2459$ & $24.1703$ & $-13.4331$ & $-25.3105$ & $-24.8325$ & 
$1.74183$ & $5.13067$\tabularnewline
\hline

\end{tabular}
\caption{Numerical values of the coefficients for Eq.~(\ref{eq:Mn-theo-FO}) for the pseudoscalar 
correlator with $n_f = 4$. (Standard fixed-order expansion).\label{tab:cPfixedorder}}
\end{table}

\begin{table}[tbh!]
\center
\scriptsize
\begin{tabular}{|c|ccccccccc|}
\hline 
 & $[\bar{C}_P]_{n,i}^{0,0}$ & $[\bar{C}_P]_{n,i}^{1,0}$ & $[\bar{C}_P]_{n,i}^{2,0}$ & 
$[\bar{C}_P]_{n,i}^{3,0}$ &
 $[\bar{C}_P]_{n,i}^{0,1}$ & $[\bar{C}_P]_{n,i}^{1,1}$ & $[\bar{C}_P]_{n,i}^{2,1}$ & 
$[\bar{C}_P]_{n,i}^{0,2}$ & $[\bar{C}_P]_{n,i}^{1,2}$\tabularnewline\hline
\multicolumn{10}{|c|}{$n=1$} \\
\hline
$i=0$ & $0.730297$ & $0$ & $0$ & $0$ & $0$ & $0$ & $0$ & $0$ & $0$\tabularnewline
$i=1$ & $1.41326$ & $0.730297$ & $0$ & $0$ & $0$ & $0$ & $0$ & $0$ & $0$\tabularnewline
$i=2$ & $3.58681$ & $2.62028$ & $1.12587$ & $0$  & $-\,2.94429$ & $-\,1.52145$ & $0$ & $0$& 
$0$\tabularnewline
$i=3$ & $-\,2.10344$ & $11.3262$ & $8.59338$ & $1.93901$ & $-\,19.4793$ & $-\,13.2609$ & 
$-\,4.69114$ & $6.13394$ & $3.16969$\tabularnewline\hline
\multicolumn{10}{|c|}{$n=2$} \\\hline
$i=0$ & $0.743002$ & $0$ & $0$ & $0$ & $0$ & $0$ & $0$ & $0$ & $0$\tabularnewline
$i=1$ & $0.738531$ & $0.743002$ & $0$ & $0$ & $0$ & $0$ & $0$ & $0$ & $0$\tabularnewline
$i=2$ & $2.55535$ & $1.96655$ & $1.14546$ & $0$ & $-\,1.53861$ & $-\,1.54792$ & $0$ & $0$ & 
$0$\tabularnewline
$i=3$ & $0.536147$ & $6.35978$ & $7.66478$ & $1.97274$ & $-\,13.0167$ & $-\,10.5778$ & $-\,4.77276$ 
& $3.20543$ & $3.22484$\tabularnewline\hline
\multicolumn{10}{|c|}{$n=3$} \\\hline
$i=0$ & $0.766734$ & $0$ & $0$ & $0$ & $0$ & $0$ & $0$ & $0$ & $0$ \tabularnewline
$i=1$ & $0.448296$ & $0.766734$ & $0$ & $0$ & $0$ & $0$ & $0$ & $0$ & $0$\tabularnewline
$i=2$ & $2.02851$ & $1.71554$ & $1.18205$ & $0$ & $-\,0.93395$ & $-\,1.59736$ & $0$ & $0$ & 
$0$\tabularnewline
$i=3$ & $1.94168$ & $4.12818$ & $7.42578$ & $2.03575$ & $-9.89039$ & $-\,9.60801$ & $-\,4.9252$ & 
$1.94573$ & $3.32784$\tabularnewline\hline
\multicolumn{10}{|c|}{$n=4$} \\\hline
$i=0$ & $0.787401$ & $0$ & $0$ & $0$ & $0$ & $0$ & $0$ & $0$ & $0$ \tabularnewline
$i=1$ & $0.267317$ & $0.787401$ & $0$ & $0$ & $0$ & $0$ & $0$ & $0$ & $0$\tabularnewline
$i=2$ & $1.624$ & $1.56872$ & $1.21391$ &  $0$ &$-\,0.55691$ & $-\,1.64042$ & $0$ & $0$& 
$0$\tabularnewline
$i=3$ & $2.58251$ & $2.65675$ & $7.3283$ & $2.09062$ & $-\,7.62431$ & $-\,9.06256$ & $-\,5.05796$ & 
$1.16023$ & $3.41754$\tabularnewline\hline

\end{tabular}
\caption{Numerical values of the coefficients for Eq.~(\ref{eq:Mn-theo-exp}) for the pseudoscalar 
current with $n_f = 4$. (Linearized expansion). \label{tab:cPtildefixedorder}}
\end{table}

\begin{table}[tbh!]
\scriptsize
\begin{tabular}{|c|ccccccccc|}
\hline 
 & $[\tilde{C}_P]_{n,i}^{0,0}$ & $[\tilde{C}_P]_{n,i}^{1,0}$ & $[\tilde{C}_P]_{n,i}^{2,0}$ & 
$[\tilde{C}_P]_{n,i}^{3,0}$ &
 $[\tilde{C}_P]_{n,i}^{0,1}$ & $[\tilde{C}_P]_{n,i}^{1,1}$ & $[\tilde{C}_P]_{n,i}^{2,1}$ & 
$[\tilde{C}_P]_{n,i}^{0,2}$ & $[\tilde{C}_P]_{n,i}^{1,2}$\tabularnewline\hline
\multicolumn{10}{|c|}{$n=1$} \\
\hline
$i=0$ & $1$ & $0$ & $0$ & $0$ & $0$ & $0$ & $0$ & $0$ & $0$\tabularnewline
$i=1$ & $1.93519$ & $1$ & $0$ & $0$ & $0$ & $0$ & $0$ & $0$ & $0$\tabularnewline
$i=2$ & $8.78182$ & $5.58796$ & $1.54167$ & $0$ & $-\,4.03164$ & $-\,2.08333$ & $0$ & $0$ & 
$0$\tabularnewline
$i=3$ & $9.22126$ & $25.7977$ & $15.8503$ & $2.65509$ & $-\,42.7996$ & $-\,26.4915$ & $-\,6.42361$ 
& 
$8.39924$ & $4.34028$\tabularnewline\hline
\multicolumn{10}{|c|}{$n=2$} \\\hline
$i=0$ & $1$ & $0$ & $0$ & $0$ & $0$ & $0$ & $0$ & $0$ & $0$\tabularnewline
$i=1$ & $0.99398$ & $1$ & $0$ & $0$ & $0$ & $0$ & $0$ & $0$ & $0$\tabularnewline
$i=2$ & $5.42718$ & $4.64676$ & $1.54167$ & $0$ & $-2.07079$ & $-2.08333$ & $0$ & $0$ & 
$0$\tabularnewline
$i=3$ & $11.733$  & $19.005$  & $14.3993$ & $2.65509$ & $-25.8023$ & $-22.5698$ & $-6.42361$ & 
$4.31416$ & $4.34028$ \tabularnewline\hline
\multicolumn{10}{|c|}{$n=3$} \\\hline
$i=0$ & $1$ & $0$ & $0$ & $0$ & $0$ & $0$ & $0$ & $0$ & $0$ \tabularnewline
$i=1$ & $0.584683$ & $1$ & $0$ & $0$ & $0$ & $0$ & $0$ & $0$ & $0$\tabularnewline
$i=2$ & $3.81501$ & $4.23746$ & $1.54167$  & $0$ & $-\,1.21809$ & $-\,2.08333$ & $0$ & $0$ & 
$0$\tabularnewline
$i=3$ & $11.0126$ & $15.8978$ & $13.7683$ & $2.65509$ & $-\,17.7717$ & $-\,20.8644$ & $-\,6.42361$ 
& 
$2.53768$ & $4.34028$\tabularnewline\hline
\multicolumn{10}{|c|}{$n=4$} \\\hline
$i=0$ & $1$ & $0$ & $0$ & $0$ & $0$ & $0$ & $0$ & $0$ & $0$ \tabularnewline
$i=1$ & $0.339493$ & $1$ & $0$ & $0$ & $0$ & $0$ & $0$ & $0$ & $0$\tabularnewline
$i=2$ & $2.74147$ & $3.99227$ & $1.54167$  & $0$ & $-\,0.707277$ & $-\,2.08333$ & $0$ & $0$ & 
$0$\tabularnewline
$i=3$ & $9.51996$ & $13.9286$ & $13.3903$ & $2.65509$ & $-\,12.512$ & $-\,19.8428$ & $-\,6.42361$ & 
$1.47349$ & $4.34028$\tabularnewline\hline

\end{tabular}
\caption{Numerical values of the coefficients for Eq.~(\ref{eq:iterative-general}) for the 
pseudoscalar current with $n_f = 4$. (Iterative linearized expansion).\label{tab:cPhat}}
\end{table}

\begin{table}[tbh!]
\center
\scriptsize
\begin{tabular}{|c|ccccccccc|}
\hline 
 & $[R_V]_{n,i}^{0,0}$ & $[R_V]_{n,i}^{1,0}$ & $[R_V]_{n,i}^{2,0}$ & $[R_V]_{n,i}^{3,0}$ & 
$[R_V]_{n,i}^{0,1}$ & $[R_V]_{n,i}^{1,1}$
 & $[R_V]_{n,i}^{2,1}$ & $[R_V]_{n,i}^{0,2}$ & $[R_V]_{n,i}^{1,2}$\tabularnewline\hline
\multicolumn{10}{|c|}{$n=1$} \\\hline
$i=0$ & $0.428571$ & $0$ & $0$ & $0$ & $0$ & $0$ & $0$ & $0$ & $0$\tabularnewline
$i=1$ & $0.0137566$ & $0.857143$ & $0$ & $0$ & $0$ & $0$ & $0$ & $0$ & $0$\tabularnewline\hline
\multirow{2}{*}{$i=2$} & $1.56736$ & $1.44418$ & $1.75$ & $0$  & $-\,0.0286596$ & $-\,1.78571$ & 
$0$ 
& $0$ & $0$\tabularnewline
& $1.72775$ & $1.32513$ & $1.67857$ & $0$& $-\,0.0263668$ & $-\,1.64286$ & $0$ & $0$ & 
$0$\tabularnewline\hline
\multirow{2}{*}{$i=3$} & $-\,4.79526$ & $2.66831$ & $9.00161$ & $3.59722$ & $-\,6.57481$ & 
$-\,8.76742$ & $-\,7.29167$ & $0.0597075$ & $3.72024$\tabularnewline
& $-\,3.53855$ & $1.29072$ & $7.81479$ & $3.26389$ & $-\,6.65629$ & $-\,7.1511$ & $-\,6.43452$ & 
$0.0505364$ & $3.14881$\tabularnewline
\hline
\multicolumn{10}{|c|}{$n=2$} \\\hline
$i=0$ & $0.592593$ & $0$ & $0$ & $0$ & $0$ & $0$ & $0$ & $0$ & $0$\tabularnewline
$i=1$ & $-\,0.302139$ & $1.18519$ & $0$ & $0$ & $0$ & $0$ & $0$ & $0$ & $0$\tabularnewline\hline
\multirow{2}{*}{$i=2$} & $0.718416$ & $1.35457$ & $2.41975$ & $0$ & $0.629455$ & $-\,2.46914$ & $0$ 
& $0$ & $0$\tabularnewline
& $1.06685$ & $1.18996$ & $2.32099$ & $0$ & $0.579099$ & $-\,2.2716$ & $0$ & $0$ & 
$0$\tabularnewline\hline
\multirow{2}{*}{$i=3$} & $-\,1.59081$ & $-\,1.60786$ & $11.1353$ & $4.97394$ & $-\,2.02404$ & 
$-\,9.44651$ &
$-\,10.0823$ & $-\,1.31137$ & $5.14403$\tabularnewline
& $0.404636$ & $-\,3.06309$ & $9.54776$ & $4.51303$ & $-\,3.35942$ & $-\,7.42572$ & $-\,8.89712$ & 
$-\,1.10994$ & $4.35391$\tabularnewline
\hline
\multicolumn{10}{|c|}{$n=3$} \\\hline
$i=0$ & $0.681818$ & $0$ & $0$ & $0$ & $0$ & $0$ & $0$ & $0$ & $0$\tabularnewline
$i=1$ & $-\,0.557447$ & $1.36364$ & $0$ & $0$ & $0$ & $0$ & $0$ & $0$ & $0$\tabularnewline\hline
\multirow{2}{*}{$i=2$} & $-\,0.119176$ & $1.13889$ & $2.78409$ & $0$ & $1.16135$ & $-\,2.84091$ & 
$0$ & $0$ & $0$\tabularnewline
& $0.369655$ & $0.949499$ & $2.67045$ & $0$ & $1.06844$ & $-\,2.61364$ & $0$ & $0$ & 
$0$\tabularnewline\hline
\multirow{2}{*}{$i=3$} & $-\,1.61506$ & $-\,5.30928$ & $11.9551$ & $5.72285$ & $2.28504$ & 
$-\,9.12039$ & $-\,11.6004$ & $-\,2.41948$ & $5.91856$\tabularnewline
& $1.3858$ & $-\,6.67953$ & $10.1636$ & $5.19255$ & $-\,0.0698462$ & $-6.9352$ & $-\,10.2367$ & 
$-\,2.04785$ & $5.00947$\tabularnewline
\hline
\end{tabular}
\caption{Numerical values for the coefficients of the standard fixed-order expansion of the ratios 
of vector moments. We display results for the vector current with $n_f = 4,(5)$ for the upper 
(lower) number.\label{tab:Rfixedorder}}
\end{table}

\begin{table}[tbh!]
\center
\scriptsize
\begin{tabular}{|c|ccccccccc|}
\hline 
 & $[R_P]_{n,i}^{0,0}$ & $[R_P]_{n,i}^{1,0}$ & $[R_P]_{n,i}^{2,0}$ & $[R_P]_{n,i}^{3,0}$ & 
$[R_P]_{n,i}^{0,1}$ & $[R_P]_{n,i}^{1,1}$
 & $[R_P]_{n,i}^{2,1}$ & $[R_P]_{n,i}^{0,2}$ & $[R_P]_{n,i}^{1,2}$\tabularnewline\hline
\multicolumn{10}{|c|}{$n=1$} \\\hline
$i=0$ & $0.571429$ & $0$ & $0$ & $0$ & $0$ & $0$ & $0$ & $0$ & $0$\tabularnewline
$i=1$ & $0.0603175$ & $1.14286$ & $0$ & $0$ & $0$ & $0$ & $0$ & $0$ & $0$\tabularnewline
$i=2$ & $3.262$ & $2.00952$ & $2.33333$ & $0$  & $-\,0.125661$ & $-\,2.38095$ & $0$ & $0$ & 
$0$\tabularnewline
$i=3$ & $6.32118$ & $6.21577$ & $12.1735$ & $4.7963$ & $-\,13.7852$ & $-\,12.0397$ & $-\,9.72222$ & 
$0.261795$ & $4.96032$\tabularnewline
\hline
\multicolumn{10}{|c|}{$n=2$} \\\hline
$i=0$ & $0.666667$ & $0$ & $0$ & $0$ & $0$ & $0$ & $0$ & $0$ & $0$\tabularnewline
$i=1$ & $-\,0.311887$ & $1.33333$ & $0$ & $0$ & $0$ & $0$ & $0$ & $0$ & $0$\tabularnewline
$i=2$ & $2.1179$ & $1.57993$ & $2.72222$ & $0$ & $0.649765$ & $-\,2.77778$ & $0$ & $0$ & 
$0$\tabularnewline
$i=3$ & $9.55945$ & $1.01989$ & $12.6416$ & $5.59568$ & $-\,7.82395$ & $-\,10.8608$ & $-\,11.3426$ 
& 
$-\,1.35368$ & $5.78704$\tabularnewline
\hline
\multicolumn{10}{|c|}{$n=3$} \\\hline
$i=0$ & $0.727273$ & $0$ & $0$ & $0$ & $0$ & $0$ & $0$ & $0$ & $0$\tabularnewline
$i=1$ & $-\,0.57611$ & $1.45455$ & $0$ & $0$ & $0$ & $0$ & $0$ & $0$ & $0$\tabularnewline
$i=2$ & $1.09403$ & $1.25182$ & $2.9697$ & $0$ & $1.20023$ & $-\,3.0303$ & $0$ & $0$ & 
$0$\tabularnewline
$i=3$ & $9.74702$ & $-\,3.08269$ & $12.8277$ & $6.10438$ & $-\,2.71009$ & $-\,9.88258$ & 
$-\,12.3737$ & $-\,2.50048$ & $6.31313$\tabularnewline
\hline

\end{tabular}
\caption{Numerical values for the coefficients of the linearized expansion of the ratios 
of pseudoscalar moments with $n_f = 4$.\label{tab:RPFOorder}}
\end{table}

\begin{table}[tbh!]
\center
\scriptsize
\begin{tabular}{|c|ccccccccc|}
\hline 
 & $[\bar{R}_V]_{n,i}^{0,0}$ & $[\bar R_V]_{n,i}^{1,0}$ & $[\bar R_V]_{n,i}^{2,0}$ & $[\bar 
R_V]_{n,i}^{3,0}$ & $[\bar R_V]_{n,i}^{0,1}$ & $[\bar R_V]_{n,i}^{1,1}$
 & $[\bar R_V]_{n,i}^{2,1}$ & $[\bar R_V]_{n,i}^{0,2}$ & $[\bar 
R_V]_{n,i}^{1,2}$\tabularnewline\hline
\multicolumn{10}{|c|}{$n=1$} \\\hline
$i=0$ & $0.654654$ & $0$ & $0$ & $0$ & $0$ & $0$ & $0$ & $0$ & $0$\tabularnewline
$i=1$ & $0.0105068$ & $0.654654$ & $0$ & $0$ & $0$ & $0$ & $0$ & $0$ & $0$\tabularnewline\hline
\multirow{2}{*}{$i=2$} & $1.19701$ & $1.0925$ & $1.00926$ & $0$ & $-\,0.0218891$ & $-\,1.36386$ & 
$0$ & $0$ & $0$\tabularnewline
& $1.31951$ & $1.00158$ & $0.954703$ & $0$ & $-\,0.020138$ & $-\,1.25475$ & $0$ & $0$ & 
$0$\tabularnewline\hline
\multirow{2}{*}{$i=3$} & $-\,3.68165$ & $0.823415$ & $5.76639$ & $1.73817$ & $-\,5.02124$ & 
$-\,6.65245$ & $-\,4.20524$ & $0.0456024$ & $2.84138$\tabularnewline
& $-\,2.72379$ & $-\,0.349776$ & $4.95174$ & $1.53813$ & $-\,5.0835$ & $-\,5.42147$ & $-\,3.6597$ & 
$0.0385979$ & $2.40494$\tabularnewline
\hline
\multicolumn{10}{|c|}{$n=2$} \\\hline
$i=0$ & $0.7698$ & $0$ & $0$ & $0$ & $0$ & $0$ & $0$ & $0$ & $0$\tabularnewline
$i=1$ & $-\,0.196245$ & $0.7698$ & $0$ & $0$ & $0$ & $0$ & $0$ & $0$ & $0$\tabularnewline\hline
\multirow{2}{*}{$i=2$} & $0.441611$ & $1.07606$ & $1.18678$ & $0$ & $0.408843$ & $-\,1.60375$ & $0$ 
& $0$ & $0$\tabularnewline
& $0.667925$ & $0.969147$ & $1.12263$ & $0$ & $0.376136$ & $-\,1.47545$ & $0$ & $0$ & 
$0$\tabularnewline\hline
\multirow{2}{*}{$i=3$} & $-\,0.92068$ & $-\,1.21163$ & $6.45905$ & $2.04389$ & $-\,1.21043$ & 
$-\,6.95338$ & $-\,4.9449$ & $-\,0.851757$ & $3.34115$\tabularnewline
& $0.433093$ & $-\,2.4104$ & $5.5185$ & $1.80867$ & $-\,2.08612$ & $-\,5.57542$ & $-\,4.3034$ & 
$-\,0.720927$ & $2.82795$\tabularnewline
\hline
\multicolumn{10}{|c|}{$n=3$} \\\hline
$i=0$ & $0.825723$ & $0$ & $0$ & $0$ & $0$ & $0$ & $0$ & $0$ & $0$\tabularnewline
$i=1$ & $-\,0.337551$ & $0.825723$ & $0$ & $0$ & $0$ & $0$ & $0$ & $0$ & $0$\tabularnewline\hline
\multirow{2}{*}{$i=2$} & $-\,0.141159$ & $1.02719$ & $1.27299$ & $0$  & $0.703232$ & $-\,1.72026$ & 
$0$ & $0$ & $0$\tabularnewline
& $0.154843$ & $0.912501$ & $1.20418$ & $0$ & $0.646973$ & $-\,1.58264$ & $0$ & $0$ & 
$0$\tabularnewline\hline
\multirow{2}{*}{$i=3$} & $-\,1.03567$ & $-2.65386$ & $6.7324$ & $2.19237$ & $1.67114$ & 
$-\,6.92913$ 
& $-\,5.30412$ & $-\,1.46507$ & $3.58387$\tabularnewline
& $0.902443$ & $-3.82647$ & $5.73411$ & $1.94007$ & $0.222185$ & $-\,5.49342$ & $-\,4.61602$ & 
$-\,1.24003$ & $3.03338$\tabularnewline
\hline
\end{tabular}
\caption{Numerical values for the coefficients of the linearized expansion of the ratios 
of vector moments. We display results for the vector current with $n_f = 4,(5)$ for the 
upper 
(lower) number.\label{tab:Rexporder}}
\end{table}

\begin{table}[tbh!]
\center
\scriptsize
\begin{tabular}{|c|ccccccccc|}
\hline 
 & $[\bar{R}_P]_{n,i}^{0,0}$ & $[\bar{R}_P]_{n,i}^{1,0}$ & $[\bar{R}_P]_{n,i}^{2,0}$ & 
$[\bar{R}_P]_{n,i}^{3,0}$ &
 $[\bar{R}_P]_{n,i}^{0,1}$ & $[\bar{R}_P]_{n,i}^{1,1}$ & $[\bar{R}_P]_{n,i}^{2,1}$ & 
$[\bar{R}_P]_{n,i}^{0,2}$ & $[\bar{R}_P]_{n,i}^{1,2}$\tabularnewline\hline
\multicolumn{10}{|c|}{$n=1$} \\
\hline
$i=0$ & $0.7559290$ & $0$ & $0$ & $0$ & $0$ & $0$ & $0$ & $0$ & $0$\tabularnewline
$i=1$ & $0.0398962$ & $0.755929$ & $0$ & $0$ & $0$ & $0$ & $0$ & $0$ & $0$\tabularnewline
$i=2$ & $2.15656$ & $1.28928$ & $1.16539$ & $0$ & $-\,0.0831172$ & $-\,1.57485$ & $0$ & $0$& 
$0$\tabularnewline
$i=3$ & $4.06725$ & $1.88674$ & $6.70126$ & $2.00706$ & $-\,9.11366$ & $-\,7.79727$ & $-\,4.85579$ 
& 
$0.173161$ & $3.28094$\tabularnewline\hline
\multicolumn{10}{|c|}{$n=2$} \\\hline
$i=0$ & $0.816497$ & $0$ & $0$ & $0$ & $0$ & $0$ & $0$ & $0$ & $0$\tabularnewline
$i=1$ & $-\,0.190991$ & $0.816497$ & $0$ & $0$ & $0$ & $0$ & $0$ & $0$ & $0$\tabularnewline
$i=2$ & $1.27461$ & $1.1585$ & $1.25877$ & $0$ & $0.397898$ & $-\,1.70103$ & $0$ & $0$ & 
$0$\tabularnewline
$i=3$ & $6.15209$ & $-\,0.379067$ & $6.87731$ & $2.16787$ & $-\,4.6981$ & $-\,7.44666$ & 
$-\,5.24486$ & $-\,0.828954$ & $3.54382$\tabularnewline\hline
\multicolumn{10}{|c|}{$n=3$} \\\hline
$i=0$ & $0.852803$ & $0$ & $0$ & $0$ & $0$ & $0$ & $0$ & $0$ & $0$ \tabularnewline
$i=1$ & $-\,0.337775$ & $0.852803$ & $0$ & $0$ & $0$ & $0$ & $0$ & $0$ & $0$\tabularnewline
$i=2$ & $0.574538$ & $1.07172$ & $1.31474$ & $0$ & $0.703697$ & $-\,1.77667$ & $0$ & $0$ & 
$0$\tabularnewline
$i=3$ & $5.94225$ & $-\,1.95744$ & $6.96991$ & $2.26427$ & $-\,1.31021$ & $-\,7.20157$ & 
$-\,5.47807$ & $-\,1.46604$ & $3.7014$\tabularnewline\hline

\end{tabular}
\caption{Numerical values for the coefficients of the linearized expansion of the ratios 
of pseudoscalar moments with $n_f = 4$. \label{tab:RPexporder}}
\end{table}

\begin{table}[tbh!]
\center
\scriptsize
\begin{tabular}{|c|ccccccccc|}
\hline 
 & $[\tilde{R}_V]_{n,i}^{0,0}$ & $[\tilde R_V]_{n,i}^{1,0}$ & $[\tilde R_V]_{n,i}^{2,0}$ & $[\tilde 
R_V]_{n,i}^{3,0}$ & $[\tilde R_V]_{n,i}^{0,1}$ & $[\tilde R_V]_{n,i}^{1,1}$
 & $[\tilde R_V]_{n,i}^{2,1}$ & $[\tilde R_V]_{n,i}^{0,2}$ & $[\tilde 
R_V]_{n,i}^{1,2}$\tabularnewline\hline
\multicolumn{10}{|c|}{$n=1$} \\\hline
$i=0$ & $1$ & $0$ & $0$ & $0$ & $0$ & $0$ & $0$ & $0$ & $0$\tabularnewline
$i=1$ & $0.0160494$ & $1$ & $0$ & $0$ & $0$ & $0$ & $0$ & $0$ & $0$\tabularnewline\hline
\multirow{2}{*}{$i=2$} & $1.86056$ & $3.66883$ & $1.54167$ & $0$ & $-\,0.0334362$ & $-\,2.08333$ & 
$0$ & $0$ & $0$\tabularnewline
& $2.04768$ & $3.52994$ & $1.45833$ & $0$ & $-\,0.0307613$ & $-\,1.91667$ & $0$ & $0$ & 
$0$\tabularnewline\hline
\multirow{2}{*}{$i=3$} & $-\,1.85046$ & $11.8662$ & $12.8916$ & $2.65509$ & $-7.80382$ & $-18.4951$ 
& $-6.42361$ & $0.0696588$ & $4.34028$\tabularnewline
& $-\,0.0174324$ & $9.52394$ & $11.4806$ & $2.34954$ & $-\,7.88822$ & $-\,15.9481$ & $-\,5.59028$ & 
$0.0589592$ & $3.67361$\tabularnewline
\hline
\multicolumn{10}{|c|}{$n=2$} \\\hline
$i=0$ & $1$ & $0$ & $0$ & $0$ & $0$ & $0$ & $0$ & $0$ & $0$\tabularnewline
$i=1$ & $-\,0.254929$ & $1$ & $0$ & $0$ & $0$ & $0$ & $0$ & $0$ & $0$\tabularnewline\hline
\multirow{2}{*}{$i=2$} & $0.0638103$ & $3.39785$ & $1.54167$ & $0$ & $0.531103$ & $-\,2.08333$ & 
$0$ 
& $0$ & $0$\tabularnewline
& $0.357801$ & $3.25896$ & $1.45833$ & $0$ & $0.488615$ & $-\,1.91667$ & $0$ & $0$ & 
$0$\tabularnewline\hline
\multirow{2}{*}{$i=3$} & $-\,2.11686$ & $9.07965$ & $12.4739$ & $2.65509$ & $0.552022$ & 
$-\,17.366$ 
& $-\,6.42361$ & $-\,1.10646$ & $4.34028$\tabularnewline
& $0.322201$ & $6.88187$ & $11.0854$ & $2.34954$ & $-\,0.755491$ & $-\,14.9093$ & $-\,5.59028$ & 
$-\,0.936512$ & $3.67361$\tabularnewline
\hline
\multicolumn{10}{|c|}{$n=3$} \\\hline
$i=0$ & $1$ & $0$ & $0$ & $0$ & $0$ & $0$ & $0$ & $0$ & $0$\tabularnewline
$i=1$ & $-\,0.408795$ & $1$ & $0$ & $0$ & $0$ & $0$ & $0$ & $0$ & $0$\tabularnewline\hline
\multirow{2}{*}{$i=2$} & $-\,0.988542$ & $3.24398$ & $1.54167$ & $0$ & $0.851656$ & $-\,2.08333$ & 
$0$ & $0$ & $0$\tabularnewline
& $-\,0.630066$ & $3.10509$ & $1.45833$ & $0$  & $0.783523$ & $-\,1.91667$ & $0$ & $0$ & 
$0$\tabularnewline\hline
\multirow{2}{*}{$i=3$} & $-\,5.11183$ & $7.46526$ & $12.2367$ & $2.65509$ & $5.43048$ & 
$-\,16.7249$ 
& $-\,6.42361$ & $-\,1.77428$ & $4.34028$\tabularnewline
& $-\,1.87845$ & $5.35334$ & $10.861$ & $2.34954$ & $3.40317$ & $-\,14.3195$ & $-\,5.59028$ & 
$-\,1.50175$ & $3.67361$\tabularnewline
\hline
\end{tabular}
\caption{Numerical values for the coefficients of the iterative linearized expansion of the ratios 
of vector moments. We display results for the vector current with $n_f = 4,(5)$ for the upper 
(lower) number.\label{tab:RITorder}}
\end{table}

\clearpage

\begin{table}[tbh!]
\scriptsize
\begin{tabular}{|c|ccccccccc|}
\hline 
 & $[\tilde{R}_P]_{n,i}^{0,0}$ & $[\tilde{R}_P]_{n,i}^{1,0}$ & $[\tilde{R}_P]_{n,i}^{2,0}$ & 
$[\tilde{R}_P]_{n,i}^{3,0}$ &
 $[\tilde{R}_P]_{n,i}^{0,1}$ & $[\tilde{R}_P]_{n,i}^{1,1}$ & $[\tilde{R}_P]_{n,i}^{2,1}$ & 
$[\tilde{R}_P]_{n,i}^{0,2}$ & $[\tilde{R}_P]_{n,i}^{1,2}$\tabularnewline\hline
\multicolumn{10}{|c|}{$n=1$} \\
\hline
$i=0$ & $1$ & $0$ & $0$ & $0$ & $0$ & $0$ & $0$ & $0$ & $0$\tabularnewline
$i=1$ & $0.0527778$ & $1$ & $0$ & $0$ & $0$ & $0$ & $0$ & $0$ & $0$\tabularnewline
$i=2$ & $2.95841$ & $3.70556$ & $1.54167$ & $0$ & $-\,0.109954$ & $-\,2.08333$ & $0$ & $0$ & 
$0$\tabularnewline
$i=3$ & $11.4629$ & $13.0982$ & $12.9483$ & $2.65509$ & $-\,12.4961$ & $-\,18.6481$ & $-\,6.42361$ 
& 
$0.22907$ & $4.34028$\tabularnewline\hline
\multicolumn{10}{|c|}{$n=2$} \\\hline
$i=0$ & $1$ & $0$ & $0$ & $0$ & $0$ & $0$ & $0$ & $0$ & $0$\tabularnewline
$i=1$ & $-\,0.233915$ & $1$ & $0$ & $0$ & $0$ & $0$ & $0$ & $0$ & $0$\tabularnewline
$i=2$ & $1.09324$ & $3.41886$ & $1.54167$ & $0$ & $0.487324$ & $-\,2.08333$ & $0$ & $0$ & 
$0$\tabularnewline
$i=3$ & $8.77473$ & $10.1858$ & $12.5063$ & $2.65509$ & $-\,3.80468$ & $-\,17.4536$ & $-\,6.42361$ 
& 
$-\,1.01526$ & $4.34028$ \tabularnewline\hline
\multicolumn{10}{|c|}{$n=3$} \\\hline
$i=0$ & $1$ & $0$ & $0$ & $0$ & $0$ & $0$ & $0$ & $0$ & $0$ \tabularnewline
$i=1$ & $-\,0.3960760$ & $1$ & $0$ & $0$ & $0$ & $0$ & $0$ & $0$ & $0$\tabularnewline
$i=2$ & $-\,0.118446$ & $3.2567$ & $1.54167$ & $0$ & $0.825158$ & $-\,2.08333$ & $0$ & $0$ & 
$0$\tabularnewline
$i=3$ & $4.92499$ & $8.38182$ & $12.2563$ & $2.65509$ & $1.76427$ & $-\,16.7779$ & $-\,6.42361$ & 
$-\,1.71908$ & $4.34028$\tabularnewline\hline

\end{tabular}
\caption{Numerical values for the coefficients of the iterative linearized expansion of the ratios 
of pseudoscalar moments with $n_f = 4$.\label{tab:RPITorder}}
\end{table}

\bibliography{../charm2}
\bibliographystyle{JHEP}

\end{document}